\documentclass[12pt]{article}

\usepackage{axodraw}
\evensidemargin=\oddsidemargin
\newcommand{\bc}{\begin{center}}
\newcommand{\ec}{\end{center}}

\def\ra{\rightarrow}
\def\la{\leftarrow}
\def\da{\downarrow}

\def\beq{\begin{equation}}
\def\eeq{\end{equation}}
\def\bet{\begin{table}}
\def\eet{\end{table}}
\def\bea{\begin{eqnarray}}
\def\eea{\end{eqnarray}}
\def\beas{\begin{eqnarray*}}
\def\eeas{\end{eqnarray*}}
\def\ba{\begin{array}}
\def\ea{\end{array}}
\def\lrover#1{\raisebox{1.6ex}{\rlap{$\leftrightarrow$}} \raisebox{ 0ex}{$#1$}}

\begin{document}

\begin{titlepage}
\date{November 6, 1995}
\title{
\hfill \mbox{\large KA-TP-8-1995}\\[-2mm]
\hfill \mbox{\large hep-ph/9511250}\\[3mm]
Complete set of Feynman rules for the MSSM -- {\em erratum} }
\author{Janusz Rosiek\thanks{e-mail: janusz.rosiek@fuw.edu.pl}\\
Institute of Theoretical Physics, Warsaw University\\ Ho{\.z}a 69,
00-681 Warsaw, POLAND}
\maketitle

\thispagestyle{empty}

\begin{abstract}
{\small This erratum contains the full corrected version of the paper
{\em Complete set of Feynman rules for the Minimal Supersymmetric
Standard Model}~\cite{PRD41}. The complete set of Feynman rules for
the R-parity conserving MSSM is listed, including the most general
form of flavour mixing.  Propagators and vertices are computed in
t'Hooft-Feynman gauge, convenient for perturbative calculations beyond
the tree level.}
\end{abstract}
\end{titlepage}

\setcounter{page}{1}

Instead of putting on the web next version of the {\em ``erratum''},
with few more errors in the original paper corrected, I decided to
resubmit the ``integrated'' version, i.e. full paper text with all
necessary corrections included - it should be easier to use in this
way. I also used this opportunity to correct the most irritating
features of the notation used in the original {\em Phys. Rev.} {\bf D}
paper, making it, wherever possible, closer to commonly used naming
conventions.  However, I kept unchanged the ``final'' matrix notation
for the interactions vertices in the mass eigenstates basis, as it
proved to be very useful in compactifying many complicated loop
calculations.

Most of the expressions for mass matrices, mixing angles and vertices
listed in~\cite{PRD41} have been checked during the calculations of
the 1-loop radiative corrections in the gauge and Higgs sectors of the
MSSM~\cite{CPR,DELTAR} and in calculations of various CP
violation/FCNC processes~\cite{POROSA,MIPORO,BCRS} The 1-loop
corrections were calculated in on-shell renormalization scheme, which
provide a very strict test of correctness of all formulae entering the
expressions for the renormalized quantities: most of the errors in
Feynman rules lead immediately to non-cancellation of the
divergencies.  Only the most exotic vertices like 4-sfermion
couplings, several rarely used 2 Higgs boson-2 sfermion couplings were
not used and did not pass this test yet.  Other vertices can be with
good probability considered as checked.

Formulae for diagonalization of mass matrices and most of the vertices
listed in~\cite{PRD41} are accessible also as the ready FORTRAN codes.
They are part of the bigger library for calculation of the 1-loop
radiative corrections in on-shell renormalization scheme to the MSSM
neutral Higgs production and decay rates.  This library can be found
at:\\[2mm]
{\sl http://www.fuw.edu.pl/\mbox{\hskip
-3pt}\raisebox{-6pt}{\large\~}rosiek/physics/neutral\_higgs.html}

\vskip 2mm

\noindent {\bf In order to avoid too many replacements in hep-ph
archive, and to speed up the process of introducing further
corrections if any were found, I will always put the most recent version of
this collection of Feynman rules on my private web page. It will be
available at:}\\[2mm]
{\sl http://www.fuw.edu.pl/\mbox{\hskip
-3pt}\raisebox{-6pt}{\large\~}rosiek/physics/prd41.html}\\[2mm]
{\bf I will not update the hep-ph version of the {\em erratum} any
more!}

\vskip 5mm
\noindent {\bf Acknowledgments}
\vskip 5mm

\noindent I wish to thank to P. Chankowski, A. Belayev, J. R. Espinosa,
J. Hagelin, M.Laine, M. Misiak, P. Slavich, M. Rauch and others who
called my attention to some of the errors listed in this erratum.
Especially, I would like to thank T. Ewerth for very careful reading of 
its integrated version.

\newpage

\begin{center}
{\large \bf Complete set of Feynman rules for the MSSM}

\vskip 1mm

{\large published in {\em Phys. Rev.} {\bf D41} (1990) 3464}

\vskip 5mm

{\large Janusz Rosiek}

\vskip 5mm

Institute of Theoretical Physics, Warsaw University\\ 
Ho{\.z}a 69, 00-681 Warsaw, POLAND

\begin{abstract}
{\small The complete set of Feynman rules for the $R$-parity
conserving MSSM is listed, including the most general form of flavour
mixing.  Propagators and vertices are computed in t'Hooft-Feynman
gauge, convenient for perturbative calculations beyond the tree
level.}
\end{abstract}
\end{center}

\setcounter{page}{1}

\section{Introduction}
\label{sec:intro}

Since many years the minimal supersymmetric extension of the standard
model (MSSM) has been the subject of intensive studies.  Practical
calculations in the MSSM are usually tedious because of its
complexity.  For easier reference, in this paper we complete all
Feynman rules for the MSSM in the t'Hooft-Feynman gauge, convenient
for calculations of loop corrections.  We put special emphasis on
including the most general form of flavour mixing allowed in the
$R$-parity conserving MSSM.

Some of the mass matrices and vertices given in the paper, mainly
related to the Higgs and -inos sectors, has been listed in other
papers (see e.g.~\cite{ref1,ref2}).  For completeness, we write down
those formulae once again, in order to collect in one place and fixed
convention the full set of rules needed to calculate any process in
the frame of the MSSM.  The spinor conventions used in the paper
follow those given in~\cite{ref1}.

The paper has the following structure.  Section~\ref{sec:general}
contains short review of the rules of constructing SUSY Yang-Mills
theories.  In section~\ref{sec:fields} we define fields and parameters
present in the MSSM Lagrangian.  The physical content of the theory -
the mass eigenstates fields are given in section~\ref{sec:eigen}.  In
section~\ref{sec:gauge} we define the gauge used throughout the paper.
Section~\ref{sec:lagr} contains the MSSM Lagrangian expressed in terms
of the physical fields.  A short summary and comments on the choice of
the parameters of the model are given in section~\ref{sec:comment}.
In~\ref{sec:app_a} we write down the MSSM Lagrangian in terms of the
initial fields, before the $SU(2)\times U(1)$ symmetry breaking.
\ref{sec:app_b} contains the full set of Feynman rules 
corresponding to the Lagrangian of section~\ref{sec:lagr}.

\section{General structure of the SUSY models}
\label{sec:general}

Supersymmetric Yang-Mills theories contain two basic types of fields -
gauge multiplets $(\lambda^{a},V_{{\mu}}^{a})$ in the adjoint
representation of a gauge group $G$ and matter multiplets
$(A_i,\psi_i)$ in some chosen representations of $G$.  By $\lambda$
and $\psi$ we denote here fermions in two-component notation, $A_i$
are complex scalar fields and $V_{\mu}^a$ are spin-1 real vector
fields (the spinor notation and conventions used in the paper are the
same as those explained in Appendix~A of ref.~\cite{ref1}).  To
construct the Lagrangian of such a theory we follow the rules given in
ref.~\cite{ref1}.  In the strictly supersymmetric case one has the
following terms (summation convention is used unless stated
otherwise):
\begin{enumerate}
\item Kinetic terms.
\item Self interaction of gauge multiplets:  three- and four-gauge
boson vertices plus additional interaction of gauginos and gauge
fields:
\beas
igf_{abc} \lambda^{a} \sigma^{\mu} \bar{\lambda}^b V_{\mu}^c
\eeas
\item Interactions of the gauge and matter multiplets ($T^{a}$ is the
Hermitian group generator in the representation corresponding to the
given multiplet):
\beas
&-gT_{ij}^{a} V_{{\mu}}^{a} (\bar{{\psi}}_{i} \bar{{\sigma}}^{{\mu}}
{\psi}_{j} +iA_{i}^{\star} {\partial}_{{\mu}} A_{j}),&
\\
&ig \sqrt{2}T_{ij}^{a} ({\lambda}^{a} {\psi}_{j} A_{i}^{\star}
-\bar{{\lambda}}^{a} \bar{{\psi}}_{i} A_{j}),&
\\
&g^{2} (T^{a} T^{b})_{ij} V_{{\mu}}^{a} V^{{\mu}b} A_{i}^{\star} A_{j}.&
\eeas
\item Self interactions of the matter multiplets.  For technical
reasons it is convenient to define the so called ``superpotential''
$W$ as at most cubic gauge-invariant polynomial which depends on
scalar fields $A_i$, but not on $A_i^{\star}$~\cite{ref1}
(alternatively, the superpotential can be defined as a function of the
superfields).  Introducing two auxiliary functions:
\beas
&F_i = {\partial}W/{\partial}A_i&\\[2mm]
&D^a = g A_i^{\star} T_{ij}^a A_j&
\eeas
one can write the scalar supersymmetric potential as:
\beas
V = \frac{1}{2} D^a D^a + F_i^{\star} F_i
\eeas
Yukawa interactions are given by:
\beas
- \frac{1}{2}\left( {\partial^{2} W\over \partial A_i \partial A_j}
\psi_i\psi_j + \mathrm{H.c.}\right)
\eeas
\end{enumerate}

In the case of semisimple groups $G=G_1 \times \ldots \times G_n$ in
the expressions above one should substitute terms of the form $gVT$ by
sums $\Sigma g_i V_i T_i$ and similarly for gauginos.  There are also
$n$ vertices $\lambda-\bar\lambda-V$ and $n$ terms $\frac{1}{2}D^2$ in
the scalar potential.  For $U(1)$ factors there are no
gaugino-gaugino-gauge interaction, and, by convention, the product $g
T_{ij}^a V_{\mu}^a$ is replaced by $\frac{1}{2}g y_i \delta_{ij}
V_{\mu} $ (no sum over $i$), where $y_i$ is the $U(1)$ quantum number
of the matter multiplet $(A_i,\psi_i)$ (similarly for gauginos).  In
general the $U(1)$ $D$-field may be shifted by the so called
Fayet-Iliopoulos term $\xi$~\cite{ref4}:
\beas
D = \frac{1}{2} g y_i A_i^{\star} A_i + {\xi}
\eeas
$\xi\neq 0$ may introduce dangerous quadratic divergences into the
theory.  In most realistic models (including MSSM) this term is
absent.

This completes the construction of a strictly supersymmetric theories.
To build models which preserve the most important feature of such
theories - absence of quadratic divergences - and which simultaneously
are experimentally acceptable, it is necessary to add to the above
Lagrangian explicit soft SUSY breaking terms.  The most general form
of appropriate expressions can be written down as~\cite{ref5}:
\beas
m_1 \Re A^2 + m_2 \Im A^2 + y(A^3 + \mathrm{ H.c.}) + m_3 (\lambda^{a}
\lambda^{a} + \mathrm{ H.c.})
\eeas
$A^2$ and $A^3$ denote symbolically all possible gauge invariant
combinations of scalar fields.  These terms split the masses of
scalars and fermions present in the SUSY multiplets and introduce new,
non-supersymmetric trilinear scalar couplings.

\section{MSSM field and coupling structure}
\label{sec:fields}

To obtain the realistic supersymmetric version of the Standard Model
one should extend the field content of the theory by adding
appropriate scalar or fermionic partners to the ordinary matter and
gauge fields.  As stated in the previous section, the superpotential
can only be constructed as a function of fields and not of their
complex conjugates.  Therefore it is not possible to give masses to
all fermions using only one Higgs doublet - at least two with opposite
$U(1)$ quantum numbers are necessary.  The full field content of the
MSSM is listed below:
\begin{enumerate}
\item  Multiplets of the gauge group $SU(3) \times SU(2) \times U(1)$:
\begin{itemize}
\item $B_{\mu},\lambda_B$ - weak hypercharge gauge fields,
coupling constant $g_1$
\item $A_{\mu}^i,\lambda_A^i$ -  weak isospin gauge
fields, coupling constant $g_2$
\item $G_{\mu}^a,\lambda_G^a$ - QCD gauge fields, coupling 
constant $g_3$
\end{itemize}
\item  Matter multiplets -  we assume that three matter
generations exist, so the index $I$ (and similarly all capital
$I,J,K\ldots$ indices in the rest of the paper) runs from $1$ to $3$
(such notation can be immediately generalized to the case of $N$
generations).
\end{enumerate}

\begin{table}[htbp]
\begin{center}
\begin{tabular}{llc}
Scalars & Fermions &  $U(1)$ charge\\[3mm]
$ L^I =  \left(
\begin{array}{c}
  \tilde{\nu}^I \\
  \tilde{e}_L^{-I} 
\end{array} 
\right)$
&
$\Psi_L^I =  \left(
\begin{array}{c}
  \nu^I \\
  e^{-I} 
\end{array} 
\right)_L$ 
& $-1$ \\[3mm]
$R^I = \tilde{e}_R^{+I}$
&
$\Psi_R^I = (e_L^{-I})^c$
& $2$ \\[3mm]
$ Q^I =  \left(
\begin{array}{c}
  \tilde{u}^{I}  \\
  \tilde{d}_{L}^{I}  
\end{array}
\right) $
& 
$ \Psi_Q^I =  \left(
\begin{array}{c}
  u^{I}  \\
  d^{I}  
\end{array}
\right)_L $
& $\frac{1}{3}$ \\[3mm]
$D^I = \tilde{d}_R^{I\star}$ 
& 
$\Psi_D^I = (d_{L}^{I})^c $ 
&  
$\frac{2}{3}$ \\[3mm]
$ U^I = \tilde{u}_{R}^{I\star}$
&
$\Psi_U^I = (u_{L}^{I})^c $
& $-\frac{4}{3}$ \\[3mm]
$ H^1 =  \left(
\begin{array}{c}
  H^{1}_{1}  \\
  H_{2}^{1}  
\end{array}
\right)$ 
&
$\Psi_H^1 =  \left(
\begin{array}{c}
  {\Psi}^1_{H1} \\
  {\Psi}^1_{H2} 
\end{array}
\right)$
& $-1$ \\[3mm]
$ H^2 =  \left(
\begin{array}{c}
  H^{2}_{1}  \\
  H_{2}^{2}  
\end{array}
\right)$
&
$ \Psi_H^2 =  \left(
\begin{array}{c}
  {\Psi}^2_{H1} \\
  {\Psi}^2_{H2} 
\end{array}
\right) $
& $1$ \\
\end{tabular}
\end{center}
\end{table}

The $SU(3)$ indices are not written explicitly.  We assume that $Q$
quarks and squarks are QCD triplets, $D$ and $U$ fields - QCD
antitriplets.

In order to define the theory we have to write down the superpotential
and introduce the soft SUSY breaking terms (without them even using
two Higgs doublets it is impossible to break spontaneously the gauge
symmetry).  The most general form of the superpotential which does not
violate gauge invariance and the SM conservation laws is:
\beas
W = \mu\epsilon_{ij} H^1_i H^2_j + \epsilon_{ij} Y_l^{IJ} H^1_i L_j^I
R^J + \epsilon_{ij} Y_d^{IJ} H^1_i Q_j^I D^J + \epsilon_{ij} Y_u^{IJ}
H^2_i Q_j^I U^J
\eeas
Some bilinear and trilinear terms (e.g. $\varepsilon_L^I
{\epsilon}_{ij} H_i^2 L_j^I$) are gauge invariant but break lepton
and/or barion number conservation, so we do not include them.  In
general, presence of such terms can be forbidden by requiring the
preservation of additional global symmetry of the model, so called
$R$-parity (for review see e.g.~\cite{rparity}):
\beas
R= (-1)^{L+3B+2S}
\eeas
Soft breaking terms can be divided into several classes:
\begin{enumerate}
\item  Mass terms for the scalar fields.
\beas
&- m_{H_1}^2 H_i^{1\star} H_i^1 - m_{H_2}^2 H_i^{2\star} H_i^2 -
(m_L^2)^{IJ} L_i^{I\star} L_i^J - (m_R^2)^{IJ} R^{I\star} R^J&
\\
&- (m_Q^2)^{IJ} Q_i^{I\star} Q_i^J - (m_D^2)^{IJ} D^{I\star} D^J -
(m_U^2)^{IJ} U^{I\star} U^J &
\eeas
\item  Mass terms for gauginos.
\beas
\frac{1}{2} M_1 \lambda_B \lambda_B  + \frac{1}{2} M_2 \lambda_A^i
\lambda_A^i  + \frac{1}{2} M_3 \lambda_G^a \lambda_G^a
+ \mathrm{H.c.}
\eeas
\item  Trilinear couplings of the scalar fields corresponding to the
Yukawa terms in the superpotential.
\beas
m_{12}^2 \epsilon_{ij} H_i^1 H_j^2 + \epsilon_{ij} A_l^{IJ} H_i^1
L_j^I R^J + \epsilon_{ij} A_d^{IJ} H_i^1 Q_j^I D^J +
\epsilon_{ij} A_u^{IJ} H_i^2 Q_j^I U^J + \mathrm{H.c.}
\eeas
\item  Trilinear couplings of the scalar fields,  different
in the from from the Yukawa terms in the superpotential (sometimes
called ``non-analytic terms'' as they involve charge conjugated Higgs
fields).  Usually such couplings are not considered as they are not
generated in the most popular SUSY-breaking models.
\beas
A_l^{'IJ} H_i^{2\star} L_i^I R^J + A_d^{'IJ} H_i^{2\star} Q_i^I D^J +
A_u^{'IJ} H_i^{1\star} Q_i^I U^J + \mathrm{H.c.}
\eeas
\end{enumerate}

In general constants $\mu$, $m_{12}^2$, Yukawa matrices, squark and
gaugino masses and the trilinear soft couplings may be complex.  One
can perform three operations which eliminate unphysical degrees of
freedom.  First, it is possible to change globally the phase of one of
the Higgs multiplets in such a way, that the constant $m_{12}^2$
becomes a real number.  Then the equations for the vacuum expectations
values of the Higgs fields involve only real parameters (at least on
the tree level).  Second, one can redefine simultaneously the phases
of all fermions in the model removing the complex phase from one of
gaugino mass parameter (see e.g.~\cite{POROSA}).  Paralelly one should
redefine the phases of other couplings to absorb the change of the
phase of the Higgs multiplet.  The last operation is the same as in
the SM - by the rotation of the fields
\begin{center}
\begin{tabular}{lp{2cm}c}
$(Q_i^I, \Psi_{Qi}^I) \ra V_{Qi}^{IJ} (Q_i^J,
\Psi_{Qi}^J)$ && (no sum over $i$)\\[3mm]
$(U^I,\Psi_U^I) \ra V_U^{IJ} (U^J,\Psi_U^J)$&&\\[3mm]
$(D^I,\Psi_D^I) \ra V_D^{IJ} (D^J,\Psi_D^J)$&&\\[3mm]
\end{tabular}
\end{center}
and similarly for leptons, one can diagonalize the matrices
$Y_l^{IJ}$, $Y_u^{IJ}$ and $Y_d^{IJ}$ obtaining Yukawa couplings of
the form $ {\epsilon}_{ij} Y_l^{I} H_{i}^{1} L_{j}^{I} R^{I} $ etc.
Simultaneous rotations of quark and squark fields lead to the
so-called super-KM basis (see e.g.~\cite{MIPORO}).  After the proper
redefinition of the parameters the matrices $V_Q$, $V_U$, $V_D$, $V_L$
and $V_R$ disappear from the Lagrangian leaving as their trace the
Kobayashi-Maskawa matrix $K$, appearing in many expressions containing
both quarks and squarks:
\beas
 K = V_{Q_1}^{\dagger} V_{Q_2}
\eeas
Of course, sfermion mass matrices in the super-KM basis may be still
non-diagonal, i.e. Yukawa and soft matrices need not to be
diagonalized by the same rotations.

In the rest of the paper, in particular in section~\ref{sec:eigen} and
in~\ref{sec:app_a} we use the {\em rotated} soft breaking parameters.
For example, if the initial Yukawa and trilinear scalar couplings were
$Y_u^{(0)}$ and $A_u^{(0)}$, respectively, the rotation of the quark
and squark fields to the super-KM basis lead to diagonal Yukawa
coupling $Y_u = V^{\dagger}_{Q1}Y_u^{(0)}V_U$ and new trilinear scalar
coupling $A_u = V^{\dagger}_{Q1}A_u^{(0)} V_U$, which are then used in
other expressions.  The procedure of the redefinition of the squark
mass matrices remains some freedom. We have chosen the redefined left
squark mass matrix parameter in such a way that the Kobayashi-Maskawa
matrix multiplies $m_Q^2$ in the up-squark mass matrix.

The full Lagrangian written in terms of the initial fields (before
$SU(2)\times U(1)$ symmetry breaking) is given in~\ref{sec:app_a}.

\section{Physical spectrum of the MSSM}
\label{sec:eigen}

In the previous section we defined the field content and all
parameters of the MSSM.  To obtain the physical spectrum of particles
present in the theory one should carry out the standard procedure of
gauge symmetry breaking via vacuum expectation values of the neutral
Higgs fields and find the eigenstates of the mass matrices for all
fields.  The VEV's of the Higgs fields satisfy the following equations
($\theta$ denotes the Weinberg angle, $s_W=\sin\theta$,
$c_W=\cos\theta$, $e = g_2 s_W = g_1 c_W$):
\beas
 <H^1> = {1\over \sqrt{2}}  \left(
\begin{array}{c}
  v_1  \\
  0
\end{array}
\right) 
\hskip 2cm
 <H^2> = {1\over \sqrt{2}}  \left(
\begin{array}{c}
  0\\
  v_2  
\end{array}
\right)
\eeas
\beas
\left[ {e^2 \over 8 s_W^2 c_W^2} \left(v_1^2 - v_2^2\right) 
+ m_{H_1}^2 + |\mu|^2 \right] v_1 &=& - m_{12}^2 v_2
\\
\left[ - {e^2 \over 8 s_W^2 c_W^2} \left(v_1^2 - v_2^2\right) 
+ m_{H_2}^2 + |\mu|^2 \right] v_2 &=& - m_{12}^2 v_1
\eeas
Parameters of the above set of equations are constrained by the
condition that $v_1$ and $v_2$ should reproduce the proper values of
the gauge boson masses.

\vskip 1mm

\noindent The physical fields of the MSSM can be identified as follows:
\begin{enumerate}
\item Gauge bosons. Eight gluons  $g_{\mu}^a$ and the photon
$F_{\mu}$ are massless, bosons $W_{\mu}^{\pm}$ and $Z_{\mu}$ have
masses
\beas
M_Z &=& {e \over 2 s_W c_W} \left(v_1^2 +v_2^2\right)^{\frac{1}{2}}
\\
M_W &=& {e \over 2 s_W} \left(v_1^2 +v_2^2\right)^{\frac{1}{2}}
\eeas
\item Charged Higgs scalars. Four charged Higgs scalars exist, two of
them with the mass
\beas
M_{H_1^{\pm}}^2 = M_W^{2} + m_{H_1}^2 + m_{H_2}^2 + 2|\mu|^2
\eeas
and the other two massless.  In the physical (unitary) gauge
$H_2^{\pm} (\equiv G^{\pm})$ are eaten by $W$ bosons and disappear
from the Lagrangian.  Fields $H_1^+$ and $H_2^+$ are related to the
initial Higgs fields by the rotation matrix $Z_H$ :
\beas
  \left(
\begin{array}{c}
  H_2^{1\star}  \\
  H_1^2
\end{array}
\right) = Z_H  \left(
\begin{array}{c}
  H_1^+  \\
  H_2^+  
\end{array}
\right)
\\
 Z_{H} = \left(v_1^2+v_2^2\right)^{-\frac{1}{2} } \left(
\begin{array}{rr}
  v_2 &-v_1  \\
  v_1 &v_2  
\end{array}
\right)
\eeas
\item  Neutral Higgs scalars.  If the Lagrangian contains only real
parameters the neutral Higgs particles have well defined CP
eigenvalues - two of them are scalars, the other two are
pseudoscalars.  This is no longer true if some parameters are complex.
Nevertheless, in both cases it is convenient to divide neutral Higgses
into two classes.
\begin{itemize}
\item[i)] ``Scalar'' particles  $H_i^0$ , $i=1,2 $, defined as:
\begin{center}
 $\sqrt{2} \Re H_i^i = Z_R^{ij} H_j^0 + v_i$ (no sum over $i$)
\end{center}
The matrix $Z_R$ and the masses of $H_i^0$ can be obtained by
diagonalizing the $M_R^2$ matrix:
\beas
Z_R^T 
\left(
\begin{array}{cc}
- m_{12}^2 {v_2\over v_1} + {e^2 v_1^2 \over 4 s_W^2 c_W^2} 
& 
m_{12}^2 - {e^2 v_1 v_2 \over 4 s_W^2 c_W^2} 
\\ 
m_{12}^2 - {e^2 v_1 v_2 \over 4 s_W^2 c_W^2} 
& 
- m_{12}^2 {v_1\over v_2} + {e^2 v_2^2 \over 4 s_W^2 c_W^2}
\end{array}
\right)
Z_R =  \left(
\begin{array}{cc}
  M^2_{H^0_1}  &0\\
  0 & M^2_{H^0_2}    
\end{array}
\right)
\eeas
\item[ii)] ``Pseudoscalar'' particles  $A_i^0$ , $i=1,2$:
\begin{center}
$ \sqrt{2} \Im H_i^i = Z_H^{ij} A_j^0$ (no sum over $i$)
\end{center}
$A_1^0 (\equiv A^0)$ has the mass $M_A^2 = m_{H_1}^2 + m_{H_2}^2 +
2|\mu|^2$, $A_2^0 (\equiv G^0)$ is the massless Goldstone boson which
disappears in the unitary gauge.  The $Z_H$ matrix is the same as in
the case of the charged Higgs bosons.
\end{itemize}

Matrix notation used in the paper is convenient in the case of
non-unitary gauge, when the Goldstone bosons are explicitly present in
the Lagrangian and enter the calculations together with the physical
Higgs particles.  In order to compare our expressions with the more
commonly used notation one should substitute:
\beas
 Z_H =  \left(
\begin{array}{cc}
    {\sin}{\beta}&-  {\cos}{\beta}\\
    {\cos}{\beta}&  {\sin}{\beta}
\end{array}
\right) 
\hskip 1cm
Z_R =  \left(
\begin{array}{cc}
    {\cos}{\alpha}&-  {\sin}{\alpha}\\
    {\sin}{\alpha}&  {\cos}{\alpha}
\end{array}
\right)
\eeas
where:
\beas
\tan\beta &=& {v_2\over v_1}\\
\tan 2\alpha &=& \tan 2\beta {M^2_A + M_Z^2 \over M_A^2 - M_Z^2}\\
\eeas
It is also worth remembering that due to the SUSY structure of the
model the Higgs boson masses fulfill two interesting tree-level
relations:
\beas
 M_{H_1^+}^2 &=& M_A^2 + M_W^2
\\
 M_{H_1^0}^2 + M_{H_2^0}^2 &=& M_A^2 + M_Z^2
\eeas
\item Matter fermions (quarks and leptons) have masses (note that
$Y_l^I, Y_d^I$ are defined as negative):
\beas
m_{\nu}^I = 0\phantom{xxx} &\hskip 2cm &  
m_e^I = - {v_1 Y_l^I \over \sqrt{2}} 
\\
m_d^I = - {v_1 Y_d^I\over \sqrt{2}}   &\hskip 2cm & 
m_u^I = {v_2 Y_u^I\over \sqrt{2}} 
\eeas
\item  Charginos.  Four  2-component spinors $(\lambda_A^1,
\lambda_A^2, \Psi_{H2}^1, \Psi_{H1}^2)$ combine to give two 
4-component Dirac fermions $\chi_1, \chi_2$ corresponding to two
physical charginos.  The chargino mixing matrices $Z_+$ and $Z_-$ are
defined by the condition:
\beas
(Z_-)^T 
\left(
\begin{array}{cc}
 M_2 & {ev_2 \over \sqrt{2}s_W}
\\  {ev_1 \over \sqrt{2}s_W} & \mu
\end{array}
\right)
Z_+ =  \left(
\begin{array}{cc}
  M_{\chi_1} & 0
\\
  0 & M_{\chi_2}  
\end{array}
\right)
\eeas
The unitary matrices $Z_-,Z_+$ are not uniquely specified - by
changing their relative phases and the ordering of the eigenvalues it
is possible to choose $M_{\chi_i}$ to be positive and $M_{\chi_2} >
M_{\chi_1}$.  The fields $\chi_i$ are related to the initial spinors
as below:
\beas
{\Psi}_{H1}^2 &=& Z_+^{2i} \kappa_i^+
\\
{\Psi}_{H2}^1 &=& Z_-^{2i} \kappa_i^- 
\hskip 3cm
{\chi}_{i} =  \left(
\begin{array}{c}
  {\kappa}^{+}_{i}  
\\
  \bar{\kappa}_{i}^{-}  
\end{array}
\right)
\\
\lambda_A^{\pm} &\equiv& {\lambda^1_A \mp i\lambda^2_A
\over \sqrt{2}} = iZ_{\pm}^{1i} \kappa_i^{\pm}
\eeas
\item  Four  2-component spinors $(\lambda_B, \lambda_A^3,
\Psi_{H1}^1, \Psi_{H2}^2)$ combine into four
Majorana fermions $\chi_i^0$, $i =1 \ldots 4 $, called neutralinos.
The formulas for mixing and mass matrices are the following:
\beas
Z_N^T
\left(
\begin{array}{cccc}
  M_1 & 0 & {-ev_{1} \over 2c_W} & {ev_{2} \over
  2c_W} \\ 
  0 & M_2 & {ev_{1} \over 2s_W} & {-ev_{2} \over 
  2s_W} \\ 
  {-ev_{1} \over 2c_W} & {ev_{1} \over 2s_W} & 0
  & -\mu \\ 
  {ev_{2} \over 2c_W} & {-ev_{2} \over 2s_W} &-\mu
  & 0
\end{array}
\right)
Z_N &=& \left(
\begin{array}{ccc}
  M_{\chi^0_1}  &&0\\
  &{\ddots}&\\
  0&& M_{\chi^0_4} 
\end{array}
\right)
\eeas
\beas
\lambda_B &=& iZ_N^{1i} {\kappa}_i^0 \\
\lambda_A^3 &=& iZ_N^{2i} {\kappa}_i^0\\
{\Psi}_{H1}^1 &=& Z_N^{3i} {\kappa}_i^0 
\hskip 2cm
{\chi}_{i}^{0} = \left(
\begin{array}{c}
  {\kappa}_{i}^{0}  \\
  \bar{{\kappa}}_{i}^{0}  
\end{array}
\right)\\ 
{\Psi}_{H2}^{2} &=& Z_N^{4i} {\kappa}_{i}^{0}
\eeas
\item  $SU(3)$ gauginos do not mix.  In four component notation one has
eight gluinos $\Lambda_G^a$ with masses $|M_3|$.
\beas
\Lambda_G^a =  \left(
\begin{array}{r}
  -i\lambda^a_G  \\
  i\bar{\lambda}_G^a  
\end{array}
\right)
\eeas
\item  Three complex scalar fields  $L_1^I$ form three sneutrino mass
eigenstates $\tilde\nu^I$ with masses given by diagonalization of a
matrix ${\cal M}_{\nu}^2$ :
\beas
 L_1^I &=& Z_{\nu}^{IJ} {\tilde\nu}^J
\\
 Z_{\nu}^{\dagger} {\cal M}_{\nu}^2 Z_{\nu} &=&  \left(
\begin{array}{ccc}
  M_{\nu_1}^2 &&0\\
  &{\ddots}&\\
  0&&M_{\nu_3}^2 
\end{array}
\right)
\\
{\cal M}_{\nu}^2 &=& {e^2(v^2_1-v^2_2) \over 8 s_W^2 c_W^2} \hat 1 +
m_L^2
\eeas
Sneutrinos are neutral but {\em complex} scalars.
\item  Fields $L_2^I$ and $R^I$ mix to give six charged selectrons
$L_i$, $i=1 \ldots 6$:
\beas
L_2^I = Z_L^{Ii\star} L^-_i \hskip 2cm R^I = Z_L^{(I+3)i} L^+_i
\eeas
\beas
 Z_L^{\dagger} \left(
\begin{array}{cc}
  \left({\cal M}_L^2\right)_{LL} & \left({\cal M}_L^2\right)_{LR} \\
  \left({\cal M}_L^2\right)_{LR}^{\dagger} & \left({\cal M}_L^2\right)_{RR} 
\end{array}
\right)Z_L =  \left(
\begin{array}{ccc}
  M_{L_1}^2 &&0\\
  &{\ddots}&\\
  0&&M_{L_6}^2 
\end{array}
\right)
\eeas
\beas
\left({\cal M}_L^2\right)_{LL} &=& {e^2(v_1^2-v_2^2)(1-2c_W^2) \over 
8 s_W^2 c_W^2} {\hat 1} + {v_1^2 Y_l^2\over 2} + (m_L^2)^T
\\
\left({\cal M}_L^2\right)_{RR} &=& - {e^2(v_1^2-v_2^2) \over 4c_W^2}
{\hat 1} + {v_1^2 Y_l^2\over 2} + m_R^2
\\
\left({\cal M}_L^2\right)_{LR} &=& {1\over \sqrt{2}} \left(v_2 (Y_l
\mu^{\star} - A_l^{'}) + v_1 A_l \right)
\eeas

\item  Fields  $Q_{1}^{I}$ and  $U^{I}$ turn into six up squarks  $U_{i}$.
\beas
Q_1^I = Z_U^{Ii} U^+_i \hskip 2cm U^I = Z_U^{(I+3)i\star} U^-_i
\eeas
\beas
Z_U^T \left(
\begin{array}{cc}
  \left({\cal M}_U^2\right)_{LL} & \left({\cal M}_U^2\right)_{LR} \\
  \left({\cal M}_U^2\right)_{LR}^{\dagger} & \left({\cal M}_U^2\right)_{RR} 
\end{array}
\right) Z_U^{\star} =  \left(
\begin{array}{ccc}
  M_{U_1}^2 &&0\\
  &{\ddots}&\\
  0&&M_{U_6}^2 
\end{array}
\right)
\eeas
\beas
\left({\cal M}_U^2\right)_{LL}  &=& - {e^2(v_1^2-v_2^2)(1-4c_W^2) 
\over 24 s_W^2 c_W^2} {\hat 1} + {v_2^2 Y_u^2 \over 2} + (Km_Q^2K^\dagger)^T
\\
\left({\cal M}_U^2\right)_{RR} &=& {e^2(v_1^2-v_2^2) \over 6c_W^2}
{\hat 1} + {v_2^2 Y_u^2 \over 2} + m_U^2
\\
\left({\cal M}_U^2\right)_{LR} &=& - {1\over \sqrt{2}} \left(v_1
(A_u^{'} + Y_u \mu^{\star}) + v_2 A_u\right)
\eeas
One should note that $Z_U$ is defined with the complex conjugate
comparing to definitions of $Z_L$ (and $Z_D$ below).  With such a
definition, all positively charged sfermion fields in the MSSM
Lagrangian in section~\ref{sec:eigen} are multiplied by $Z_X^{ij}$,
negatively charged by $Z_X^{ij\star}$. This makes easier to control
correctness of various calculations including complex parameters.

\item  Finally one has six down-squarks  $D_i$ composed from fields
$Q_2^I$ and $D^I$:
\beas
Q_2^I = Z_D^{Ii\star} D^-_i \hskip 2cm D^I = Z_D^{(I+3)i} D^+_i
\eeas
\beas
Z_D^{\dagger} \left(
\begin{array}{cc}
  \left({\cal M}_D^2\right)_{LL} & \left({\cal M}_D^2\right)_{LR} \\
  \left({\cal M}_D^2\right)_{LR}^{\dagger} & \left({\cal M}_D^2\right)_{RR} 
\end{array}
\right)Z_D =  \left(
\begin{array}{ccc}
  M_{D_1}^2 &&0\\
  &{\ddots}&\\
  0&&M_{D_6}^2 
\end{array}
\right)
\eeas
\beas
\left({\cal M}_D^2\right)_{LL}  &=& - {e^2(v_1^2-v_2^2)(1+2c_W^2) 
\over 24 s_W^2 c_W^2} {\hat 1} + {v_1^2 Y_d^2 \over 2} + (m_Q^2)^T
\\
\left({\cal M}_D^2\right)_{RR} &=& - {e^2(v_1^2-v_2^2) \over 12c_W^2}
{\hat 1} + {v_1^2 Y_d^2 \over 2} + m_D^2
\\
\left({\cal M}_D^2\right)_{LR} &=& {1\over \sqrt{2}} \left(v_2 
(Y_d \mu^{\star} - A_d^{'}) + v_1 A_d \right)
\eeas
\end{enumerate}

We have now completely defined all the physical fields existing in the
MSSM:\\
\begin{tabular}{llll}
Photon & $F_{\mu}$ & \\
Gauge bosons & $Z^0_{\mu}, W_{\mu}^{\pm}$ & \\
Gluons & $g_{\mu}^a$ & $a=1 \ldots 8$ \\
Gluinos & $\Lambda_G^a$ & $a=1 \ldots 8$ & (Majorana spinors)\\
Charginos & $\chi_i$ & $i=1,2$ & (Dirac spinors)\\
Neutralinos & $\chi_i^0$ & $i =1\ldots 4$ & (Majorana spinors)\\
Neutrinos & $\nu^I$ & $I=1 \ldots 3$ & (Dirac spinors)\\
Electrons & $e^I$ & $I=1 \ldots 3$ & (Dirac spinors)\\
Quarks & $u^I, d^I$ & $I =1 \ldots 3$ & (Dirac spinors)\\
Sneutrinos & ${\tilde\nu}^I$ & $I=1 \ldots 3$ \\
Selectrons & $ L_i^{\pm}$ & $i=1 \ldots 6$  \\
Squarks & $ U^{\pm}_i, D^{\pm}_i$ & $i =1 \ldots 6$  \\
Higgs particles: &&&\\
\hskip 1cm charged & $H_1^{\pm}$ ($\equiv H^\pm$) &\\
\hskip 1cm neutral ``scalar'' &  $H_1^0, H_2^0$ ($\equiv H,h$)&\\
\hskip 1cm neutral ``pseudoscalar'' &  $ A^0_1$ ($\equiv A^0$)&\\
\end{tabular}

Not for all the possible values of input parameters one can obtain
reasonable sets of particle masses.  For instance, for some choices of
the Higgs sector data the $SU(2)\times U(1)$ symmetry would not be
broken or, on the opposite, incorrect values of squark sector
parameters can lead to negative values of their masses and in
consequence to the color symmetry breaking~\cite{QCD_BREAK}

\section{Choice of the gauge}
\label{sec:gauge}

As long as one considers various processes in the spontaneously broken
gauge theory in the tree approximation, the most natural and preferred
choice is the unitary gauge in which the unphysical Goldstone bosons
are absent from the Lagrangian and Feynman rules.  When one wants to
calculate higher order corrections, one must include the ghost loops
suitable for the given gauge.  In such case it is much more efficient
to use the t'Hooft-Feynman gauge, in which the Goldstone fields appear
explicitly in the calculations, but ghost vertices are relatively
simple.  For our model the appropriate choice for the gauge fixing
terms is:
\beas
L_{GF} &=& - {1\over 2\kappa} \left({\partial}^{\mu} G_{\mu}^a
\right)^2 - {1\over 2\xi} \left({\partial}^{\mu} A_{\mu}^3 + \xi M_Z
c_W G^0 \right)^2 - {1\over 2\xi} \left({\partial}^{\mu} B_{\mu} -
\xi M_Z s_W G^0 \right)^2
\\
&-& {1\over 2{\xi}} \left({\partial}^{{\mu}} A_{{\mu}}^{1} + {i\over
\sqrt{2}} \xi M_W (G^+ - G^-)\right)^2 - {1\over 2\xi} 
\left({\partial}^{\mu} A_{\mu}^2 - {1\over \sqrt{2}} \xi M_W
(G^+ + G^-)\right)^2
\\
&=& - {1\over 2\kappa} \left({\partial}^{\mu} G_{\mu}^a \right)^2 -
{1\over 2\xi} \left({\partial}^{\mu} Z_{\mu} \right)^2 - {1\over 2\xi}
\left({\partial}^{\mu} F_{\mu} \right)^2 - \frac{1}{\xi} 
\left({\partial}^{\mu} W^+_{\mu}\right) \left({\partial}^{\mu}
W^-_{\mu} \right)
\\
&-& M_Z G^0 {\partial}^{\mu} Z_{\mu} - iM_W (G^+ {\partial}^{\mu}
W^-_{\mu} - G^- {\partial}^{\mu} W^+_{\mu}) - \frac{1}{2} \xi M_Z^2
(G^0)^2 - \xi M_W^2 G^+ G^-
\eeas

In some calculations it may be convenient to use even more complicated
version of the above expression, with different gauge fixing
parameters for various gauge groups (see e.g.~\cite{CPR}).

\section{The interaction Lagrangian}
\label{sec:lagr}

Although we consider only the minimal extension of the standard model,
the full set of Feynman rules for such a theory in the gauge described
in section~\ref{sec:gauge} is very complicated.  In this section we
write down the interaction part of the MSSM Lagrangian. The
propagators and vertices suitable for the chosen gauge are collected
in~\ref{sec:app_b}.  Of course, one can obtain from them rules for
tree calculations in the unitary gauge by setting $H_{2}^{{\pm}}$ and
$A_{2}^{0}$ to zero and neglecting the ghost terms.

It is convenient to divide all terms in the Lagrangian into classes
corresponding to the different types of particles taking part in the
interactions (the quark, squark and gluino vertices which contain the
QCD coupling constant $g_3$ are collected together as the separate
class).

We start from two technical remarks explaining the notation used
through the rest of the paper.  First, the expression ``$ + \mathrm{
H.c.}$'' always refers only to the line in which it was used.  Second,
after the diagonalization of the superpotential the Yukawa matrices
$Y_l^{IJ}, Y_u^{IJ}, Y_d^{IJ}$ change into $Y_l^{I} {\delta}^{IJ},
Y_u^{I} {\delta}^{IJ}, Y_d^{I} {\delta}^{IJ}$ and simultaneously sums
of the type $A^{IJ} Y_l^{JK} B^{KL}$ convert into $A^{IJ} Y_l^{J}
B^{JL}$ etc., containing the capital indices $I,J,K\ldots$ more than
twice; nevertheless, one should always use the summation convention in
those cases.  This should not lead to any misunderstandings.

\newcounter{lpvert}
\setcounter{lpvert}{0}
\newcommand{\lpv}{\addtocounter{lpvert}{1}\arabic{lpvert}.~}

\parindent 0pt

\vskip 3mm

\lpv Interactions of gauge bosons and superscalars.  

i) quark-squark-gauge interactions (the color indices are not written
explicitly):
\beas
&-& \frac{2}{3} e\bar{u}^I {\gamma}^{\mu} u^I F_{\mu}
- \frac{2}{3}ie(U_{i}^{-} \lrover{\partial}^{\mu} U_{i}^{+}) F_{\mu}
+ \frac{1}{3}e\bar{d}^I {\gamma}^{\mu} d^I F_{\mu}
+ \frac{1}{3}ie(D_{i}^{+} \lrover{\partial}^{\mu} D_{i}^{-}) F_{\mu}
\\
&-& {e \over 2s_Wc_W} \bar{u}^I {\gamma}^{\mu} (P_L - \frac{4}{3}
s_W^2) u^I Z_{\mu}
- {ie \over 2s_Wc_W} (Z_U^{Ii\star} Z_U^{Ij} - \frac{4}{3}
s_W^2{\delta}^{ij}) (U_{i}^{-} \lrover{\partial}^{\mu} U_{j}^{+})
Z_{\mu}
\\
&+& {e \over 2s_Wc_W} \bar{d}^I {\gamma}^{\mu} (P_L - \frac{2}{3}
s_W^2) d^I Z_{\mu}
+ {ie \over 2s_Wc_W} (Z_D^{Ii} Z_D^{Ij\star} - \frac{2}{3}
s_W^2{\delta}^{ij})(D_{i}^{+} \lrover{\partial}^{\mu} D_{j}^{-})
Z_{\mu}
\\
&-& {e \over \sqrt{2}s_W} K^{JI\star} \bar{d}^I \gamma^{\mu} P_L u^J
W_{\mu}^-
- {ie \over \sqrt{2}s_W} Z_D^{Ii} Z_U^{Jj} K^{JI\star} (D_{i}^{+}
\lrover{\partial}^{\mu} U_{j}^{+})W_{\mu}^{-} + \mathrm{H.c.}
\\
&+& \frac{4}{9}e^2 F_{\mu} F^{\mu} U_{i}^{-} U_{i}^{+}
+ {2e^2 \over 3s_Wc_W}(Z_U^{Ii\star} Z_U^{Ij} - \frac{4}{3}
{\delta}^{ij} s_W^2) Z_{\mu} F^{\mu} U_{i}^{-} U_{j}^{+}
\\
&+& {e^2\over 3c_W^2} \left[\frac{4}{3}{\delta}^{ij} s_W^2 + {3-8s_W^2
\over 4s_W^2} Z_U^{Ii\star} Z_U^{Ij} \right]Z_{\mu} Z^{\mu} U_{i}^{-}
U_{j}^{+}
\\
&+& \frac{1}{9}e^2 F_{\mu} F^{\mu} D_{i}^{+} D_{i}^{-}
+ {e^2 \over 3s_Wc_W}(Z_D^{Ii} Z_D^{Ij\star} - \frac{2}{3}
\delta^{ij}s_W^2) Z_{\mu} F^{\mu} D_{i}^{+} D_{j}^{-}
\\
&+& {e^2\over 3c_W^2} \left[\frac{1}{3}{\delta}^{ij} s_W^2 + {3-4s_W^2
\over 4s_W^2} Z_D^{Ii} Z_D^{Ij\star}\right] Z_{\mu} Z^{\mu} D_{i}^{+}
D_{j}^{-}
\\
&+& {e^2 \over 2s_W^2} Z_U^{Ii\star} Z_U^{Ij} W_{\mu}^{+} W^{-{\mu}}
U_{i}^{-} U_{j}^{+}
+ {e^2 \over 2s_W^2} Z_D^{Ii} Z_D^{Ij\star} W_{\mu}^{+} W^{-{\mu}}
D_{i}^{+} D_{j}^{-}
\\
&-& {e^2\sqrt{2} \over 6s_Wc_W} Z_D^{Ii} Z_U^{Jj} K^{JI\star}
D_{i}^{+} U_{j}^{+} (Z^{\mu} s_W - F^{\mu}c_W) W_{\mu}^{-} +
\mathrm{H.c.}
\eeas

ii) lepton-slepton-gauge interactions
\beas
&-& {e \over 2s_Wc_W} \bar{\nu}^I \gamma^{\mu} P_L \nu^I Z_{\mu}
- {ie \over 2s_Wc_W} ({\tilde\nu}^{I\star} \lrover{\partial}^{\mu}
{\tilde\nu}^I) Z_{\mu}
+ e\bar{e}^I {\gamma}^{\mu} e^I F_{\mu}
+ ie(L_{i}^{+} \lrover{\partial}^{\mu} L_{i}^{-})F_{\mu}
\\
&+& {e \over 2s_Wc_W} \bar{e}^I {\gamma}^{\mu} (P_L - 2s_W^2) e^I
Z_{\mu}
+ {ie \over 2s_Wc_W} (Z_L^{Ii} Z_L^{Ij\star} - 2s_W^2 \delta^{ij})
(L_{i}^{+} \lrover{\partial}^{\mu} L_{j}^{-})Z_{\mu}
\\
&-& {e \over \sqrt{2}s_W} \bar{\nu}^I {\gamma}^{\mu} P_L e^I
W_{\mu}^{+}
- {ie \over \sqrt{2}s_W} Z_{\nu}^{IJ} Z_L^{Ii}
(L_{i}^{+} \lrover{\partial}^{\mu} {\tilde\nu}^J) W_{\mu}^{-} +
\mathrm{H.c.}
\\
&+& {e^2 \over 4s_W^2c_W^2} Z_{\mu} Z^{\mu} {\tilde\nu}^{I\star}
{\tilde\nu}^I
+ e^2 F_{\mu} F^{\mu} L_{i}^{-} L_{i}^{+}
+ {e^2\over s_Wc_W} \left(Z_L^{Ii\star} Z_L^{Ij} - 2 \delta^{ij} s_W^2
\right) F_{\mu} Z^{\mu} L_{i}^{-} L_{j}^{+}
\\
&+& {e^2\over c_W^2} \left({\delta}^{ij} s_W^2 + {1-4s_W^2 \over
4s_W^2} Z_L^{Ii\star} Z_L^{Ij} \right)Z_{\mu} Z^{\mu} L_{i}^{-}
L_{j}^{+}
\\
&+& {e^2 \over 2s_W^2} W_{\mu}^{+} W^{-{\mu}} {\tilde\nu}^{I\star}
{\tilde\nu}^I
+ {e^2 \over 2s_W^2} Z_L^{Ii\star} Z_L^{Ij} W_{\mu}^{+} W^{-{\mu}}
L_{i}^{-} L_{j}^{+}
\\
&+& {e^2 \over \sqrt{2}s_Wc_W} Z_{\nu}^{IJ} Z_L^{Ii} {\tilde\nu}^J
L_{i}^{+} W_{\mu}^{-} (Z^{\mu} s_W - F^{\mu} c_W) + \mathrm{H.c.}
\eeas

\lpv Interactions of the Higgs particles and gauge bosons.  For more
concise notation, we define the auxiliary matrices
\bea
A_M^{ij} = Z_R^{1i} Z_H^{1j} - Z_R^{2i} Z_H^{2j} 
\hskip 2cm 
C_R^i = v_1 Z_R^{1i} + v_2 Z_R^{2i}
\eea
The Higgs-gauge interaction has the form:
\beas 
&+& {e \over 2s_Wc_W} A_M^{ij} (H^0_i \lrover{\partial}^{\mu} A^0_j)
Z_{\mu} + {ie\over 2 s_W c_W} (H^+_i \lrover{\partial}^{\mu} H^-_i) (
(c_W^2 - s_W^2) Z_{\mu} + 2 s_W c_W F_{\mu})
\\
&-& {ie \over 2s_W} A_M^{ij} (H^0_i \lrover{\partial}^{\mu} H^-_j)
W_{\mu}^{+} - {e \over 2s_W} (A^0_i \lrover{\partial}^{\mu} H^-_i)
W_{\mu}^{+} + \mathrm{H.c.}
\\
&+& {e^2 \over 2s_W^2} C_R^i (W_{\mu}^{+} W^{-{\mu}} + {1 \over
  2c_W^2} Z_{\mu} Z^{\mu}) H^0_i
- \left( {e M_W\over c_W} (Z^{\mu} s_W - F^{\mu}c_W) W_{\mu}^{+}
H_{2}^{-} + \mathrm{H.c.}\right)
\\
&+& {e^2\over 4 s_W^2 c_W^2} \left( (c_W^2 - s_W^2)^2 Z_{\mu} Z^{\mu}
+ 4 s_W c_W (c_W^2 - s_W^2) Z_{\mu} F^{\mu} + 4 s_W^2 c_W^2 F_{\mu}
F^{\mu}\right) H^+_i H^-_i
\\
&+& {e^2 \over 4s_W^2} (W_{\mu}^{+} W^{-{\mu}} + {1
\over 2c_W^2} Z_{\mu} Z^{\mu})(H^0_i H^0_i + A^0_i A^0_i)
+ {e^2 \over 2s_W^2} W_{\mu}^{+} W^{-{\mu}} H^+_i H^-_i
\\
&+& {e^2 \over 2s_Wc_W} A_M^{ij} (Z^{\mu} s_W - F^{\mu} c_W)
W_{\mu}^{+} H^-_j H^0_i - {ie^2 \over 2s_Wc_W} (Z^{\mu} s_W -
F^{\mu}c_W) W_{\mu}^+ H^-_i A^0_i + \mathrm{H.c.}
\eeas

\lpv  Higgs-lepton and Higgs-quark interactions. 
\beas 
&+& {1\over \sqrt{2}} Y_l^I Z_R^{1i} \bar{e}^I e^I H^0_i - {i\over
\sqrt{2}} Y_l^I Z_H^{1i} \bar{e}^I \gamma_5 e^I A^0_i - Y_l^I Z_H^{1i}
(\bar{e}^I P_L \nu^I H^-_i + \mathrm{H.c.})
\\
&+& {1\over \sqrt{2}} Y_d^I Z_R^{1i} \bar{d}^I d^I H^0_i - {1\over
\sqrt{2}} Y_u^I Z_R^{2i} \bar{u}^I u^I H^0_i - {i\over \sqrt{2}} Y_d^I
Z_H^{1i} \bar{d}^I \gamma_5 d^I A^0_i + {i\over \sqrt{2}} Y_u^I
Z_H^{2i} \bar{u}^I \gamma_5 u^I A^0_i
\\
&+& \bar{d}^I (-Y_d^I Z_H^{1i} P_L + Y_u^J Z_H^{2i} P_R) K^{JI\star}
u^J H^-_i + \mathrm{H.c.}  
\eeas

\lpv Interactions of the charginos and neutralinos with the gauge
bosons:
\beas 
&-& e\bar{\chi}_i \gamma^{\mu} \chi_i F_{\mu}
- {e \over 2s_Wc_W} \bar{\chi}_i \gamma^{\mu} \left(Z_+^{1i\star}
Z_+^{1j} P_L + Z_-^{1i} Z_-^{1j\star} P_R + (c_W^2 - s_W^2)
\delta^{ij} \right) \chi_j Z_{\mu}
\\
&+& {e \over s_W} \bar{\chi}_j \gamma^{\mu} \left[(Z_N^{2i}
Z_+^{1j\star} - {1\over \sqrt{2}} Z_N^{4i} Z_+^{2j\star}) P_L +
(Z_N^{2i\star} Z_-^{1j} + {1\over \sqrt{2}} Z_N^{3i\star} Z_-^{2j})
P_R\right] \chi^0_i W^+_{\mu} + \mathrm{H.c.}
\\
&+& {e \over 4 s_W c_W} \bar{\chi}_i^0 {\gamma}^{\mu} \left(
(Z_N^{4i\star} Z_N^{4j} - Z_N^{3i\star} Z_N^{3j}) P_L - (Z_N^{4i}
Z_N^{4j\star} - Z_N^{3i} Z_N^{3j\star}) P_R\right)\chi^0_j Z_{\mu}
\eeas

\lpv  Interactions of the charginos and neutralinos with the
superscalars.

i) interactions with squarks (superscript ``C'' denotes the char\-ge
con\-ju\-ga\-ted spi\-nor):
\beas 
&+& U^-_i \bar{\chi}^0_j \left[ \left( {-e \over \sqrt{2}s_Wc_W}
Z_U^{Ii\star} (\frac{1}{3}Z_N^{1j} s_W + Z_N^{2j} c_W) - Y_u^I
Z_U^{(I+3)i\star} Z_N^{4j} \right) P_L\right.
\\
&&\mbox{\hspace{3cm}} + \left. \left( {2e\sqrt{2} \over 3c_W}
Z_U^{(I+3)i\star} Z_N^{1j\star} - Y_u^I Z_U^{Ii\star} Z_N^{4j\star}
\right) P_R \right] u^I + \mathrm{H.c.}
\\
&+& D^+_i \bar{\chi}^0_j \left[ \left( {-e \over \sqrt{2}s_Wc_W}
    Z_D^{Ii} (\frac{1}{3}Z_N^{1j} s_W - Z_N^{2j} c_W) + Y_d^I
    Z_D^{(I+3)i} Z_N^{3j} \right) P_L\right.
\\
&&\mbox{\hspace{3cm}} + \left. \left( {-e\sqrt{2} \over 3c_W}
Z_D^{(I+3)i} Z_N^{1j\star} + Y_d^I Z_D^{Ii} Z_N^{3j\star} \right)
P_R\right] d^I + \mathrm{H.c.}
\\
&+& U^+_i \bar{d}^I \left[( {-e \over s_W} Z_U^{Ji} Z_+^{1j\star} +
Y_u^J Z_U^{(J+3)i} Z_+^{2j\star}) P_R - Y_d^I Z_U^{Ji} Z_-^{2j}
P_L\right] K^{JI\star} \chi_j^C + \mathrm{H.c.}
\\
&-& D^+_i \bar{\chi}_j \left[( {e \over s_W} Z_D^{Ii} Z_-^{1j} + Y_d^I
Z_D^{(I+3)i} Z_-^{2j}) P_L - Y_u^J Z_D^{Ii} Z_+^{2j\star}
P_R\right]K^{JI\star} u^J + \mathrm{H.c.}
\eeas

ii) interactions with sleptons
\beas
&+& {e \over \sqrt{2}s_Wc_W} Z_{\nu}^{IJ\star} (Z_N^{1i} s_W -
Z_N^{2i} c_W){\tilde\nu}^{J\star} \bar{\chi}^0_i P_L \nu^I +
\mathrm{H.c.}
\\
&+& \bar{\chi}^0_j \left[ \left( {e \over
\sqrt{2}s_Wc_W} Z_L^{Ii} (Z_N^{1j} s_W + Z_N^{2j} c_W) + Y_l^I
Z_L^{(I+3)i} Z_N^{3j} \right) P_L\right.
\\
&& \mbox{\hspace{3cm}} + \left. \left( {-e\sqrt{2} \over c_W}
Z_L^{(I+3)i} Z_N^{1j\star} + Y_l^I Z_L^{Ii} Z_N^{3j\star} \right)
P_R\right] e^I L^+_i + \mathrm{H.c.}
\\
&-& \bar{\chi}_i^C ( {e \over s_W} Z_+^{1i} P_L + Y_l^I Z_-^{2i\star}
P_R) Z_{\nu}^{IJ\star} e^I {\tilde\nu}^{J\star} + \mathrm{H.c.}
\\
&-& ( {e \over s_W} Z_L^{Ii} Z_-^{1j} + Y_l^I Z_L^{(I+3)i}
Z_-^{2j})\bar{\chi}_j P_L \nu^I L_i^+ + \mathrm{H.c.}
\eeas

\lpv Interactions of the charginos and neutralinos with the Higgs
particles.
\beas
&+& {e \over 2s_Wc_W} \bar{\chi}^0_i \left[(Z_R^{1k} Z_N^{3j} -
Z_R^{2k} Z_N^{4j})(Z_N^{1i} s_W - Z_N^{2i} c_W)P_L\right.
\\
&&\mbox{\hspace{3cm}} + \left.(Z_R^{1k} Z_N^{3i\star} - Z_R^{2k}
Z_N^{4i\star}) (Z_N^{1j\star} s_W - Z_N^{2j\star} c_W) P_R\right]
\chi^0_j H^0_k
\\
&-& {ie \over 2s_Wc_W} \bar{\chi}^0_i \left[(Z_H^{1k} Z_N^{3j} -
Z_H^{2k} Z_N^{4j}) (Z_N^{1i} s_W - Z_N^{2i} c_W) P_L \right.
\\
&& \mbox{\hspace{3cm}} - \left. (Z_H^{1k} Z_N^{3i\star} - Z_H^{2k}
Z_N^{4i\star}) (Z_N^{1j\star} s_W - Z_N^{2j\star} c_W) P_R\right]
\chi^0_j A^0_k
\\
&-& {e \over \sqrt{2}s_W} \bar{\chi}_i \left[(Z_R^{1k} Z_-^{2i}
Z_+^{1j} + Z_R^{2k} Z_-^{1i} Z_+^{2j}) P_L + (Z_R^{1k} Z_-^{2j\star}
Z_+^{1i\star} + Z_R^{2k} Z_-^{1j\star} Z_+^{2i\star}) P_R\right]
\chi_j H^0_k
\\
&+& {ie \over \sqrt{2}s_W} \bar{\chi}_{i} \left[(Z_H^{1k} Z_-^{2i}
Z_+^{1j} + Z_H^{2k} Z_-^{1i} Z_+^{2j}) P_L - (Z_H^{1k} Z_-^{2j\star}
Z_+^{1i\star} + Z_H^{2k} Z_-^{1j\star} Z_+^{2i\star}) P_R \right]
{\chi}_j A^0_k
\\
&+& {e \over s_Wc_W} \bar{\chi}_{j} \left[Z_H^{1k} \left({1\over
\sqrt{2}}Z_-^{2j} (Z_N^{1i} s_W + Z_N^{2i} c_W) - Z_-^{1j}
Z_N^{3i} c_W\right)P_L\right.
\\
&&\mbox{\hspace{1.5cm}} - \left.  Z_H^{2k} \left({1\over \sqrt{2}}
Z_+^{2j\star} (Z_N^{1i\star} s_W + Z_N^{2i\star} c_W) + Z_+^{1j\star}
Z_N^{4i\star} c_W\right) P_R\right] \chi^0_i H^+_k + \mathrm{H.c.}
\eeas

\lpv Self-interactions of the gauge bosons.
\beas
&-& ie \left(W^+_{\mu} W^-_{\nu} (\partial^{\mu} F^{\nu} -
\partial^{\nu} F^{\mu}) + F_{\mu} (W_{\nu}^{-} \partial^{\nu}
W^{+\mu} - W^+_{\nu} \partial^{\nu} W^{-\mu} + W^+_{\nu}
\lrover{\partial}^{\mu} W^{-\nu}) \right)
\\
&-& {ie c_W\over s_W} \left(W_{\mu}^{+} W_{\nu}^{-} ({\partial}^{\mu}
Z^{\nu} - Z^{\nu} F^{\mu}) + Z_{\mu} (W_{\nu}^{-} {\partial}^{\nu}
W^{+{\mu}} - W_{\nu}^{+} {\partial}^{\nu} W^{-{\mu}} + W_{\nu}^{+}
\lrover{\partial}^{\mu} W^{-{\nu}}) \right)
\\
&+& {e^2 \over 2s_W^2} (g^{{\mu}{\lambda}} g^{{\nu}{\rho}} -
g^{{\mu}{\nu}} g^{{\lambda}{\rho}})W_{\mu}^{+} W_{{\lambda}}^{+}
W_{\nu}^{-} W_{{\rho}}^{-}
+ e^2 (g^{{\mu}{\lambda}} g^{{\nu}{\rho}} - g^{{\mu}{\nu}}
g^{{\lambda}{\rho}})F_{\mu} F_{\nu} W_{{\lambda}}^{+}
W_{{\rho}}^{-} 
\\
&+& {e^2c_W^2\over s_W^2} (g^{\mu\lambda} g^{\nu\rho} - g^{\mu\nu}
g^{\lambda\rho}) Z_{\mu} Z_{\nu} W_{\lambda}^{+} W_{\rho}^{-}
+ {e^2c_W\over s_W} (g^{\mu\lambda} g^{\nu\rho} + g^{\mu\rho}
g^{\nu\lambda} - 2g^{\mu\nu} g^{\lambda\rho}) Z_{\mu} F_{\nu}
W_{\lambda}^{+} W_{\rho}^{-}
\eeas

\lpv  Ghost terms.
\beas
&-& {ie\over s_W} (Z_{\mu} c_W + F_{\mu} s_W) [(\partial^{\mu}
\bar{\eta}^-) \eta^- - (\partial^{\mu} \bar{\eta}^+) \eta^+]
\\
&+& {ie\over s_W} W_{\mu}^{+} [(\partial^{\mu} \bar{\eta}_Z c_W +
\partial^{\mu} \bar{\eta}_F s_W) \eta^- - (\partial^{\mu}
\bar{\eta}^+) (\eta_Z c_W + \eta_F s_W)]
\\
&+& {ie\over s_W} W_{\mu}^{-} [-( \partial^{\mu} \bar{\eta}_Z c_W +
\partial^{\mu} \bar{\eta}_F s_W) \eta^+ + (\partial^{\mu}
\bar{\eta}^-) (\eta_Z c_W  + \eta_F s_W)]
\\
& -& {{\xi} e^2 \over 4s_W^2} v_i Z_R^{ij} \left( {1 \over c_W^2}
\bar{\eta}_Z {\eta}_Z + \bar{\eta}^{+} {\eta}^{+} +
\bar{\eta}^{-} {\eta}^{-} \right) H^0_j
+ {ie\xi M_W \over 2s_W} (\bar{\eta}^{-} {\eta}^{-} -
\bar{\eta}^{+} {\eta}^{+}) A^0_2
\\
& +& {\xi eM_W \over 2s_Wc_W}(\bar{\eta}_Z {\eta}^{-} - (c_W^2 -
s_W^2) \bar{\eta}^+ \eta_Z - \bar{\eta}^{+} 2 s_W c_W {\eta}_F) H^+_2
\\
& +& {\xi eM_W \over 2s_Wc_W}(\bar{\eta}_Z {\eta}^{+} - (c_W^2 -
s_W^2) \bar{\eta}^- \eta_Z - \bar{\eta}^{-} 2 s_W c_W {\eta}_F) H^-_2
\eeas

\lpv  Scalar potential of the Higgs particles.  This is the first part
of the huge and complicated quartic potential of the 27 scalar fields
appearing in the theory.  To streamline notation we define four
further auxiliary matrices:
\beas
 A_H^{ij} &=& Z_H^{1i} Z_H^{1j} - Z_H^{2i} Z_H^{2j}
\hskip 2cm
 A_R^{ij} = Z_R^{1i} Z_R^{1j} - Z_R^{2i} Z_R^{2j}
\\
 A_P^{ij} &=& Z_R^{1i} Z_H^{2j} + Z_R^{2i} Z_H^{1j}
\hskip 2cm
 B_R^{i} = v_1 Z_R^{1i} - v_2 Z_R^{2i}
\eeas
The Higgs potential is expressed as:
\beas
&-& {e^2 \over 8s_W^2c_W^2} A_R^{ij} B_R^k H^0_i H^0_j H^0_k - {e^2
\over 8s_W^2c_W^2} A_H^{ij} B_R^k A^0_i A^0_j H^0_k
\\
&-& \left( {e^2 \over 4s_W^2c_W^2} A_H^{ij} B_R^k + {eM_W \over 2s_W}
(A_P^{kj} \delta^{1i} + A_P^{ki} \delta^{1j})\right) H^+_i H^-_j H^0_k
\\
&+& {ieM_W \over 2s_W} {\epsilon}_{ij} {\delta}^{1k} H^+_i H^-_j A^0_k
+ {ie^2 \over 4s_W^2} A_P^{ij} {\epsilon}_{kl} H^0_i A^0_j H^+_k H^-_l
\\
&-& {e^2 \over 32s_W^2c_W^2} A_R^{ij} A_R^{kl} H^0_i H^0_j H^0_k H^0_l
- {e^2 \over 32s_W^2c_W^2} A_H^{ij} A_H^{kl} A^0_i A^0_j A^0_k A^0_l
\\
&-& {e^2 \over 16s_W^2c_W^2} A_R^{ij} A_H^{kl} H^0_i H^0_j A^0_k A^0_l
- {e^2 \over 8s_W^2c_W^2} A_H^{ij} A_H^{kl} H^+_i H^+_k H^-_j H^-_l
\\
&-& {e^2 \over 4s_W^2} \left( {1 \over 2c_W^2} A_R^{ij} A_H^{kl} +
A_P^{ik} A_P^{jl}\right) H^0_i H^0_j H^+_k H^-_l
\\
&-& {e^2 \over 4s_W^2} \left( {1 \over 2c_W^2} A_H^{ij} A_H^{kl} +
\epsilon_{ik} \epsilon_{jl}\right) A^0_i A^0_j H^+_k H^-_l
\eeas

It is easy to see that the couplings $H^+_i H^-_j A^0_k$ and $H^+_i
H^-_j A^0_k H^0_l$ disappear in the unitary gauge.

\lpv  Interactions of the sleptons and Higgs bosons.  

i) three scalar (two sleptons and one Higgs) couplings:
\beas
&-& {e^2 \over 4s_W^2c_W^2} B_R^i {\tilde\nu}^{I\star} {\tilde\nu}^I
H^0_i
\\
&+& Z_{\nu}^{IJ} \left( {-\sqrt{2}e^2 \over 4s_W^2} C_R^j Z_L^{Ii} +
  (A_l^{IK} Z_L^{(K+3)i} + {v_1\over \sqrt{2}} (Y_l^I)^2 Z_L^{Ii})
  Z_H^{1j}\right.
\\
&& \hspace{4cm} \left. + (A_l^{'IK} Z_L^{(K+3)i} - \mu^{\star} Y_l^I
  Z_L^{(I+3)i}) Z_H^{2j} \right){\tilde\nu}^J L^+_i H^-_j +
\mathrm{H.c.}
\\
&+& {i\over \sqrt{2}} \left((A_l^{IJ\star} Z_L^{Ij} Z_L^{(J+3)i\star}
  - A_l^{IJ} Z_L^{Ii\star} Z_L^{(J+3)j})Z_H^{1k} + (A_l^{'IJ\star}
  Z_L^{Ij} Z_L^{(J+3)i\star}\right.
\\
&&\hspace{1cm} \left.  - A_l^{'IJ} Z_L^{Ii\star} Z_L^{(J+3)j})Z_H^{2k}
  + Y_l^I (\mu^{\star} Z_L^{Ii\star} Z_L^{(I+3)j} - \mu Z_L^{Ij}
  Z_L^{(I+3)i\star}) Z_H^{2k} \right)L_{i}^{-} L_{j}^{+} A^0_k
\\
&+& \left( {e^2 \over 2c_W^2} B_R^k \left({\delta}^{ij} + {1-4s_W^2
      \over 2s_W^2} Z_L^{Ii\star} Z_L^{Ij}\right) - (Y_l^I)^2 v_1
  Z_R^{1k} (Z_L^{Ii\star} Z_L^{Ij} + Z_L^{(I+3)i\star}
  Z_L^{(I+3)j})\right.
\\
&&\hspace{4cm} - {1\over \sqrt{2}}Z_R^{1k} (A_l^{IJ\star} Z_L^{Ij}
Z_L^{(J+3)i\star} + A_l^{IJ} Z_L^{Ii\star} Z_L^{(J+3)j})
\\
&&\hspace{4cm} + {1\over \sqrt{2}}Z_R^{2k} (A_l^{'IJ\star} Z_L^{Ij}
Z_L^{(J+3)i\star} + A_l^{'IJ} Z_L^{Ii\star} Z_L^{(J+3)j})
\\
&&\hspace{4cm}\left. - {1\over \sqrt{2}} Y_l^I Z_R^{2k} (\mu^{\star}
  Z_L^{Ii\star} Z_L^{(I+3)j} + \mu Z_L^{Ij} Z_L^{(I+3)i\star})\right)
L^-_i L^+_j H^0_k \eeas

ii) two slepton--two Higgs couplings:
\beas 
&-& {e^2 \over 8s_W^2c_W^2} A_R^{ij} {\tilde\nu}^{I\star}
{\tilde\nu}^I H^0_i H^0_j - {e^2 \over 8s_W^2c_W^2} A_H^{ij}
{\tilde\nu}^{I\star} {\tilde\nu}^I A^0_i A^0_j
\\
&+& Z_{\nu}^{KJ\star} Z_{\nu}^{KI} (e^2 {c_W^2 - s_W^2\over 4 s_W^2
  c_W^2} A_H^{ij} - (Y_l^K)^2 Z_H^{1i} Z_H^{1j}){\tilde\nu}^{J\star}
{\tilde\nu}^I H^-_i H^+_j
\\
&+& {i\over \sqrt{2}}Z_L^{Ii} Z_{\nu}^{IJ} \left( {e^2 \over 2s_W^2}
  A_H^{jk} - (Y_l^I)^2 Z_H^{1j} Z_H^{1k} \right){\tilde\nu}^J L^+_i
H^-_j A^0_k + \mathrm{H.c.}
\\
&+& {1\over \sqrt{2}}Z_L^{Ii} Z_{\nu}^{IJ} \left( {-e^2 \over 2s_W^2}
  (Z_H^{1j} Z_R^{1k} + Z_H^{2j} Z_R^{2k}) + (Y_l^I)^2 Z_H^{1j}
  Z_R^{1k} \right){\tilde\nu}^J L^+_i H^-_j H^0_k + \mathrm{H.c.}
\\
&+& \left( {e^2 \over 4c_W^2} A_H^{kl} \left({\delta}^{ij} + {1-4s_W^2
      \over 2s_W^2} Z_L^{Ii\star} Z_L^{Ij}\right)\right.
\\
&&\hspace{2cm}\left. - \frac{1}{2}(Y_l^I)^2 Z_H^{1k} Z_H^{1l}
  (Z_L^{Ii\star} Z_L^{Ij} + Z_L^{(I+3)i\star} Z_L^{(I+3)j})\right)
L^-_i L^+_j A^0_k A^0_l
\\
&+& \left( {e^2 \over 4c_W^2} A_R^{kl} \left({\delta}^{ij} + {1-4s_W^2
      \over 2s_W^2} Z_L^{Ii\star} Z_L^{Ij}\right)\right.
\\
&&\hspace{2cm} \left. - \frac{1}{2}(Y_l^I)^2 Z_R^{1k} Z_R^{1l}
  (Z_L^{Ii\star} Z_L^{Ij} + Z_L^{(I+3)i\star} Z_L^{(I+3)j})\right)
L^-_i L^+_j H^0_k H^0_l
\\
&+& \left( {e^2 \over 2c_W^2} A_H^{kl} \left({\delta}^{ij} - {1+2s_W^2
      \over 2s_W^2} Z_L^{Ii\star} Z_L^{Ij}\right) - (Y_l^I)^2 Z_H^{1k}
  Z_H^{1l} Z_L^{(I+3)i\star} Z_L^{(I+3)j} \right) L^-_i L^+_j H^-_k
H^+_l \eeas

\lpv  Interactions of the squarks and Higgs bosons:

i) three scalar couplings:
\beas
&+& {i\over \sqrt{2}} \left((A_u^{IJ} Z_U^{Ij} Z_U^{(J+3)i\star} -
A_u^{IJ\star} Z_U^{Ii\star} Z_U^{(J+3)j})Z_H^{2k} + (A_u^{'IJ\star}
Z_U^{Ii\star} Z_U^{(J+3)j}\right.
\\
&&\hspace{1cm} - \left. A_u^{'IJ} Z_U^{Ij} Z_U^{(J+3)i\star})Z_H^{1k}
+ Y_u^I (\mu Z_U^{Ii\star} Z_U^{(I+3)j} - \mu^{\star} Z_U^{Ij}
Z_U^{(I+3)i\star}) Z_H^{1k} \right) U^-_i U^+_j A^0_k
\\
&+& \left( {-e^2 \over 3c_W^2} \left({\delta}^{ij} + {3-8s_W^2 \over
4s_W^2} Z_U^{Ii\star} Z_U^{Ij}\right)  B_R^k \right .
- (Y_u^I)^2 v_2 Z_R^{2k} (Z_U^{Ii\star} Z_U^{Ij} + Z_U^{(I+3)i\star}
Z_U^{(I+3)j})
\\
&&\hspace{4cm} + {1\over \sqrt{2}} Z_R^{2k} (A_u^{IJ\star}
Z_U^{Ii\star} Z_L^{(J+3)j} + A_u^{IJ} Z_U^{Ij} Z_U^{(J+3)i\star})
\\
&&\hspace{4cm} + {1\over \sqrt{2}}Z_R^{1k} (A_u^{'IJ\star}
Z_U^{Ii\star} Z_U^{(J+3)j} + A_u^{'IJ} Z_U^{Ij} Z_U^{(J+3)i\star})
\\
&&\hspace{4cm} \left. + {1\over \sqrt{2}} Y_u^I Z_R^{1k} (\mu^{\star}
  Z_U^{Ij} Z_U^{(I+3)i\star} + \mu Z_U^{Ii\star} Z_U^{(I+3)j})\right)
U^-_i U^+_j H^0_k
\\
&+& {i\over \sqrt{2}} \left((A_d^{IJ\star} Z_D^{Ij} Z_D^{(J+3)i\star}
  - A_d^{IJ} Z_D^{Ii\star} Z_D^{(J+3)j})Z_H^{1k} + (A_d^{'IJ\star}
  Z_D^{Ij} Z_D^{(J+3)i\star}\right.
\\
&&\hspace{1cm} \left.  - A_d^{'IJ} Z_D^{Ii\star} Z_D^{(J+3)j})Z_H^{2k}
  + Y_d^I (\mu^{\star} Z_D^{Ii\star} Z_D^{(I+3)j} - \mu Z_D^{Ij}
  Z_D^{(I+3)i\star}) Z_H^{2k} \right) D^-_i D^+_j A^0_k
\\
&+& \left( {e^2 \over 6c_W^2} \left({\delta}^{ij} + {3-4s_W^2 \over
      2s_W^2} Z_D^{Ii\star} Z_D^{Ij}\right) B_R^k - (Y_d^I)^2 v_1
  Z_R^{1k} (Z_D^{Ii\star} Z_D^{Ij} + Z_D^{(I+3)i\star} Z_D^{(I+3)j})
\right.
\\
&&\hspace{4cm} - {1\over \sqrt{2}}Z_R^{1k} (A_d^{IJ\star} Z_D^{Ij}
Z_D^{(J+3)i\star} + A_d^{IJ} Z_D^{Ii\star} Z_D^{(J+3)j})
\\
&&\hspace{4cm} + {1\over \sqrt{2}}Z_R^{2k} (A_d^{'IJ\star} Z_D^{Ij}
Z_D^{(J+3)i\star} + A_d^{'IJ} Z_D^{Ii\star} Z_D^{(J+3)j})
\\
&&\hspace{4cm} \left. - {1\over \sqrt{2}} Y_d^I Z_R^{2k} (\mu^{\star}
  Z_D^{Ii\star} Z_D^{(I+3)j} + \mu Z_D^{Ij} Z_D^{(I+3)i\star}) \right)
D^-_i D^+_j H^0_k
\\
&+& \left[{1\over \sqrt{2}} \left( {-e^2 \over 2s_W^2} C_R^k + v_1
    (Y_d^I)^2 Z_H^{1k} + v_2 (Y_u^J)^2 Z_H^{2k} \right) K^{JI\star}
  Z_D^{Ij} Z_U^{Ji} \right.
\\
&& \hspace{12mm} -{\sqrt{2}M_Ws_W \over e}{\delta}^{1k} Y_u^J Y_d^I
K^{JI\star} Z_D^{(I+3)j} Z_U^{(J+3)i}
\\
&& \hspace{12mm} + \left(Z_H^{1k} \mu Y_u^J K^{JI\star} + (Z_H^{1k}
  A_u^{'KJ\star} - Z_H^{2k} A_u^{KJ\star}) K^{KI\star}\right)
Z_U^{(J+3)i} Z_D^{Ij}
\\
&& \hspace{12mm} \left. + \left((Z_H^{1k} A_d^{KI} + Z_H^{2k}
    A_d^{'KI}) K^{JK\star} - Z_H^{2k} \mu^{\star} Y_d^I
    K^{JI\star}\right) Z_U^{Ji} Z_D^{(I+3)j} \right] U^+_i D^+_j H^-_k
+ \mathrm{H.c.}
\eeas

ii) four scalar couplings:
\beas
&+& \left({-e^2 \over 6c_W^2} A_H^{kl} \left({\delta}^{ij} + {3-8s_W^2
\over 4s_W^2} Z_U^{Ii\star} Z_U^{Ij}\right) \right.
\\
&&\hspace{2cm} \left. - \frac{1}{2}(Y_u^I)^2 Z_H^{2k} Z_H^{2l}
(Z_U^{Ii\star} Z_U^{Ij} + Z_U^{(I+3)i\star} Z_U^{(I+3)j})\right) U^-_i
U^+_j A^0_k A^0_l
\\
&+& \left({-e^2 \over 6c_W^2} A_R^{kl} \left({\delta}^{ij} + {3-8s_W^2
\over 4s_W^2} Z_U^{Ii\star} Z_U^{Ij}\right) \right.
\\
&&\hspace{2cm} \left. - \frac{1}{2} (Y_u^I)^2 Z_R^{2k} Z_R^{2l}
(Z_U^{Ii\star} Z_U^{Ij} + Z_U^{(I+3)i\star} Z_U^{(I+3)j})\right) U^-_i
U^+_j H^0_k H^0_l
\\
&+& \left({-e^2 \over 3c_W^2} A_H^{kl} \left({\delta}^{ij} - {3+2s_W^2
\over 4s_W^2} Z_U^{Ii\star} Z_U^{Ij}\right) \right.
\\
&&\hspace{1cm} \left. - (Y_u^I)^2 Z_H^{2k} Z_H^{2l} Z_U^{(I+3)i\star}
Z_U^{(I+3)j} - (Y_d^I)^2 Z_H^{1k} Z_H^{1l} K^{JI\star} K^{KI}
Z_U^{Ki\star} Z_U^{Jj} \right) U^-_i U^+_j H^-_k H^+_l
\\
&+& \left({e^2 \over 12c_W^2} A_H^{kl} \left({\delta}^{ij} + {3-4s_W^2
\over 2s_W^2} Z_D^{Ii\star} Z_D^{Ij}\right) \right. 
\\
&&\hspace{2cm} \left. - \frac{1}{2} (Y_d^I)^2 Z_H^{1k} Z_H^{1l}
(Z_D^{Ii\star} Z_D^{Ij} + Z_D^{(I+3)i\star} Z_D^{(I+3)j})\right) D^-_i
D^+_j A^0_k A^0_l
\\
&+& \left({e^2 \over 12c_W^2} A_R^{kl} \left({\delta}^{ij} + {3-4s_W^2
\over 2s_W^2} Z_D^{Ii\star} Z_D^{Ij}\right) \right.
\\
&&\hspace{2cm} \left. - \frac{1}{2} (Y_d^I)^2 Z_R^{1k} Z_R^{1l}
(Z_D^{Ii\star} Z_D^{Ij} + Z_D^{(I+3)i\star} Z_D^{(I+3)j})\right) D^-_i
D^+_j H^0_k H^0_l
\\
&+& \left({e^2 \over 6c_W^2} A_H^{kl} \left({\delta}^{ij} - {3-2s_W^2
\over 2s_W^2} Z_D^{Ii\star} Z_D^{Ij}\right) \right.
\\
&&\hspace{1cm} \left. - (Y_d^I)^2 Z_H^{1k} Z_H^{1l} Z_D^{(I+3)i\star}
Z_D^{(I+3)j} - (Y_u^K)^2 Z_H^{2k} Z_H^{2l} K^{KI\star} K^{KJ}
Z_D^{Ji\star} Z_D^{Ij} \right) D^-_i D^+_j H^-_k H^+_l
\\
&+& {1\over \sqrt{2}} K^{JI\star} \left({-e^2 \over 2s_W^2}(Z_H^{1k}
Z_R^{1l} + Z_H^{2k} Z_R^{2l})Z_U^{Ji} Z_D^{Ij} - A_{P}^{lk} Y_u^J
Y_d^I Z_U^{(J+3)i} Z_D^{(I+3)j}\right.
\\
&&\hspace{2cm} \left. + \left( (Y_u^J)^2 Z_H^{2k} Z_R^{2l} + (Y_d^I)^2
Z_H^{1k} Z_R^{1l}\right) Z_U^{Ji} Z_D^{Ij} \right) U^+_i D^+_j H^-_k
H^0_l + \mathrm{H.c.}
\\
&+& {i\over \sqrt{2}} K^{JI\star} \left({e^2 \over 2s_W^2} A_H^{kl}
Z_U^{Ji} Z_D^{Ij} - \epsilon_{kl} Y_u^J Y_d^I Z_U^{(J+3)i}
Z_D^{(I+3)j} \right.
\\
&&\hspace{2cm} \left. + \left( (Y_u^J)^2 Z_H^{2k} Z_H^{2l} - (Y_d^I)^2
Z_H^{1k} Z_H^{1l}\right) Z_U^{Ji} Z_D^{Ij} \right) U^+_i D^+_j H^-_k
A^0_l + \mathrm{H.c.}
\eeas

\lpv  Four-scalar interactions of four sleptons or two sleptons and 
two squarks.
\beas
&-& {e^2 \over 8s_W^2c_W^2} {\tilde\nu}^{I\star} {\tilde\nu}^{J\star}
{\tilde\nu}^I {\tilde\nu}^J
- \left( {- e^2 \over 2c_W^2} \left({\delta}^{ij} + {1-4s_W^2 \over
2s_W^2} Z_L^{Ki\star} Z_L^{Kj}\right){\delta}^{IJ} \right.
\\
&&\hspace{2cm}\left. + \left( {e^2 \over 2s_W^2} Z_L^{Ki\star}
Z_L^{Lj} + Y_l^K Y_l^L Z_L^{(K+3)i\star} Z_L^{(L+3)j}\right)
Z_{\nu}^{LI} Z_{\nu}^{KJ\star} \right){\tilde\nu}^{J\star}
{\tilde\nu}^I L^-_i L^+_j
\\
&-& {e^2 \over 3c_W^2} \left({\delta}^{ij} + {3-8s_W^2 \over 4s_W^2}
Z_U^{Ji\star} Z_U^{Jj}\right){\tilde\nu}^{I\star} {\tilde\nu}^I U^-_i
U^+_j
\\
&+& {e^2 \over 6c_W^2} \left({\delta}^{ij} + {3-4s_W^2 \over 2s_W^2}
Z_D^{Ji\star} Z_D^{Jj}\right){\tilde\nu}^{I\star} {\tilde\nu}^I D^-_i
D^+_j
\\
&-& Z_{\nu}^{JI} Z_U^{Li\star} K^{LK} \left( {e^2 \over 2s_W^2}
Z_D^{Kj\star} Z_L^{Jk} + Y_l^J Y_d^K Z_D^{(K+3)j\star} Z_L^{(J+3)k}
\right){\tilde\nu}^I U^-_i D^-_j L^+_k + \mathrm{H.c.}
\\
&-& \left({e^2(1+8s_W^2) \over 8s_W^2c_W^2} Z_L^{Ii\star}
Z_L^{Jj\star} Z_L^{Jl} Z_L^{Ik} + {e^2 \over 2c_W^2} {\delta}^{jl}
({\delta}^{ik} - 3Z_L^{Ii\star} Z_L^{Ik})\right.
\\
&&\hspace{3cm}\left. + Y_l^I Y_l^J Z_L^{Jj\star} Z_L^{Ik}
Z_L^{(I+3)i\star} Z_L^{(J+3)l} \right)L_{i}^{-} L_{j}^{-} L_{k}^{+}
L_{l}^{+}
\\
&+& {e^2 \over 6c_W^2} \left( {3+12s_W^2 \over 2s_W^2} Z_L^{Ii\star}
Z_L^{Ij} Z_U^{Jk\star} Z_U^{Jl} - 6\delta^{kl} Z_L^{Ii\star} Z_L^{Ij}
- 5 \delta^{ij} Z_U^{Ik\star} Z_U^{Il} + 4 \delta^{ij} \delta^{kl}
\right) L^-_i L^+_j U^-_k U^+_l
\\
&+& \left[ {e^2 \over 6c_W^2} \left( -{3 \over 2s_W^2} Z_L^{Ii\star}
Z_L^{Ij} Z_D^{Jk\star} Z_D^{Jl} + 3 \delta^{kl} Z_L^{Ii\star} Z_L^{Ij}
+ \delta^{ij} Z_D^{Ik\star} Z_D^{Il} - 2 \delta^{ij} \delta^{kl}
\right) \right.
\\
&&\hspace{2cm}\left.  - Y_l^I Y_d^J (Z_L^{(I+3)i\star} Z_L^{Ij}
Z_D^{Jk\star} Z_D^{(J+3)l} + Z_L^{Ii\star} Z_L^{(I+3)j}
Z_D^{(J+3)k\star} Z_D^{Jl})\right] L^-_i L^+_j D^-_k D^+_l
\eeas

\lpv Vertices containing the strong coupling constant $g_3$.  
There are two basic types of such interactions - couplings of the
quarks and squarks with the gluons and gluinos and four squarks
interactions.  By $Y^a$ we denote the matrices of the $SU(3)$
generators in $\mathbf{3}$ representation (generally $a, b, c
\ldots$ are indices corresponding to $\mathbf{8}$ (adjoint)
representation, $\alpha, \beta, \gamma \dots$ - to $\mathbf{3}$
(basic) representation of the QCD gauge group).

\beas
&-& g_3 \bar{d}^I Y^a {\gamma}^{\mu} d^I g_{\mu}^a
- ig_3 (D^+_i Y^a \lrover{\partial}^{\mu} D^-_i) g_{\mu}^a 
+ g_3^2 D^+_i Y^a Y^b D^-_i g_{\mu}^a g^{b\mu} 
\\
&-& \frac{2}{3}eg_3 D^+_i Y^a D^-_i g_{\mu}^a F^{\mu}
+ {eg_3 \over s_Wc_W} (-Z_D^{Ii} Z_D^{Ij\star} + \frac{2}{3}
\delta^{ij} s_W^2) D^+_i Y^a D^-_j g_{\mu}^a Z^{\mu}
\\
&-& g_3 \bar{u}^I Y^a {\gamma}^{\mu} u^I g_{\mu}^a
- ig_3 (U^-_i Y^a \lrover{\partial}^{\mu} U^+_i) g_{\mu}^a 
+ g_3^2 U^-_i Y^a Y^b U^+_i g_{\mu}^a g^{b\mu} 
\\
&+& \frac{4}{3} e g_3 U^-_i Y^a U^+_i g_{\mu}^a F^{\mu}
+ {eg_3 \over s_Wc_W}(Z_U^{Ii\star} Z_U^{Ij} - \frac{4}{3}
\delta^{ij}s_W^2) U^-_i Y^a U^+_j g_{\mu}^a Z^{\mu}
\\
&+& {eg_3\sqrt{2} \over s_W} Z_U^{Jj} Z_D^{Ii} K^{JI\star} D^+_i Y^a
U^+_j g_{\mu}^a W^{-{\mu}} + \mathrm{H.c.}
\\
&+& g_3 \sqrt{2}U^-_i Y^a \bar{\Lambda}_{G}^{a} (-Z_U^{Ii\star}P_L +
Z_U^{(I+3)i\star} P_R) u^I + \mathrm{H.c.}
\\
&+& g_3 \sqrt{2}D^+_i Y^a \bar{\Lambda}_{G}^{a} (-Z_D^{Ii}P_L +
Z_D^{(I+3)i} P_R) d^I + \mathrm{H.c.}
\\
&+& \frac{1}{2}ig_3 f_{abc} \bar{\Lambda}_G^a \gamma^{\mu} \Lambda_G^b
g^c_{\mu} + g_3 f_{abc} (\partial^{\mu} \bar{\eta}_G^a) \eta_G^b
g^c_{\mu}
\\
&+& \frac{1}{2} g_3 f_{abc} (\partial_{\mu} g^a_{\nu} - \partial_{\nu}
g_{\mu}^a) g^{b\mu} g^{c\nu} - \frac{1}{4} g_3^2 f_{abc} f_{ade}
g^b_{\mu} g^{d\mu} g^c_{\nu} g^{e\nu} \eeas

In the next terms it is necessary to write down explicitly the color
indices $\alpha, \beta, \gamma \ldots$ We also define following
abbreviations:
\begin{center}
\begin{tabular}{lp{1cm}l}
$R_U^{ij}=Z_U^{Ii\star}Z_U^{Ij}$ && $R_D^{ij}=Z_D^{Ii}Z_D^{Ij\star}$ \\
$X_U^{ij}=\delta^{ij} - 2R_U^{ij}$ && $X_D^{ij}=\delta^{ij} - 2R_D^{ij}$ \\
$Y_U^{ij}=5R_U^{ij}-4\delta^{ij}$ && $Y_D^{ij}=2\delta^{ij} - R_D^{ij}$\\
$V_U^{ijkl}=Y_u^I Y_u^J Z_U^{(I+3)i\star} Z_U^{Ij} Z_U^{Jk\star}
Z_U^{(J+3)l}$ && $V_D^{ijkl}= Y_d^I Y_d^J Z_D^{(I+3)i} Z_D^{Ij\star}
Z_D^{Jk} Z_D^{(J+3)l\star}$
\end{tabular}
\end{center}
\beas &-&{1\over 4}\left\{\left[{g_3^2 \over 6} (3X_U^{il}X_U^{kj} -
    X_U^{ij}X_U^{kl}) +{e^2\over 4 s_W^2} R_U^{ij}R_U^{kl} +{e^2\over
      36 c_W^2}Y_U^{ij}Y_U^{kl} + V_U^{ijkl} + V_U^{klij}\right]
  \delta_{\alpha\beta} \delta_{\gamma\delta} \right.
\\
&+& \left. \left[{g_3^2 \over 6} (3X_U^{ij}X_U^{kl} -
    X_U^{il}X_U^{kj}) +{e^2\over 4 s_W^2} R_U^{il}R_U^{kj} + {e^2\over
      36 c_W^2}Y_U^{il}Y_U^{kj} + V_U^{ilkj} + V_U^{kjil}\right]
  \delta_{\alpha\delta} \delta_{\beta\gamma}\right\} U^-_{i\alpha}
U^-_{k\gamma} U^+_{j\beta} U^+_{l\delta}
\\
&-&{1\over 4}\left\{\left[{g_3^2 \over 6} (3X_D^{il}X_D^{kj} -
    X_D^{ij}X_D^{kl}) +{e^2\over 4 s_W^2} R_D^{ij}R_D^{kl} +{e^2\over
      36 c_W^2}Y_D^{ij}Y_D^{kl} + V_D^{ijkl} + V_D^{klij}\right]
  \delta_{\alpha\beta} \delta_{\gamma\delta} \right.
\\
&+&\left. \left[{g_3^2 \over 6} (3X_D^{ij}X_D^{kl} - X_D^{il}X_D^{kj})
    +{e^2\over 4 s_W^2} R_D^{il}R_D^{kj} + {e^2\over 36
      c_W^2}Y_D^{il}Y_D^{kj} + V_D^{ilkj} + V_D^{kjil}\right]
  \delta_{\alpha\delta} \delta_{\beta\gamma}\right\} D^-_{i\alpha}
D^-_{k\gamma} D^+_{j\beta} D^+_{l\delta}
\\
&-&\left[{g_3^2 \over 6} X_U^{ij}X_D^{lk} (3\delta_{\alpha\delta}
  \delta_{\beta\gamma} - \delta_{\alpha\beta} \delta_{\gamma\delta})
  +{e^2\over 4s_W^2}(2 K^{KI} K^{LJ\star} \delta_{\alpha\gamma}
  \delta_{\beta\delta} - \delta^{IJ}\delta^{KL} \delta_{\alpha\beta}
  \delta_{\gamma\delta}) Z_U^{Ki\star} Z_U^{Lj} Z_D^{Jl} Z_D^{Ik\star}
\right.
\\
&&\hskip 30mm +{e^2\over 36c_W^2} Y_U^{ij} Y_D^{lk}
\delta_{\alpha\beta} \delta_{\gamma\delta}
+(Y_d^I Y_d^J K^{KI} K^{LJ\star} Z_U^{Ki\star} Z_U^{Lj}
Z_D^{(I+3)l\star} Z_D^{(J+3)k} 
\\
&&\hskip 30mm +\left. Y_u^I Y_u^J K^{JL} K^{IK\star} Z_U^{(I+3)j}
Z_U^{(J+3)i\star} Z_D^{Ll\star} Z_D^{Kk}) \delta_{\alpha\delta}
\delta_{\beta\gamma}\right] U^-_{i\alpha} U^+_{j\beta}
D^-_{k\gamma} D^+_{l\delta}
\eeas

\parindent 5mm

\section{Summary}
\label{sec:comment}

The model described in the previous sections contains a large number
of free parameters which considerably limit its predictive power.
There are some commonly used ways to reduce the number of free
constants in this theory.  The most often employed method is to obtain
values of the parameters at the electroweak scale by RGE running from
the coupling constants generated at high scale by some SUSY breaking
scenario.  Usually such theories are much more unified and contain
typically only few free numbers.  Of course, there are many
constraints originating from experimental data.  Obviously, the masses
of the superpartners are bounded from below by negative result of
direct SUSY searches in collider experiments.  One can find also many
indirect limits: for example, if the masses of the superscalars are
not very high, then the non-diagonal soft Yukawa couplings $A_d^{IJ},
A_d^{'IJ}$ etc.  can lead by loop corrections to too strong flavor
changing neutral currents in the quark sector.

It is worth remembering that although in the superpotential one can
have only three matrices of Yukawa couplings (four if right neutrino
exists), six such matrices for the soft SUSY breaking terms are
allowed, three additional ones describing interactions of the
superscalars with the complex conjugated Higgs doublets.  Those three
new couplings can affect various processes, in particular they are
present in the formulas for the scalar mass matrices.  In most SUSY
breaking scenarios such terms are not generated, but in principle they
are not forbidden by MSSM symmetries.

\parindent 0pt

\vskip 3mm

\noindent {\bf Acknowledgments.}\\[2mm]
The author would like to thank professor S.  Pokorski for many helpful
discussions.

\renewcommand{\thesection}{Appendix~\Alph{section}}
\renewcommand{\thesubsection}{\Alph{section}.\arabic{subsection}}
\setcounter{section}{0}

\section{MSSM Lagrangian before gauge symmetry breaking}
\label{sec:app_a}

In this Appendix we write down the Lagrangian of the MSSM in terms of
the initial fields, before the $SU(2)$ symmetry breaking, but already
after the redefinition of the coupling constants described in
section~\ref{sec:fields}.  Expressions given below should help the
reader to compare the conventions used here with other papers and
eventually check calculations of the vertices expressed in terms of
the mass eigenstate fields.  In the formulas below we use the
2-component fermion notation.  Transition to the 4-fermion notation
can be done by substitution:
\beas
u^I = \left(
\begin{array}{c}
  {\Psi}^I_{Q1}  \\
  \bar{\Psi}_U^I  
\end{array}
\right) 
\hskip 1cm
d^I = \left(
\begin{array}{c}
  {\Psi}^I_{Q2}  \\
  \bar{\Psi}_D^I  
\end{array}
\right)
\eeas
and similarly for the leptons.

After the diagonalization of the Yukawa couplings the superpotential
has the form ($K$ is the Kobayashi-Maskawa matrix, $\epsilon_{12} =
-\epsilon_{21} = -1$):
\beas
W &=& \mu \epsilon_{ij} H_i^1 H_j^2 + Y_l^I \epsilon_{ij} H_i^1 L_j^I
R^I
\\
&-& Y_u^I (H_1^2 K^{IJ} Q_2^J - H_2^2 Q_1^I) U^I - Y_d^I (H_1^1 Q_2^I
- H_2^1 K^{JI\star} Q_1^J) D^I
\eeas

\parindent 0cm

The MSSM Lagrangian contains the following types of interactions:

1.  Gauge boson-gaugino, gauge boson-gauge boson.
\beas
&&ig_3 f_{abc} {\lambda}_G^a {\sigma}^{\mu} \bar{\lambda}_G^b
G_{\mu}^c + ig_2 {\epsilon}_{ijk} {\lambda}_A^i {\sigma}^{\mu}
\bar{\lambda}_A^j A_{\mu}^k
\\
&+& \frac{1}{2} g_3 f_{abc} ({\partial}_{\mu} G_{\nu}^a -
\partial_{\nu}  G_{\mu}^a) G^{b{\mu}} G^{c\nu} - \frac{1}{4} g_3^2 
f_{abc} f_{ade} G_{\mu}^b G^{d{\mu}} G_{\nu}^c G^{e\nu}
\\
&+& \frac{1}{2} g_2 {\epsilon}_{ijk} (\partial_{\mu} A_{\nu}^i -
\partial_{\nu} A_{\mu}^i) A^{j\mu} A^{k\nu} - \frac{1}{4} g_2^2 
\epsilon_{ijk} \epsilon_{imn} A_{\mu}^j A^{m\mu} A_{\nu}^k A^{n\nu}
\eeas

2. Quark-squark-gauge.  $T^i$ are the $SU(2)$ generators,
$T^i=\frac{1}{2}{\tau}^i$, where $\tau^i$ denote the Pauli matrices.
$Y^a$ and $\bar{Y}^a$ are the $SU(3)$ generators in the $\mathbf{3}$
and $\mathbf{\bar{3}}$ representations, respectively.  $SU(3)$ and,
wherever possible, also $SU(2)$ indices are suppressed.  Finally,
$A_{\mu}^{\pm} = A_{\mu}^1 \pm iA_{\mu}^2=\sqrt{2} W_{\mu}^{\mp}$.
\beas
&-& \bar{{\Psi}}_{Q}^I (g_3 Y^a G_{\mu}^a + g_2 T^3 A_{\mu}^3 +
\frac{1}{6}g_1 B_{\mu} )\bar{{\sigma}}^{\mu} {\Psi}_{Q}^I -
iQ^{I\star} (g_3 Y^a G_{\mu}^a + g_2 T^3 A_{\mu}^3 +
\frac{1}{6}g_1 B_{\mu} )\lrover{\partial}_{Q}^{\mu} Q^I
\\
&-& \frac{1}{2} g_2 K^{JI\star} (\bar{\Psi}_{Q2}^I
\bar{{\sigma}}^{\mu} {\Psi}_{Q1}^J + iQ_2^{I\star}
\lrover{\partial}^{\mu} Q_1^J )A_{\mu}^{+} + \mathrm{H.c.}
\\
&+& i \sqrt{2}Q^{I\star} (g_3 Y^a {\lambda}_G^a + g_2 T^3
{\lambda}_A^3 + \frac{1}{6}g_1 {\lambda}_B) {\Psi}_{Q}^I +
\mathrm{H.c.}
\\
&+& ig_2 (K^{JI\star} Q_2^{I\star} {\lambda}_A^{-} {\Psi}_{Q1}^J +
K^{JI} Q_1^{J\star} {\lambda}_A^{+} {\Psi}_{Q2}^I ) + \mathrm{H.c.}
\\
&+& (Q_1,KQ_2 )^{I\star} (g_3 Y^a G_{\mu}^a + g_2 T^i A_{\mu}^i +
\frac{1}{6}g_1 B_{\mu} )(g_3 Y^b G^{b{\mu}} + g_2 T^j A^{j{\mu}} +
\frac{1}{6}g_1 B^{\mu} ) \left(
\begin{array}{c}
  Q_1 \\ 
  KQ_2
\end{array}
\right)^I
\\
&-& \bar{{\Psi}}_U^I (g_3 \bar{Y}^a G_{\mu}^a - \frac{2}{3} g_1
B_{\mu}) \bar{\sigma}^{\mu} {\Psi}_U^I - iU^{I\star} (g_3\bar{Y}^a
G_{\mu}^a - \frac{2}{3}g_1 B_{\mu} )\lrover{\partial}_U^{\mu} U^I
\\
&+& U^{I\star} (g_3 \bar{Y}^a G_{\mu}^a - \frac{2}{3}g_1 B_{\mu}) (g_3
\bar{Y}^b G^{b\mu} - \frac{2}{3}g_1 B^{\mu}) U^I
\\
&-& \bar{\Psi}_D^I (g_3 \bar{Y}^a G_{\mu}^a + \frac{1}{3} g_1 B_{\mu})
\bar{\sigma}^{\mu} {\Psi}_D^I - iD^{I\star} (g_3 \bar{Y}^a G_{\mu}^a +
\frac{1}{3} g_1 B_{\mu}) \lrover{\partial}_D^{\mu} D^I
\\
&+& D^{I\star} (g_3 \bar{Y}^a G_{\mu}^a + \frac{1}{3}g_1 B_{\mu}) (g_3
\bar{Y}^b G^{b{\mu}} + \frac{1}{3}g_1 B^{\mu} )D^I
\\
&+& i \sqrt{2}U^{I\star} (g_3 \bar{Y}^a {\lambda}_G^a -
\frac{2}{3} g_1 {\lambda}_B) {\Psi}_U^I + i \sqrt{2}D^{I\star}
(g_3 \bar{Y}^a {\lambda}_G^a + \frac{1}{3}g_1 \lambda_B) {\Psi}_D^I +
\mathrm{H.c.}
\eeas

3.  Lepton-slepton-gauge.
\beas
&-& \bar{{\Psi}}_{L}^I (g_2 T^i A_{\mu}^i - \frac{1}{2}g_1 B_{\mu})
\bar{{\sigma}}^{\mu} {\Psi}_{L}^I - iL^{I\star} (g_2 T^i A_{\mu}^i -
\frac{1}{2}g_1 B_{\mu} )\lrover{\partial}_{L}^{\mu} L^I
\\
&+& L^{I\star} (g_2 T^i A_{\mu}^i - \frac{1}{2}g_1 B_{\mu}) (g_2 T^j
A^{j\mu} - \frac{1}{2}g_1 B^{\mu}) L^I
\\
&-& g_1 \bar{\Psi}_R^I B_{\mu} \bar{\sigma}^{\mu} \Psi_R^I - ig_1
B_{\mu} (R^{I\star} \lrover{\partial}^{\mu} R^I ) + g_1^2 R^{I\star}
R^I B_{\mu} B^{\mu}
\\
&+& i \sqrt{2}L^{I\star} (g_2 T^i {\lambda}_A^i - \frac{1}{2} g_1
\lambda_B) {\Psi}_L^I + i \sqrt{2}g_1 R^{I\star} \lambda_B \Psi_R^I 
+ \mathrm{H.c.}
\eeas

4. Higgs boson-Higgsino-gauge.
\beas
&-& \bar{\Psi}_{H1} (g_2 T^i A_{\mu}^i - \frac{1}{2}g_1 B_{\mu})
\bar{\sigma}^{\mu} \Psi_{H1} - iH^{1\star} (g_2 T^i A_{\mu}^i -
\frac{1}{2}g_1 B_{\mu}) \lrover{\partial}_H^{\mu} H^1
\\
&+& H^{1\star} (g_2 T^i A_{\mu}^i - \frac{1}{2}g_1 B_{\mu}) (g_2 T^j
A^{j\mu} - \frac{1}{2}g_1 B^{\mu}) H^1
\\
&-& \bar{\Psi}_{H2} (g_2 T^i A_{\mu}^i + \frac{1}{2}g_1 B_{\mu})
\bar{\sigma}^{\mu} \Psi_{H2} - iH^{2\star} (g_2 T^i A_{\mu}^i +
\frac{1}{2}g_1 B_{\mu}) \lrover{\partial}_H^{\mu} H^2
\\
&+& H^{2\star} (g_2 T^i A_{\mu}^i + \frac{1}{2}g_1 B_{\mu}) (g_2 T^j
A^{j\mu} + \frac{1}{2}g_1 B^{\mu}) H^2
\\
&+& i \sqrt{2} H^{1\star} (g_2 T^i \lambda_A^i - \frac{1}{2}g_1
\lambda_B) \Psi_{H1} + i \sqrt{2} H^{2\star} (g_2 T^i \lambda_A^i
+ \frac{1}{2}g_1 \lambda_B) {\Psi}_{H2} + \mathrm{H.c.}
\eeas

5. Yukawa couplings.
\beas
&-& \mu\epsilon_{ij} \Psi_{H1}^i \Psi_{H2}^j - Y_l^I \epsilon_{ij}
\Psi_{H1}^i \Psi_{Lj}^I R^I - Y_l^I \epsilon_{ij} \Psi_{H1}^i
\Psi_{R}^I L_j^I - Y_l^I \epsilon_{ij} \Psi_{R}^I \Psi_{Lj}^I H_i^1 +
\mathrm{ H.c.}
\\
&-& Y_d^I (- \Psi_{H1}^1 \Psi_{Q2}^I + K^{JI\star} \Psi_{H2}^1
\Psi_{Q1}^J )D^I - Y_d^I (- \Psi_{H1}^1 Q_2^I + K^{JI\star}
\Psi_{H2}^1 Q_1^J) \Psi_D^I + \mathrm{H.c.}
\\
&-& Y_d^I (-H_1^1 \Psi_{Q2}^I + K^{JI\star} H_2^1 \Psi_{Q1}^J)
\Psi_D^I - Y_u^I (-K^{IJ} H_1^2 \Psi_{Q2}^J + H_2^2 \Psi_{Q1}^I)
\Psi_U^I +\mathrm{H.c.}
\\
&-& Y_u^I (-K^{IJ} \Psi_{H1}^2 \Psi_{Q2}^J + \Psi_{H2}^2 \Psi_{Q1}^I
)U^I - Y_u^I (-K^{IJ} \Psi_{H1}^2 Q_2^J + \Psi_{H2}^2 Q_1^I )\Psi_U^I
+ \mathrm{H.c.}
\eeas

6. The scalar potential (in the Lagrangian $-V$ appears):
\beas
V = \frac{1}{2}(D_G^a D_G^a + D_A^i D_A^i + D_B D_B) + F_i^{\star} F_i
\eeas

For the $F_{H1}^{\star} F_{H1}$ and $F_{H2}^{\star} F_{H2}$ terms we
explicitly write down $SU(3)$ indices in places where their
contraction can be ambiguous ($\bar{Y}^a=-(Y^a)^T$).
\beas
D_G^a &=& g_3 (Q_i^{I\star} Y^a Q_i^I + D^{I\star}\bar{Y}^a D^I +
U^{I\star} \bar{Y}^a U^I)
\\
D_A^i &=& g_2 [(Q_1,KQ_2 )^{I\star} T^i \left(
\begin{array}{c}
  Q_1 \\
  KQ_2  
\end{array}
\right)^I + L^{I\star} T^i L^I + H^{1\star} T^i H^1 + H^{2\star} T^i H^2]
\\
D_B &=& \frac{1}{2}g_1 (\frac{1}{3}Q_i^{I\star} Q_i^I +
\frac{2}{3}D^{I\star} D^I - \frac{4}{3}U^{I\star} U^I -
L_i^{I\star} L_i^I + 2R^{I\star} R^I - H_i^{1\star} H_i^1 +
H_i^{2\star} H_i^2 )
\eeas
\beas
\\
F_{H1}^{\star} F_{H1} &=& |\mu|^2 H_i^{2\star} H_i^2 + Y_l^I Y_l^J
L_i^{I\star} L_i^J R^{I\star} R^J + Y_d^I Y_d^J (K^{KI} K^{LJ\star}
Q_{1\alpha}^{K\star} Q_{1\beta}^L + Q_{2{\alpha}}^{I\star}
Q_{2\beta}^J) D_{\alpha}^{I\star} D_{\beta}^J
\\
&+& (Y_l^I \mu^{\star} H_i^{2\star} L_i^I R^I + Y_d^I \mu^{\star}
(K^{JI\star} H_1^{2\star} Q_1^J + H_2^{2\star} Q_2^I) D^I +
\mathrm{H.c.})
\\
&+& (Y_l^I Y_d^J (K^{KJ\star} L_1^{I\star} Q_1^K + L_2^{I\star} Q_2^J)
R^{I\star} D^J + \mathrm{H.c.})
\\
F_{H2}^{\star} F_{H2} &=& |\mu|^2 H_i^{1\star} H_i^1 + Y_u^I Y_u^J
(K^{IK\star} K^{JL} Q_{2{\alpha}}^{K\star} Q_{2{\beta}}^{L} +
Q_{1{\alpha}}^{I\star} Q_{1{\beta}}^J) U_{\alpha}^{I\star} U_{\beta}^J
\\
&-& (Y_u^I \mu^{\star} (K^{IJ} H_2^{1\star} Q_2^J + H_1^{1\star}
Q_1^I) U^I + \mathrm{H.c.})
\\
F_{Li}^{I\star} F_{Li}^I &=& (Y_l^I )^2 H_i^{1\star} H_i^1 R^{I\star}
R^I
\\
F_{R}^{I\star} F_{R}^I &=& (Y_l^I )^2 \epsilon_{ij} \epsilon_{kl}
H_i^{1\star} H_{k}^1 L_j^{I\star} L_{l}^I
\\
F_{Qi}^{I\star} F_{Qi}^I &=& (Y_d^I )^2 H_i^{1\star} H_i^1 D^{I\star}
D^I + (Y_u^I )^2 H_i^{2\star} H_i^2 U^{I\star} U^I + (Y_u^I Y_d^J
K^{IJ} U^I D^{J\star} H_i^{1\star} H_i^2 + \mathrm{H.c.})
\\
F_U^{I\star} F_U^I &=& (Y_u^I )^2 [K^{IJ\star} K^{IK} H_1^{2\star}
H_1^2 Q_2^{J\star} Q_2^K + H_2^{2\star} H_2^2 Q_1^{I\star} Q_1^I -
(K^{IJ\star} H_1^{2\star} H_2^2 Q_2^{J\star} Q_1^I + \mathrm{H.c.})]
\\
F_D^{I\star} F_D^I &=& (Y_d^I )^2 [H_1^{1\star} H_1^1 Q_2^{I\star}
Q_2^I + K^{JI} K^{KI\star} H_2^{1\star} H_2^1 Q_1^{J\star} Q_1^K -
(K^{JI\star} H_1^{1\star} H_2^1 Q_2^{I\star} Q_1^J + \mathrm{H.c.})]
\eeas

7.  The soft SUSY breaking terms.
\beas
&-& m_{H_1 }^2 H_i^{1\star} H_i^1 - m_{H_2}^2 H_i^{2\star} H_i^2 -
(m_L^2)^{IJ} L_i^{I\star} L_i^J - (m_R^2)^{IJ} R^{I\star} R^J
\\
&-& (m_Q^2)^{IJ} [Q_2^{I\star} Q_2^J + K^{KI} K^{LJ\star} Q_1^{K\star}
Q_1^{L}] - (m_D^2 )^{IJ} D^{I\star} D^J - (m_U^2 )^{IJ} U^{I\star} U^J
\\
&+& \frac{1}{2} M_1 \lambda_B \lambda_B + \frac{1}{2} M_2
\lambda_A^i \lambda_A^i + \frac{1}{2} M_3 \lambda_G^a \lambda_G^a
+ \mathrm{H.c.}
\\
&+& m_{12}^2 \epsilon_{ij} H_i^1 H_j^2 + \epsilon_{ij} A_l^{IJ}
H_i^1 L_j^I R^J + A_l^{'IJ} H_i^{2\star} L_i^I R^J + \mathrm{H.c.}
\\
&+& A_u^{IJ} (- K^{IK} H_1^2 Q_2^k + H_2^2 Q_1^I )U^J + A_u^{'IJ}
(K^{IK} H_2^{1\star} Q_2^k + H_1^{1\star} Q_1^I )U^J + \mathrm{H.c.}
\\
&+& A_d^{IJ} (- H_1^1 Q_2^I + K^{KI\star} H_2^1 Q_1^k) D^J + A_d^{'IJ}
(H_2^{2\star} Q_2^I + K^{KI\star} H_1^{2\star} Q_1^k )D^J +
\mathrm{H.c.}
\eeas

\section{Feynman rules}
\label{sec:app_b}

\setcounter{lpvert}{0}

\parindent 0cm

In this appendix we complete Feynman rules for the interactions
described in section~\ref{sec:lagr}. For simplicity we write down the
propagators and ghost terms for $\xi= 1$ - extension to the general
case is straightforward~.  $\xi$-dependence of ghost vertices and
Goldstone boson masses can be found in section~\ref{sec:gauge} and
section~\ref{sec:lagr} point 7. All vertices are properly symmetrized.

\subsection{Propagators}

1. Scalar particles (Higgs boson or superscalar):

\begin{tabular}{ll}
\begin{picture}(60,20)(0,0)
\DashLine(0,10)(60,10){5}
\end{picture}
&
\raisebox{10\unitlength}{
\begin{minipage}{5cm}
\beas 
{i \over p^2-m^2} 
\eeas
\end{minipage}
}
\end{tabular}
\vspace*{3mm}

2. Vector bosons:

\begin{tabular}{ll}
\begin{picture}(60,20)(0,0)
\Photon(0,10)(60,10){3}{4}
\Text(0,0)[]{$\mu$}
\Text(60,0)[]{$\nu$}
\end{picture}
&
\raisebox{10\unitlength}{
\begin{minipage}{5cm}
\beas 
{-ig_{\mu\nu} \over p^2-m^2} 
\eeas
\end{minipage}
}
\end{tabular}
\vspace*{3mm}

3. Fermions. Some of the MSSM fermions are Majorana spinors, what
introduce additional complications into the calculations. The detailed
discussion of the Feynman rules for the Majorana fermions can be found
e.g. in~\cite{denner}.

\begin{tabular}{ll}
\begin{picture}(60,20)(0,0)
\ArrowLine(0,10)(60,10)
\end{picture}
&
\raisebox{10\unitlength}{
\begin{minipage}{5cm}
\beas 
{i \over p_{\mu}\gamma^{\mu}  - m} 
\eeas
\end{minipage}
}
\end{tabular}
\vspace*{3mm}

4. Ghosts:

\begin{tabular}{ll}
\begin{picture}(60,20)(0,0)
\ZigZag(0,10)(60,10){3}{4}
\end{picture}
&
\raisebox{10\unitlength}{
\begin{minipage}{5cm}
\beas 
{i \over p^2  - m^2} 
\eeas
\end{minipage}
}
\end{tabular}
\vspace*{3mm}

The propagators of the quarks, squarks and the color ghosts should be
multiplied by the factor $\delta^{ab}$, where $a$ and $b$ are as usual
color indices.

\subsection{Vertices}

\lpv Quark-squark-gauge boson.

\input q_sq_gauge

\lpv Lepton-slepton-gauge boson.

\begin{tabular}{ll}
\begin{picture}(150,70)(0,-10)
\Photon(10,0)(60,0){3}{4}
\Text(0,0)[c]{$\gamma_{\mu}$}
\ArrowLine(60,0)(110,0)
\Text(120,0)[c]{$e^J$}
\ArrowLine(60,50)(60,0)
\Text(65,45)[l]{$e^I$}
\Vertex(60,0){2}
\end{picture}
&
\raisebox{35\unitlength}{
\begin{minipage}{5cm}
\lefteqn{
ie\gamma^{\mu} \delta^{IJ}
}
\end{minipage}
}
\end{tabular}

\begin{tabular}{ll}
\begin{picture}(150,70)(0,-10)
\Photon(10,0)(60,0){3}{4}
\Text(0,0)[c]{$Z^0_{\mu}$}
\ArrowLine(60,0)(110,0)
\Text(120,0)[c]{$e^J$}
\ArrowLine(60,50)(60,0)
\Text(65,45)[l]{$e^I$}
\Vertex(60,0){2}
\end{picture}
&
\raisebox{35\unitlength}{
\begin{minipage}{5cm}
\lefteqn{
{ie \over 2s_Wc_W} {\gamma}^{\mu} (P_L - 2s_W^2) {\delta}^{IJ}
}
\end{minipage}
}
\end{tabular}

\begin{tabular}{ll}
\begin{picture}(150,70)(0,-10)
\Photon(10,0)(60,0){3}{4}
\Text(0,0)[c]{$Z^0_{\mu}$}
\ArrowLine(60,0)(110,0)
\Text(120,0)[c]{$\nu^J$}
\ArrowLine(60,50)(60,0)
\Text(65,45)[l]{$\nu^I$}
\Vertex(60,0){2}
\end{picture}
&
\raisebox{35\unitlength}{
\begin{minipage}{5cm}
\lefteqn{
- {ie \over 2s_Wc_W} {\gamma}^{\mu} P_L{\delta}^{IJ}
}
\end{minipage}
}
\end{tabular}

\begin{tabular}{ll}
\begin{picture}(150,70)(0,-10)
\Photon(10,0)(60,0){3}{4}
\Text(0,0)[c]{$W_{\mu}$}
\ArrowLine(60,0)(110,0)
\Text(120,0)[c]{$\nu^J$}
\ArrowLine(60,50)(60,0)
\Text(65,45)[l]{$e^I$}
\Vertex(60,0){2}
\end{picture}
&
\raisebox{35\unitlength}{
\begin{minipage}{5cm}
\lefteqn{
- {ie \over \sqrt{2}s_W} {\gamma}^{\mu} P_L{\delta}^{IJ}
}
\end{minipage}
}
\end{tabular}

\begin{tabular}{ll}
\begin{picture}(150,70)(0,-10)
\Photon(10,0)(60,0){3}{4}
\Text(0,0)[c]{$\gamma_{\mu}$}
\DashArrowLine(60,0)(110,0){5}
\Text(120,0)[c]{$L^-_i$}
\Text(90,10)[c]{$k$}
\DashArrowLine(60,50)(60,0){5}
\Text(65,45)[l]{$L^-_j$}
\Text(55,30)[r]{$p$}
\Vertex(60,0){2}
\end{picture}
&
\raisebox{35\unitlength}{
\begin{minipage}{5cm}
\lefteqn{
ie \delta^{ij} (p + k)^{\mu} 
}
\end{minipage}
}
\end{tabular}

\begin{tabular}{ll}
\begin{picture}(150,70)(0,-10)
\Photon(10,0)(60,0){3}{4}
\Text(0,0)[c]{$Z^0_{\mu}$}
\DashArrowLine(60,0)(110,0){5}
\Text(120,0)[c]{$\tilde\nu_I$}
\Text(90,10)[c]{$k$}
\DashArrowLine(60,50)(60,0){5}
\Text(65,45)[l]{$\tilde\nu_J$}
\Text(55,30)[r]{$p$}
\Vertex(60,0){2}
\end{picture}
&
\raisebox{35\unitlength}{
\begin{minipage}{5cm}
\lefteqn{
- {ie \over 2s_Wc_W} \delta^{IJ} (p + k)^{\mu}
}
\end{minipage}
}
\end{tabular}

\begin{tabular}{ll}
\begin{picture}(150,70)(0,-10)
\Photon(10,0)(60,0){3}{4}
\Text(0,0)[c]{$Z^0_{\mu}$}
\DashArrowLine(60,0)(110,0){5}
\Text(120,0)[c]{$L^-_i$}
\Text(90,10)[c]{$k$}
\DashArrowLine(60,50)(60,0){5}
\Text(65,45)[l]{$L^-_j$}
\Text(55,30)[r]{$p$}
\Vertex(60,0){2}
\end{picture}
&
\raisebox{35\unitlength}{
\begin{minipage}{5cm}
\lefteqn{
{ie \over 2s_Wc_W} (Z^{Ii}_{L} Z^{Ij\star}_{L} - 2s_W^2 \delta^{ij})
(p + k)^{\mu}
}
\end{minipage}
}
\end{tabular}

\begin{tabular}{ll}
\begin{picture}(150,70)(0,-10)
\Photon(10,0)(60,0){3}{4}
\Text(0,0)[c]{$W_{\mu}$}
\DashArrowLine(60,0)(110,0){5}
\Text(120,0)[c]{$L^-_i$}
\Text(90,10)[c]{$k$}
\DashArrowLine(60,50)(60,0){5}
\Text(65,45)[l]{$\tilde\nu_J$}
\Text(55,30)[r]{$p$}
\Vertex(60,0){2}
\end{picture}
&
\raisebox{35\unitlength}{
\begin{minipage}{5cm}
\lefteqn{
- {ie \over \sqrt{2}s_W} Z_{\nu}^{IJ} Z_L^{Ii} (p + k)^{\mu}
}
\end{minipage}
}
\end{tabular}

\vskip 5mm

\begin{tabular}{ll}
\begin{picture}(150,110)(0,0)
\Photon(10,70)(60,70){3}{4}
\Text(0,70)[c]{$W_{\mu}$}
\Photon(60,70)(110,70){3}{4}
\Text(120,70)[c]{$W_{\nu}$}
\DashArrowLine(60,120)(60,70){5}
\Text(65,115)[l]{$\tilde\nu_J$}
\DashArrowLine(60,70)(60,20){5}
\Text(65,25)[l]{$\tilde\nu_I$}
\Vertex(60,70){2}
\end{picture}
&
\raisebox{68\unitlength}{
\begin{minipage}{5cm}
\lefteqn{
  {ie^{2} \over 2s_W^2} {\delta}^{IJ} g^{\mu\nu} 
}
\end{minipage}
}
\end{tabular}

\begin{tabular}{ll}
\begin{picture}(150,110)(0,0)
\Photon(10,70)(60,70){3}{4}
\Text(0,70)[c]{$Z^0_{\mu}$}
\Photon(60,70)(110,70){3}{4}
\Text(120,70)[c]{$Z^0_{\nu}$}
\DashArrowLine(60,120)(60,70){5}
\Text(65,115)[l]{$\tilde\nu_J$}
\DashArrowLine(60,70)(60,20){5}
\Text(65,25)[l]{$\tilde\nu_I$}
\Vertex(60,70){2}
\end{picture}
&
\raisebox{68\unitlength}{
\begin{minipage}{5cm}
\lefteqn{
{ie^{2} \over 2s_W^2c_W^2} {\delta}^{IJ} g^{\mu\nu}
}
\end{minipage}
}
\end{tabular}

\begin{tabular}{ll}
\begin{picture}(150,110)(0,0)
\Photon(10,70)(60,70){3}{4}
\Text(0,70)[c]{$\gamma_{\mu}$}
\Photon(60,70)(110,70){3}{4}
\Text(120,70)[c]{$\gamma_{\nu}$}
\DashArrowLine(60,120)(60,70){5}
\Text(65,115)[l]{$L^-_j$}
\DashArrowLine(60,70)(60,20){5}
\Text(65,25)[l]{$L^-_i$}
\Vertex(60,70){2}
\end{picture}
&
\raisebox{68\unitlength}{
\begin{minipage}{5cm}
\lefteqn{
2ie^2 \delta^{ij} g^{\mu\nu}
}
\end{minipage}
}
\end{tabular}

\begin{tabular}{ll}
\begin{picture}(150,110)(0,0)
\Photon(10,70)(60,70){3}{4}
\Text(0,70)[c]{$\gamma_{\mu}$}
\Photon(60,70)(110,70){3}{4}
\Text(120,70)[c]{$Z^0_{\nu}$}
\DashArrowLine(60,120)(60,70){5}
\Text(65,115)[l]{$L^-_j$}
\DashArrowLine(60,70)(60,20){5}
\Text(65,25)[l]{$L^-_i$}
\Vertex(60,70){2}
\end{picture}
&
\raisebox{68\unitlength}{
\begin{minipage}{5cm}
\lefteqn{
{ie^2\over s_W c_W} \left(Z_{L}^{Ii} Z_{L}^{Ij\star} - 2 \delta^{ij}
s_W^2\right) g^{\mu\nu} 
}
\end{minipage}
}
\end{tabular}

\begin{tabular}{ll}
\begin{picture}(150,110)(0,0)
\Photon(10,70)(60,70){3}{4}
\Text(0,70)[c]{$Z^0_{\mu}$}
\Photon(60,70)(110,70){3}{4}
\Text(120,70)[c]{$Z^0_{\nu}$}
\DashArrowLine(60,120)(60,70){5}
\Text(65,115)[l]{$L^-_j$}
\DashArrowLine(60,70)(60,20){5}
\Text(65,25)[l]{$L^-_i$}
\Vertex(60,70){2}
\end{picture}
&
\raisebox{68\unitlength}{
\begin{minipage}{5cm}
\lefteqn{
{2ie^2\over c_W^2} \left[{\delta}^{ij} s_W^2 + {1-4s_W^2 \over 4s_W^2}
Z_{L}^{Ii} Z_{L}^{Ij\star}\right] g^{\mu\nu}
}
\end{minipage}
}
\end{tabular}

\begin{tabular}{ll}
\begin{picture}(150,110)(0,0)
\Photon(10,70)(60,70){3}{4}
\Text(0,70)[c]{$W_{\mu}$}
\Photon(60,70)(110,70){3}{4}
\Text(120,70)[c]{$W_{\nu}$}
\DashArrowLine(60,120)(60,70){5}
\Text(65,115)[l]{$L^-_j$}
\DashArrowLine(60,70)(60,20){5}
\Text(65,25)[l]{$L^-_i$}
\Vertex(60,70){2}
\end{picture}
&
\raisebox{68\unitlength}{
\begin{minipage}{5cm}
\lefteqn{
{ie^2 \over 2 s_W^2} Z_{L}^{Ii} Z_{L}^{Ij\star} g^{\mu\nu} 
}
\end{minipage}
}
\end{tabular}

\begin{tabular}{ll}
\begin{picture}(150,110)(0,0)
\Photon(10,70)(60,70){3}{4}
\Text(0,70)[c]{$W_{\mu}$}
\Photon(60,70)(110,70){3}{4}
\Text(120,70)[c]{$\gamma_{\nu}$}
\DashArrowLine(60,120)(60,70){5}
\Text(65,115)[l]{$\tilde\nu_J$}
\DashArrowLine(60,70)(60,20){5}
\Text(65,25)[l]{$L^-_i$}
\Vertex(60,70){2}
\end{picture}
&
\raisebox{68\unitlength}{
\begin{minipage}{5cm}
\lefteqn{
- {ie^{2} \over \sqrt{2}s_W} Z_{\nu}^{IJ} Z_L^{Ii} g^{\mu\nu}
}
\end{minipage}
}
\end{tabular}

\begin{tabular}{ll}
\begin{picture}(150,110)(0,0)
\Photon(10,70)(60,70){3}{4}
\Text(0,70)[c]{$W_{\mu}$}
\Photon(60,70)(110,70){3}{4}
\Text(120,70)[c]{$Z^0_{\nu}$}
\DashArrowLine(60,120)(60,70){5}
\Text(65,115)[l]{$\tilde\nu_J$}
\DashArrowLine(60,70)(60,20){5}
\Text(65,25)[l]{$L^-_i$}
\Vertex(60,70){2}
\end{picture}
&
\raisebox{68\unitlength}{
\begin{minipage}{5cm}
\lefteqn{
{ie^{2} \over \sqrt{2}c_W} Z_{\nu}^{IJ} Z_L^{Ii} g^{\mu\nu}
}
\end{minipage}
}
\end{tabular}

\lpv Higgs particle-gauge boson.

\input h_gauge

\lpv Chargino- and neutralino-gauge boson.

\begin{tabular}{ll}
\begin{picture}(150,70)(0,-10)
\Photon(10,0)(60,0){3}{4}
\Text(0,0)[c]{$\gamma_{\mu}$}
\ArrowLine(60,0)(110,0)
\Text(120,0)[c]{$\chi_i$}
\ArrowLine(60,50)(60,0)
\Text(65,45)[l]{$\chi_j$}
\Vertex(60,0){2}
\end{picture}
&
\raisebox{35\unitlength}{
\begin{minipage}{5cm}
\lefteqn{
-ie\gamma^{\mu} \delta^{ij}
}
\end{minipage}
}
\end{tabular}

\begin{tabular}{ll}
\begin{picture}(150,70)(0,-10)
\Photon(10,0)(60,0){3}{4}
\Text(0,0)[c]{$Z^0_{\mu}$}
\ArrowLine(60,0)(110,0)
\Text(120,0)[c]{$\chi_i$}
\ArrowLine(60,50)(60,0)
\Text(65,45)[l]{$\chi_j$}
\Vertex(60,0){2}
\end{picture}
&
\raisebox{35\unitlength}{
\begin{minipage}{5cm}
\lefteqn{
- {ie \over 2s_Wc_W} \gamma^{\mu} \left(Z_+^{1i\star} Z_+^{1j} P_L +
Z_-^{1i} Z_-^{1j\star} P_R + \delta^{ij} (c_W^2 - s_W^2)\right) 
}
\end{minipage}
}
\end{tabular}

\begin{tabular}{ll}
\begin{picture}(150,70)(0,-10)
\Photon(10,0)(60,0){3}{4}
\Text(0,0)[c]{$Z^0_{\mu}$}
\ArrowLine(60,0)(110,0)
\Text(120,0)[c]{$\chi^0_i$}
\ArrowLine(60,50)(60,0)
\Text(65,45)[l]{$\chi^0_j$}
\Vertex(60,0){2}
\end{picture}
&
\raisebox{35\unitlength}{
\begin{minipage}{5cm}
\lefteqn{
{ie \over 2 s_W c_W} {\gamma}^{\mu} \left( (Z_N^{4i\star} Z_N^{4j} -
Z_N^{3i\star} Z_N^{3j}) P_L - (Z_N^{4i} Z_N^{4j\star} - Z_N^{3i}
Z_N^{3j\star}) P_R\right)
}
\end{minipage}
}
\end{tabular}

\begin{tabular}{ll}
\begin{picture}(150,70)(0,-10)
\Photon(10,0)(60,0){3}{4}
\Text(0,0)[c]{$W_{\mu}$}
\ArrowLine(60,0)(110,0)
\Text(120,0)[c]{$\chi_j$}
\ArrowLine(60,50)(60,0)
\Text(65,45)[l]{$\chi^0_i$}
\Vertex(60,0){2}
\end{picture}
&
\raisebox{15\unitlength}{
\begin{minipage}{5cm}
\lefteqn{
{ie \over s_W} \gamma^{\mu} \left[\left (Z^{2i}_N Z_+^{1j\star} -
{1\over \sqrt{2}}Z^{4i}_N Z_+^{2j\star} \right) P_L \right.  
}
\lefteqn{
+\left. \left(Z_N^{2i\star} Z_-^{1j} + {1\over\sqrt{2}} Z_N^{3i\star}
Z_-^{2j}\right) P_R\right] 
}
\end{minipage}
}
\end{tabular}

\vskip 2mm

\lpv Chargino- and neutralino-quark and squark.

\begin{tabular}{ll}
\begin{picture}(150,70)(0,-10)
\DashArrowLine(10,0)(60,0){5}
\Text(0,0)[c]{$U_i^-$}
\ArrowLine(60,0)(110,0)
\Text(120,0)[c]{$\chi_j^0$}
\ArrowLine(60,50)(60,0)
\Text(65,45)[l]{$u^I$}
\Vertex(60,0){2}
\end{picture}
&
\raisebox{15\unitlength}{
\begin{minipage}{5cm}
\lefteqn{
i \left[ \left( {-e \over \sqrt{2}s_Wc_W} Z^{Ii\star}_{U}
(\frac{1}{3}Z^{1j}_{N} s_W + Z^{2j}_{N} c_W) - Y_u^{I}
Z_{U}^{(I+3)i\star} Z_{N}^{4j} \right) P_L \right.
}
\lefteqn{
+ \left. \left( {2\sqrt{2}e \over 3c_W} Z_{U}^{(I+3)i\star}
Z_{N}^{1j\star} - Y_u^{I} Z_{U}^{Ii\star} Z_{N}^{4j\star} \right) P_R
\right]
}
\end{minipage}
}
\end{tabular}

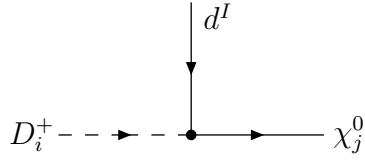
\begin{tabular}{ll}
\begin{picture}(150,70)(0,-10)
\DashArrowLine(10,0)(60,0){5}
\Text(0,0)[c]{$D_i^+$}
\ArrowLine(60,0)(110,0)
\Text(120,0)[c]{$\chi_j^0$}
\ArrowLine(60,50)(60,0)
\Text(65,45)[l]{$d^I$}
\Vertex(60,0){2}
\end{picture}
&
\raisebox{15\unitlength}{
\begin{minipage}{5cm}
\lefteqn{
i \left[ \left( {-e \over \sqrt{2}s_Wc_W} Z^{Ii}_{D}
(\frac{1}{3}Z^{1j}_{N} s_W - Z^{2j}_{N} c_W) + Y_d^{I} Z_{D}^{(I+3)i}
Z_{N}^{3j} \right) P_L\right.  
}
\lefteqn{
 + \left.\left( {-e\sqrt{2} \over 3c_W} Z_{D}^{(I+3)i} Z_{N}^{1j\star}
 + Y_d^{I} Z_{D}^{Ii} Z_{N}^{3j\star} \right) P_R \right]
}
\end{minipage}
}
\end{tabular}

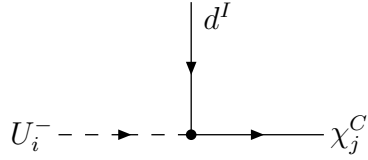
\begin{tabular}{ll}
\begin{picture}(150,70)(0,-10)
\DashArrowLine(10,0)(60,0){5}
\Text(0,0)[c]{$U_i^-$}
\ArrowLine(60,0)(110,0)
\Text(120,0)[c]{$\chi_j^C$}
\ArrowLine(60,50)(60,0)
\Text(65,45)[l]{$d^I$}
\Vertex(60,0){2}
\end{picture}
&
\raisebox{35\unitlength}{
\begin{minipage}{5cm}
\lefteqn{
i \left[( {-e \over s_W} Z_U^{Ji\star} Z_+^{1j} + Y_u^J
Z_U^{(J+3)i\star} Z_+^{2j}) P_L - Y_d^I Z_U^{Ji\star} Z_-^{2j\star}
P_R\right] K^{JI} 
}
\end{minipage}
}
\end{tabular}

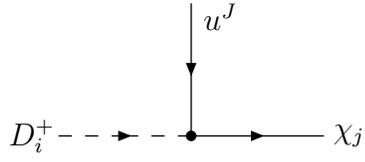
\begin{tabular}{ll}
\begin{picture}(150,70)(0,-10)
\DashArrowLine(10,0)(60,0){5}
\Text(0,0)[c]{$D_i^+$}
\ArrowLine(60,0)(110,0)
\Text(120,0)[c]{$\chi_j$}
\ArrowLine(60,50)(60,0)
\Text(65,45)[l]{$u^J$}
\Vertex(60,0){2}
\end{picture}
&
\raisebox{35\unitlength}{
\begin{minipage}{5cm}
\lefteqn{
i \left[-( {e \over s_W} Z_D^{Ii} Z_-^{1j} + Y_d^I Z_D^{(I+3)i}
Z_-^{2j}) P_L + Y_u^J Z_D^{Ii} Z_+^{2j\star} P_R\right] K^{JI\star} 
}
\end{minipage}
}
\end{tabular}

\lpv Charginos and neutralinos-leptons and sleptons.

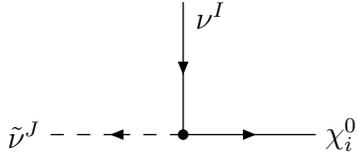
\begin{tabular}{ll}
\begin{picture}(150,70)(0,-10)
\DashArrowLine(60,0)(10,0){5}
\Text(0,0)[c]{$\tilde\nu^J$}
\ArrowLine(60,0)(110,0)
\Text(120,0)[c]{$\chi^0_i$}
\ArrowLine(60,50)(60,0)
\Text(65,45)[l]{$\nu^I$}
\Vertex(60,0){2}
\end{picture}
&
\raisebox{35\unitlength}{
\begin{minipage}{5cm}
\lefteqn{
{ie \over \sqrt{2}s_Wc_W} Z_{\nu}^{IJ\star} (Z_N^{1i} s_W - Z_N^{2i} c_W) P_L 
}
\end{minipage}
}
\end{tabular}

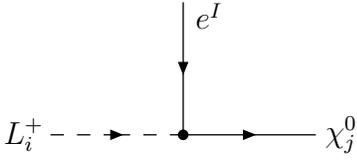
\begin{tabular}{ll}
\begin{picture}(150,70)(0,-10)
\DashArrowLine(10,0)(60,0){5}
\Text(0,0)[c]{$L_i^+$}
\ArrowLine(60,0)(110,0)
\Text(120,0)[c]{$\chi^0_j$}
\ArrowLine(60,50)(60,0)
\Text(65,45)[l]{$e^I$}
\Vertex(60,0){2}
\end{picture}
&
\raisebox{15\unitlength}{
\begin{minipage}{5cm}
\lefteqn{
i \left[ \left( {e \over \sqrt{2}s_Wc_W} Z_L^{Ii} (Z_N^{1j} s_W +
Z_N^{2j} c_W) + Y_l^I Z_L^{(I+3)i} Z_N^{3j} \right) P_L \right.
}
\lefteqn{
+ \left.\left( {-e\sqrt{2} \over c_W} Z_L^{(I+3)i} Z_N^{1j\star} +
Y_l^I Z_{L}^{Ii} Z_{N}^{3j\star} \right) P_R \right] 
}
\end{minipage}
}
\end{tabular}

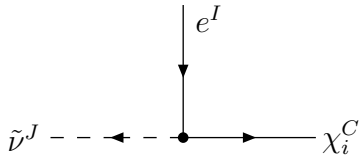
\begin{tabular}{ll}
\begin{picture}(150,70)(0,-10)
\DashArrowLine(60,0)(10,0){5}
\Text(0,0)[c]{$\tilde\nu^J$}
\ArrowLine(60,0)(110,0)
\Text(120,0)[c]{$\chi_i^C$}
\ArrowLine(60,50)(60,0)
\Text(65,45)[l]{$e^I$}
\Vertex(60,0){2}
\end{picture}
&
\raisebox{35\unitlength}{
\begin{minipage}{5cm}
\lefteqn{
- i \left( {e \over s_W} Z_+^{1i} P_L + Y_l^I Z_-^{2i\star} P_R
\right) Z_{\nu}^{IJ\star} 
}
\end{minipage}
}
\end{tabular}

\begin{tabular}{ll}
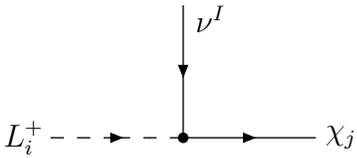
\begin{picture}(150,70)(0,-10)
\DashArrowLine(10,0)(60,0){5}
\Text(0,0)[c]{$L_i^+$}
\ArrowLine(60,0)(110,0)
\Text(120,0)[c]{$\chi_j$}
\ArrowLine(60,50)(60,0)
\Text(65,45)[l]{$\nu^I$}
\Vertex(60,0){2}
\end{picture}
&
\raisebox{35\unitlength}{
\begin{minipage}{5cm}
\lefteqn{
- i\left( {e \over s_W} Z_L^{Ii} Z_-^{1j} + Y_l^I Z_L^{(I+3)i}
Z_-^{2j} \right) P_L
}
\end{minipage}
}
\end{tabular}

\lpv Chargino- and neutralino-Higgs boson.

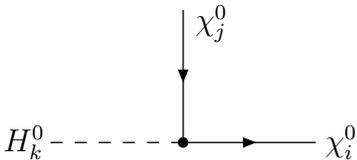
\begin{tabular}{ll}
\begin{picture}(150,70)(0,-10)
\DashLine(60,0)(10,0){5}
\Text(0,0)[c]{$H^0_k$}
\ArrowLine(60,0)(110,0)
\Text(120,0)[c]{$\chi^0_i$}
\ArrowLine(60,50)(60,0)
\Text(65,45)[l]{$\chi^0_j$}
\Vertex(60,0){2}
\end{picture}
&
\raisebox{5\unitlength}{
\begin{minipage}{5cm}
\lefteqn{
{ie \over 2 s_W c_W} \left[ \left( (Z_R^{1k} Z_N^{3j} - Z_R^{2k}
Z_N^{4j} ) (Z_N^{1i} s_W - Z_N^{2i} c_W)\right.\right.
} 
\lefteqn{
\left. + (Z_R^{1k} Z_N^{3i} - Z_R^{2k} Z_N^{4i} ) (Z_N^{1j} s_W - Z_N^{2j}
c_W)\right) P_L 
}
\lefteqn{
+ \left( (Z_R^{1k} Z_N^{3i\star} - Z_R^{2k} Z_N^{4i\star} )
(Z_N^{1j\star} s_W - Z_N^{2j\star} c_W)\right.  
}
\lefteqn{
\left.\left. +(Z_R^{1k} Z_N^{3j\star} - Z_R^{2k} Z_N^{4j\star} )
(Z_N^{1i\star} s_W - Z_N^{2i\star}c_W)\right) P_R\right]
} 
\end{minipage}
}
\end{tabular}

\begin{tabular}{ll}
\begin{picture}(150,70)(0,-10)
\DashLine(60,0)(10,0){5}
\Text(0,0)[c]{$A^0_k$}
\ArrowLine(60,0)(110,0)
\Text(120,0)[c]{$\chi^0_i$}
\ArrowLine(60,50)(60,0)
\Text(65,45)[l]{$\chi^0_j$}
\Vertex(60,0){2}
\end{picture}
&
\raisebox{5\unitlength}{
\begin{minipage}{5cm}
\lefteqn{
{e \over 2s_Wc_W} \left[ \left( (Z_H^{1k} Z_N^{3j} - Z_H^{2k}
Z_N^{4j}) (Z_N^{1i} s_W - Z_N^{2i} c_W)\right.\right.  
}
\lefteqn{
\left. + (Z_H^{1k} Z_N^{3i} - Z_H^{2k} Z_N^{4i} )
(Z_N^{1j} s_W - Z_N^{2j} c_W)\right)P_L
}
\lefteqn{
- \left((Z_H^{1k} Z_N^{3i\star} - Z_H^{2k} Z_N^{4i\star} )
(Z_N^{1j\star} s_W - Z_N^{2j\star} c_W)\right.
}
\lefteqn{
\left.\left. +(Z_H^{1k} Z_N^{3j\star} - Z_H^{2k} Z_N^{4j\star} )
(Z_N^{1i\star} s_W - Z_N^{2i\star}c_W)\right)P_R\right]
}
\end{minipage}
}
\end{tabular}

\begin{tabular}{ll}
\begin{picture}(150,70)(0,-10)
\DashLine(60,0)(10,0){5}
\Text(0,0)[c]{$H^0_k$}
\ArrowLine(60,0)(110,0)
\Text(120,0)[c]{$\chi_i$}
\ArrowLine(60,50)(60,0)
\Text(65,45)[l]{$\chi_j$}
\Vertex(60,0){2}
\end{picture}
&
\raisebox{15\unitlength}{
\begin{minipage}{5cm}
\lefteqn{
- {ie \over \sqrt{2}s_W} \left[(Z_R^{1k} Z_-^{2i} Z_+^{1j} + Z_R^{2k}
Z_-^{1i} Z_+^{2j}) P_L \right.  
}
\lefteqn{
\left.  + (Z_R^{1k} Z_-^{2j\star} Z_+^{1i\star} + Z_R^{2k}
Z_-^{1j\star}  Z_+^{2i\star}) P_R \right] 
}
\end{minipage}
}
\end{tabular}

\begin{tabular}{ll}
\begin{picture}(150,70)(0,-10)
\DashLine(60,0)(10,0){5}
\Text(0,0)[c]{$A^0_k$}
\ArrowLine(60,0)(110,0)
\Text(120,0)[c]{$\chi_i$}
\ArrowLine(60,50)(60,0)
\Text(65,45)[l]{$\chi_j$}
\Vertex(60,0){2}
\end{picture}
&
\raisebox{15\unitlength}{
\begin{minipage}{5cm}
\lefteqn{
{e \over \sqrt{2}s_W}\left[ - (Z_H^{1k} Z_-^{2i} Z_+^{1j} + Z_H^{2k}
Z_-^{1i} Z_+^{2j}) P_L \right.
}
\lefteqn{
\left. + (Z_H^{1k} Z_-^{2j\star} Z_+^{1i\star} + Z_H^{2k} 
Z_-^{1j\star} Z_+^{2i\star}) P_R \right] 
}
\end{minipage}
}
\end{tabular}

\begin{tabular}{ll}
\begin{picture}(150,70)(0,-10)
\DashArrowLine(10,0)(60,0){5}
\Text(0,0)[c]{$H^+_k$}
\ArrowLine(60,0)(110,0)
\Text(120,0)[c]{$\chi_j$}
\ArrowLine(60,50)(60,0)
\Text(65,45)[l]{$\chi^0_i$}
\Vertex(60,0){2}
\end{picture}
&
\raisebox{15\unitlength}{
\begin{minipage}{5cm}
\lefteqn{
{ie \over s_Wc_W} \left[Z_H^{1k} \left({1\over \sqrt{2}} Z_-^{2j}
(Z_N^{1i} s_W + Z_N^{2i} c_W) - Z_-^{1j} Z_N^{3i} c_W\right) P_L
\right.
}
\lefteqn{
\left. - Z_H^{2k} \left({1\over \sqrt{2}} Z_+^{2j\star} (Z_N^{1i\star}
s_W + Z_N^{2i\star} c_W) + Z_+^{1j\star} Z_N^{4i\star} c_W\right)
P_R \right]
}
\end{minipage}
}
\end{tabular}

\lpv Leptons and quarks-Higgs particles.

\begin{tabular}{ll}
\begin{picture}(150,70)(0,-10)
\DashLine(60,0)(10,0){5}
\Text(0,0)[c]{$H^0_i$}
\ArrowLine(60,0)(110,0)
\Text(120,0)[c]{$e^I$}
\ArrowLine(60,50)(60,0)
\Text(65,45)[l]{$e^I$}
\Vertex(60,0){2}
\end{picture}
&
\raisebox{35\unitlength}{
\begin{minipage}{5cm}
\lefteqn{
{i\over \sqrt{2}} Y_l^I Z_R^{1i}
}
\end{minipage}
}
\end{tabular}

\begin{tabular}{ll}
\begin{picture}(150,70)(0,-10)
\DashLine(60,0)(10,0){5}
\Text(0,0)[c]{$A^0_i$}
\ArrowLine(60,0)(110,0)
\Text(120,0)[c]{$e^I$}
\ArrowLine(60,50)(60,0)
\Text(65,45)[l]{$e^I$}
\Vertex(60,0){2}
\end{picture}
&
\raisebox{35\unitlength}{
\begin{minipage}{5cm}
\lefteqn{
{1\over \sqrt{2}} Y_l^I Z_H^{1i} \gamma_5
}
\end{minipage}
}
\end{tabular}

\begin{tabular}{ll}
\begin{picture}(150,70)(0,-10)
\DashArrowLine(10,0)(60,0){5}
\Text(0,0)[c]{$H^-_i$}
\ArrowLine(60,0)(110,0)
\Text(120,0)[c]{$e^I$}
\ArrowLine(60,50)(60,0)
\Text(65,45)[l]{$\nu^I$}
\Vertex(60,0){2}
\end{picture}
&
\raisebox{35\unitlength}{
\begin{minipage}{5cm}
\lefteqn{
- i  Y_l^I Z_H^{1i} P_L 
}
\end{minipage}
}
\end{tabular}

\begin{tabular}{ll}
\begin{picture}(150,70)(0,-10)
\DashLine(60,0)(10,0){5}
\Text(0,0)[c]{$H^0_i$}
\ArrowLine(60,0)(110,0)
\Text(120,0)[c]{$d^I$}
\ArrowLine(60,50)(60,0)
\Text(65,45)[l]{$d^I$}
\Vertex(60,0){2}
\end{picture}
&
\raisebox{35\unitlength}{
\begin{minipage}{5cm}
\lefteqn{
{i\over \sqrt{2}} Y_d^I Z_R^{1i}
}
\end{minipage}
}
\end{tabular}

\begin{tabular}{ll}
\begin{picture}(150,70)(0,-10)
\DashLine(60,0)(10,0){5}
\Text(0,0)[c]{$H^0_i$}
\ArrowLine(60,0)(110,0)
\Text(120,0)[c]{$u^I$}
\ArrowLine(60,50)(60,0)
\Text(65,45)[l]{$u^I$}
\Vertex(60,0){2}
\end{picture}
&
\raisebox{35\unitlength}{
\begin{minipage}{5cm}
\lefteqn{
-{i\over \sqrt{2}} Y_u^I Z_R^{2i}
}
\end{minipage}
}
\end{tabular}

\begin{tabular}{ll}
\begin{picture}(150,70)(0,-10)
\DashLine(60,0)(10,0){5}
\Text(0,0)[c]{$A^0_i$}
\ArrowLine(60,0)(110,0)
\Text(120,0)[c]{$d^I$}
\ArrowLine(60,50)(60,0)
\Text(65,45)[l]{$d^I$}
\Vertex(60,0){2}
\end{picture}
&
\raisebox{35\unitlength}{
\begin{minipage}{5cm}
\lefteqn{
{1\over \sqrt{2}} Y_d^I Z_H^{1i} \gamma_5
}
\end{minipage}
}
\end{tabular}

\begin{tabular}{ll}
\begin{picture}(150,70)(0,-10)
\DashLine(60,0)(10,0){5}
\Text(0,0)[c]{$A^0_i$}
\ArrowLine(60,0)(110,0)
\Text(120,0)[c]{$u^I$}
\ArrowLine(60,50)(60,0)
\Text(65,45)[l]{$u^I$}
\Vertex(60,0){2}
\end{picture}
&
\raisebox{35\unitlength}{
\begin{minipage}{5cm}
\lefteqn{
 - {1\over \sqrt{2}} Y_u^I Z_H^{2i} \gamma_5
}
\end{minipage}
}
\end{tabular}

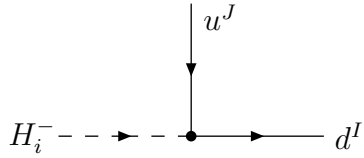
\begin{tabular}{ll}
\begin{picture}(150,70)(0,-10)
\DashArrowLine(10,0)(60,0){5}
\Text(0,0)[c]{$H^-_i$}
\ArrowLine(60,0)(110,0)
\Text(120,0)[c]{$d^I$}
\ArrowLine(60,50)(60,0)
\Text(65,45)[l]{$u^J$}
\Vertex(60,0){2}
\end{picture}
&
\raisebox{35\unitlength}{
\begin{minipage}{5cm}
\lefteqn{
i (-Y_d^I Z_H^{1i} P_L + Y_u^J Z_H^{2i} P_R) K^{JI\star} 
}
\end{minipage}
}
\end{tabular}

\lpv Self-interactions of gauge bosons.

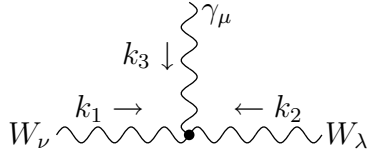
\begin{tabular}{ll}
\begin{picture}(150,70)(0,-10)
\Photon(10,0)(60,0){3}{4}
\Text(0,0)[c]{$W_{\nu}$}
\Text(30,10)[c]{$k_1\ra$}
\Photon(60,0)(110,0){3}{4}
\Text(120,0)[c]{$W_{\lambda}$}
\Text(90,10)[c]{$\la k_2$}
\Photon(60,50)(60,0){3}{4}
\Text(65,45)[l]{$\gamma_{\mu}$}
\Text(55,30)[r]{$k_3\da$}
\Vertex(60,0){2}
\end{picture}
&
\raisebox{35\unitlength}{
\begin{minipage}{5cm}
\lefteqn{
ie[g^{\nu\lambda} (k_1 -k_2)^{\mu} + g^{\lambda\mu} (k_2 -k_3)^{\nu} +
g^{\mu\nu} (k_3 - k_1)^{\lambda}]
} 
\end{minipage}
}
\end{tabular}

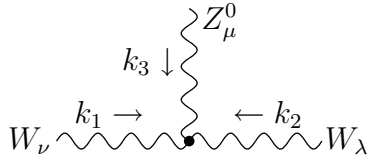
\begin{tabular}{ll}
\begin{picture}(150,70)(0,-10)
\Photon(10,0)(60,0){3}{4}
\Text(0,0)[c]{$W_{\nu}$}
\Text(30,10)[c]{$k_1\ra$}
\Photon(60,0)(110,0){3}{4}
\Text(120,0)[c]{$W_{\lambda}$}
\Text(90,10)[c]{$\la k_2$}
\Photon(60,50)(60,0){3}{4}
\Text(65,45)[l]{$Z^0_{\mu}$}
\Text(55,30)[r]{$k_3\da$}
\Vertex(60,0){2}
\end{picture}
&
\raisebox{35\unitlength}{
\begin{minipage}{5cm}
\lefteqn{
i{ec_W\over s_W}[g^{\nu\lambda} (k_1 -k_2)^{\mu} + g^{\lambda\mu} (k_2 -k_3)^{\nu} +
g^{\mu\nu} (k_3 - k_1)^{\lambda}]
} 
\end{minipage}
}
\end{tabular}

\vskip 5mm

\begin{tabular}{ll}
\begin{picture}(150,110)(0,0)
\Photon(10,70)(60,70){3}{4}
\Text(0,70)[c]{$\gamma_{\alpha}$}
\Photon(60,70)(110,70){3}{4}
\Text(120,70)[c]{$\gamma_{\beta}$}
\Photon(60,120)(60,70){3}{4}
\Text(65,115)[l]{$W_{\mu}$}
\Photon(60,70)(60,20){3}{4}
\Text(65,25)[l]{$W_{\nu}$}
\Vertex(60,70){2}
\end{picture}
&
\raisebox{68\unitlength}{
\begin{minipage}{5cm}
\lefteqn{
- ie^2 (2g^{\alpha\beta} g^{\mu\nu} - g^{\alpha\mu} g^{\beta\nu} -
g^{\alpha\nu} g^{\beta\mu})
}
\end{minipage}
}
\end{tabular}

\begin{tabular}{ll}
\begin{picture}(150,110)(0,0)
\Photon(10,70)(60,70){3}{4}
\Text(0,70)[c]{$\gamma_{\alpha}$}
\Photon(60,70)(110,70){3}{4}
\Text(120,70)[c]{$Z^0_{\beta}$}
\Photon(60,120)(60,70){3}{4}
\Text(65,115)[l]{$W_{\mu}$}
\Photon(60,70)(60,20){3}{4}
\Text(65,25)[l]{$W_{\nu}$}
\Vertex(60,70){2}
\end{picture}
&
\raisebox{68\unitlength}{
\begin{minipage}{5cm}
\lefteqn{
- {ie^2 c_W\over s_W} (2g^{\alpha\beta} g^{\mu\nu} - g^{\alpha\mu}
g^{\beta\nu} - g^{\alpha\nu} g^{\beta\mu}) 
}
\end{minipage}
}
\end{tabular}

\begin{tabular}{ll}
\begin{picture}(150,110)(0,0)
\Photon(10,70)(60,70){3}{4}
\Text(0,70)[c]{$Z^0_{\alpha}$}
\Photon(60,70)(110,70){3}{4}
\Text(120,70)[c]{$Z^0_{\beta}$}
\Photon(60,120)(60,70){3}{4}
\Text(65,115)[l]{$W_{\mu}$}
\Photon(60,70)(60,20){3}{4}
\Text(65,25)[l]{$W_{\nu}$}
\Vertex(60,70){2}
\end{picture}
&
\raisebox{68\unitlength}{
\begin{minipage}{5cm}
\lefteqn{
- {ie^2 c_W^2\over s_W^2} (2g^{\alpha\beta} g^{\mu\nu} - g^{\alpha\mu}
g^{\beta\nu} - g^{\alpha\nu} g^{\beta\mu}) 
}
\end{minipage}
}
\end{tabular}

\begin{tabular}{ll}
\begin{picture}(150,110)(0,0)
\Photon(10,70)(60,70){3}{4}
\Text(0,70)[c]{$W^+_{\alpha}$}
\Text(30,80)[c]{$\ra$}
\Photon(60,70)(110,70){3}{4}
\Text(120,70)[c]{$W^+_{\beta}$}
\Text(90,80)[c]{$\la$}
\Photon(60,120)(60,70){3}{4}
\Text(65,115)[l]{$W^-_{\mu}$}
\Text(70,90)[c]{$\da$}
\Photon(60,70)(60,20){3}{4}
\Text(65,25)[l]{$W^-_{\nu}$}
\Text(70,50)[c]{$\uparrow$}
\Vertex(60,70){2}
\end{picture}
&
\raisebox{68\unitlength}{
\begin{minipage}{5cm}
\lefteqn{
{ie^2 \over s_W^2} (2g^{\alpha\beta} g^{\mu\nu} - g^{\alpha\mu}
g^{\beta\nu} - g^{\alpha\nu} g^{\beta\mu}) 
}
\end{minipage}
}
\end{tabular}

\lpv Ghost terms.

\begin{tabular}{ll}
\begin{picture}(150,70)(0,-10)
\Photon(10,0)(60,0){3}{4}
\Text(0,0)[c]{$\gamma_{\mu}$}
\ZigZag(60,0)(110,0){3}{3}
\Text(120,0)[c]{$\eta^{\pm}$}
\Text(90,10)[c]{$\ra p$}
\ZigZag(60,50)(60,0){3}{3}
\Text(65,45)[l]{$\eta^{\pm}$}
\Text(65,25)[l]{$\da$}
\Vertex(60,0){2}
\end{picture}
&
\raisebox{35\unitlength}{
\begin{minipage}{5cm}
\lefteqn{
\pm ie p^{\mu}
}
\end{minipage}
}
\end{tabular}

\begin{tabular}{ll}
\begin{picture}(150,70)(0,-10)
\Photon(10,0)(60,0){3}{4}
\Text(0,0)[c]{$Z^0_{\mu}$}
\ZigZag(60,0)(110,0){3}{3}
\Text(120,0)[c]{$\eta^{\pm}$}
\Text(90,10)[c]{$\ra p$}
\ZigZag(60,50)(60,0){3}{3}
\Text(65,45)[l]{$\eta^{\pm}$}
\Text(65,25)[l]{$\da$}
\Vertex(60,0){2}
\end{picture}
&
\raisebox{35\unitlength}{
\begin{minipage}{5cm}
\lefteqn{
\pm {iec_W\over s_W} p^{\mu}
}
\end{minipage}
}
\end{tabular}

\begin{tabular}{ll}
\begin{picture}(150,70)(0,-10)
\Photon(10,0)(60,0){3}{4}
\Text(0,0)[c]{$W^{\pm}_{\mu}$}
\ZigZag(60,0)(110,0){3}{3}
\Text(115,0)[l]{$\eta^+(\eta_F)$}
\Text(90,10)[c]{$\ra p$}
\ZigZag(60,50)(60,0){3}{3}
\Text(65,45)[l]{$\eta_F(\eta^+)$}
\Text(65,25)[l]{$\da$}
\Vertex(60,0){2}
\end{picture}
&
\raisebox{35\unitlength}{
\begin{minipage}{5cm}
\lefteqn{
 - ie p^{\mu}
}
\end{minipage}
}
\end{tabular}

\begin{tabular}{ll}
\begin{picture}(150,70)(0,-10)
\Photon(10,0)(60,0){3}{4}
\Text(0,0)[c]{$W^{\pm}_{\mu}$}
\ZigZag(60,0)(110,0){3}{3}
\Text(115,0)[l]{$\eta_F(\eta^-)$}
\Text(90,10)[c]{$\ra p$}
\ZigZag(60,50)(60,0){3}{3}
\Text(65,45)[l]{$\eta^-(\eta_F)$}
\Text(65,25)[l]{$\da$}
\Vertex(60,0){2}
\end{picture}
&
\raisebox{35\unitlength}{
\begin{minipage}{5cm}
\lefteqn{
ie p^{\mu}
}
\end{minipage}
}
\end{tabular}

\begin{tabular}{ll}
\begin{picture}(150,70)(0,-10)
\Photon(10,0)(60,0){3}{4}
\Text(0,0)[c]{$W^{\pm}_{\mu}$}
\ZigZag(60,0)(110,0){3}{3}
\Text(115,0)[l]{$\eta^+(\eta_Z)$}
\Text(90,10)[c]{$\ra p$}
\ZigZag(60,50)(60,0){3}{3}
\Text(65,45)[l]{$\eta_Z(\eta^+)$}
\Text(65,25)[l]{$\da$}
\Vertex(60,0){2}
\end{picture}
&
\raisebox{35\unitlength}{
\begin{minipage}{5cm}
\lefteqn{
 - {ie c_W\over s_W} p^{\mu}
}
\end{minipage}
}
\end{tabular}

\begin{tabular}{ll}
\begin{picture}(150,70)(0,-10)
\Photon(10,0)(60,0){3}{4}
\Text(0,0)[c]{$W^{\pm}_{\mu}$}
\ZigZag(60,0)(110,0){3}{3}
\Text(115,0)[l]{$\eta_Z(\eta^-)$}
\Text(90,10)[c]{$\ra p$}
\ZigZag(60,50)(60,0){3}{3}
\Text(65,45)[l]{$\eta^-(\eta_Z)$}
\Text(65,25)[l]{$\da$}
\Vertex(60,0){2}
\end{picture}
&
\raisebox{35\unitlength}{
\begin{minipage}{5cm}
\lefteqn{
{ie c_W\over s_W} p^{\mu}
}
\end{minipage}
}
\end{tabular}

\begin{tabular}{ll}
\begin{picture}(150,70)(0,-10)
\DashLine(10,0)(60,0){5}
\Text(0,0)[c]{$H^0_j$}
\ZigZag(60,0)(110,0){3}{3}
\Text(115,0)[l]{$\eta_Z$}
\Text(90,10)[c]{$\ra$}
\ZigZag(60,50)(60,0){3}{3}
\Text(65,45)[l]{$\eta_Z$}
\Text(65,25)[l]{$\da$}
\Vertex(60,0){2}
\end{picture}
&
\raisebox{35\unitlength}{
\begin{minipage}{5cm}
\lefteqn{
- {ie^2 \over 4 s_W^2 c_W^2} v_i Z_R^{ij}
}
\end{minipage}
}
\end{tabular}

\begin{tabular}{ll}
\begin{picture}(150,70)(0,-10)
\DashLine(10,0)(60,0){5}
\Text(0,0)[c]{$H^0_j$}
\ZigZag(60,0)(110,0){3}{3}
\Text(115,0)[l]{$\eta^{\pm}$}
\Text(90,10)[c]{$\ra$}
\ZigZag(60,50)(60,0){3}{3}
\Text(65,45)[l]{$\eta^{\pm}$}
\Text(65,25)[l]{$\da$}
\Vertex(60,0){2}
\end{picture}
&
\raisebox{35\unitlength}{
\begin{minipage}{5cm}
\lefteqn{
- {ie^2 \over 4 s_W^2} v_i Z_R^{ij}
}
\end{minipage}
}
\end{tabular}

\begin{tabular}{ll}
\begin{picture}(150,70)(0,-10)
\DashLine(10,0)(60,0){5}
\Text(0,0)[c]{$A^0_2$}
\ZigZag(60,0)(110,0){3}{3}
\Text(115,0)[l]{$\eta^{\pm}$}
\Text(90,10)[c]{$\ra$}
\ZigZag(60,50)(60,0){3}{3}
\Text(65,45)[l]{$\eta^{\pm}$}
\Text(65,25)[l]{$\da$}
\Vertex(60,0){2}
\end{picture}
&
\raisebox{35\unitlength}{
\begin{minipage}{5cm}
\lefteqn{
\pm {e M_W \over 2s_W}
}
\end{minipage}
}
\end{tabular}

\begin{tabular}{ll}
\begin{picture}(150,70)(0,-10)
\DashArrowLine(10,0)(60,0){5}
\Text(0,0)[c]{$H^{\pm}_2$}
\ZigZag(60,0)(110,0){3}{3}
\Text(115,0)[l]{$\eta^{\pm}$}
\Text(90,10)[c]{$\ra$}
\ZigZag(60,50)(60,0){3}{3}
\Text(65,45)[l]{$\eta_F$}
\Text(65,25)[l]{$\da$}
\Vertex(60,0){2}
\end{picture}
&
\raisebox{35\unitlength}{
\begin{minipage}{5cm}
\lefteqn{
 \mp ie M_W
}
\end{minipage}
}
\end{tabular}

\begin{tabular}{ll}
\begin{picture}(150,70)(0,-10)
\DashArrowLine(10,0)(60,0){5}
\Text(0,0)[c]{$H^{\pm}_2$}
\ZigZag(60,0)(110,0){3}{3}
\Text(115,0)[l]{$\eta_Z$}
\Text(90,10)[c]{$\ra$}
\ZigZag(60,50)(60,0){3}{3}
\Text(65,45)[l]{$\eta^{\mp}$}
\Text(65,25)[l]{$\da$}
\Vertex(60,0){2}
\end{picture}
&
\raisebox{35\unitlength}{
\begin{minipage}{5cm}
\lefteqn{
{ie M_W \over 2s_Wc_W} 
}
\end{minipage}
}
\end{tabular}

\begin{tabular}{ll}
\begin{picture}(150,70)(0,-10)
\DashArrowLine(10,0)(60,0){5}
\Text(0,0)[c]{$H^{\pm}_2$}
\ZigZag(60,0)(110,0){3}{3}
\Text(115,0)[l]{$\eta^{\pm}$}
\Text(90,10)[c]{$\ra$}
\ZigZag(60,50)(60,0){3}{3}
\Text(65,45)[l]{$\eta_Z$}
\Text(65,25)[l]{$\da$}
\Vertex(60,0){2}
\end{picture}
&
\raisebox{35\unitlength}{
\begin{minipage}{5cm}
\lefteqn{
- ie M_W {c_W^2- s_W^2\over 2 s_W c_W}
}
\end{minipage}
}
\end{tabular}

\lpv Slepton-Higgs boson.

\begin{tabular}{ll}
\begin{picture}(150,70)(0,-10)
\DashLine(10,0)(60,0){5}
\Text(0,0)[c]{$H^0_i$}
\DashArrowLine(60,0)(110,0){5}
\Text(120,0)[c]{$\tilde\nu_I$}
\DashArrowLine(60,50)(60,0){5}
\Text(65,45)[l]{$\tilde\nu_J$}
\Vertex(60,0){2}
\end{picture}
&
\raisebox{35\unitlength}{
\begin{minipage}{5cm}
\lefteqn{
- {ie^2 \over 4s_W^2c_W^2} B_R^i \delta^{IJ} 
}
\end{minipage}
}
\end{tabular}

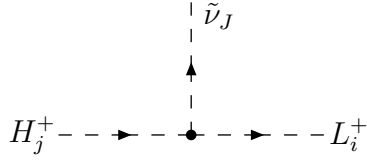
\begin{tabular}{ll}
\begin{picture}(150,70)(0,-10)
\DashArrowLine(10,0)(60,0){5}
\Text(0,0)[c]{$H^+_j$}
\DashArrowLine(60,0)(110,0){5}
\Text(120,0)[c]{$L^+_i$}
\DashArrowLine(60,0)(60,50){5}
\Text(65,45)[l]{$\tilde\nu_J$}
\Vertex(60,0){2}
\end{picture}
&
\raisebox{15\unitlength}{
\begin{minipage}{5cm}
\lefteqn{
iZ_{\nu}^{IJ\star} \left(- {\sqrt{2}e^2 \over 4s_W^2} C_R^j
Z_L^{Ii\star} + Z_H^{1j} ({v_1 \over \sqrt{2}} (Y_l^I)^2 Z_L^{Ii\star}
+ A_l^{IK\star} Z_L^{(K+3)i\star})\right.  
}
\lefteqn{ 
\left.
+ Z_{H}^{2j} (A_l^{'IK\star} Z_L^{(K+3)i\star} - \mu Y_l^{I}
Z_L^{(I+3)i\star})\right) 
}
\end{minipage}
}
\end{tabular}

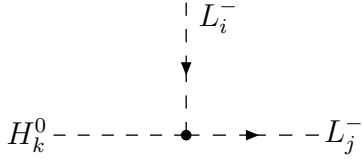
\begin{tabular}{ll}
\begin{picture}(150,70)(0,-10)
\DashLine(10,0)(60,0){5}
\Text(0,0)[c]{$H^0_k$}
\DashArrowLine(60,0)(110,0){5}
\Text(120,0)[c]{$L^-_j$}
\DashArrowLine(60,50)(60,0){5}
\Text(65,45)[l]{$L^-_i$}
\Vertex(60,0){2}
\end{picture}
&
\raisebox{-25\unitlength}{
\begin{minipage}{5cm}
\lefteqn{
i \left( {e^2 \over 2c_W^2} B_R^k \left(\delta^{ij} + {1-4s_W^2 \over
2s_W^2} Z_L^{Ii\star} Z_L^{Ij}\right) \right.  
}
\lefteqn{ 
 - (Y_l^I)^2 v_1 Z_R^{1k} (Z_L^{Ii\star} Z_L^{Ij} + Z_L^{(I+3)i\star}
 Z_L^{(I+3)j})
}
\lefteqn{
- {1\over \sqrt{2}}Z_R^{1k} (A_l^{IJ\star} Z_L^{Ij} Z_L^{(J+3)i\star}
+ A_l^{IJ} Z_L^{Ii\star} Z_L^{(J+3)j} ) 
}
\lefteqn{
 + {1\over \sqrt{2}}Z_R^{2k} (A_l^{'IJ\star} Z_L^{Ij} Z_L^{(J+3)i\star}
+ A_l^{'IJ} Z_L^{Ii\star} Z_L^{(J+3)j})
}
\lefteqn{
\left.  - {1\over \sqrt{2}} Y_l^{I} Z_R^{2k} (\mu^{\star}
Z_L^{Ii\star} Z_L^{(I+3)j} + \mu Z_L^{Ij} Z_L^{(I+3)i\star})\right)
}
\end{minipage}
}
\end{tabular}

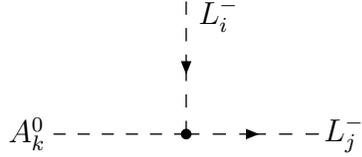
\begin{tabular}{ll}
\begin{picture}(150,70)(0,-10)
\DashLine(10,0)(60,0){5}
\Text(0,0)[c]{$A^0_k$}
\DashArrowLine(60,0)(110,0){5}
\Text(120,0)[c]{$L^-_j$}
\DashArrowLine(60,50)(60,0){5}
\Text(65,45)[l]{$L^-_i$}
\Vertex(60,0){2}
\end{picture}
&
\raisebox{0\unitlength}{
\begin{minipage}{5cm}
\lefteqn{
- {1\over \sqrt{2}} \left((A_l^{IJ\star} Z_L^{Ij} Z_L^{(J+3)i\star} -
A_l^{IJ} Z_L^{Ii\star} Z_L^{(J+3)j})Z_H^{1k} \right.
}
\lefteqn{ 
+ (A_l^{'IJ\star} Z_L^{Ij} Z_L^{(J+3)i\star} 
- A_l^{'IJ} Z_L^{Ii\star} Z_L^{(J+3)j})Z_H^{2k}
}
\lefteqn{
\left.  + Y_l^{I} (\mu^{\star} Z_L^{Ii\star} Z_L^{(I+3)j} 
- \mu Z_L^{Ij} Z_L^{(I+3)i\star} )Z_H^{2k} \right)
}
\end{minipage}
}
\end{tabular}

\vskip 5mm

\begin{tabular}{ll}
\begin{picture}(150,110)(0,0)
\DashLine(10,70)(60,70){5}
\Text(0,70)[c]{$H^0_i$}
\DashLine(60,70)(110,70){5}
\Text(120,70)[c]{$H^0_j$}
\DashArrowLine(60,120)(60,70){5}
\Text(65,115)[l]{$\tilde\nu_I$}
\DashArrowLine(60,70)(60,20){5}
\Text(65,25)[l]{$\tilde\nu_J$}
\Vertex(60,70){2}
\end{picture}
&
\raisebox{68\unitlength}{
\begin{minipage}{5cm}
\lefteqn{
- {ie^2 \over 4s_W^2c_W^2} A_R^{ij} \delta^{IJ} 
}
\end{minipage}
}
\end{tabular}

\begin{tabular}{ll}
\begin{picture}(150,110)(0,0)
\DashLine(10,70)(60,70){5}
\Text(0,70)[c]{$A^0_i$}
\DashLine(60,70)(110,70){5}
\Text(120,70)[c]{$A^0_j$}
\DashArrowLine(60,120)(60,70){5}
\Text(65,115)[l]{$\tilde\nu_I$}
\DashArrowLine(60,70)(60,20){5}
\Text(65,25)[l]{$\tilde\nu_J$}
\Vertex(60,70){2}
\end{picture}
&
\raisebox{68\unitlength}{
\begin{minipage}{5cm}
\lefteqn{
- {ie^2 \over 4s_W^2c_W^2} A_H^{ij} \delta^{IJ} 
}
\end{minipage}
}
\end{tabular}

\begin{tabular}{ll}
\begin{picture}(150,110)(0,0)
\DashArrowLine(10,70)(60,70){5}
\Text(0,70)[c]{$H^-_i$}
\DashArrowLine(60,70)(110,70){5}
\Text(120,70)[c]{$H^-_j$}
\DashArrowLine(60,120)(60,70){5}
\Text(65,115)[l]{$\tilde\nu_I$}
\DashArrowLine(60,70)(60,20){5}
\Text(65,25)[l]{$\tilde\nu_J$}
\Vertex(60,70){2}
\end{picture}
&
\raisebox{68\unitlength}{
\begin{minipage}{5cm}
\lefteqn{
iZ_{\nu}^{KJ\star} Z_{\nu}^{KI} \left(e^2 {c_W^2 - s_W^2 \over 4 s_W^2 c_W^2}
 A_H^{ij} - (Y_l^K)^2 Z_H^{1i} Z_H^{1j}\right)
}
\end{minipage}
}
\end{tabular}

\begin{tabular}{ll}
\begin{picture}(150,110)(0,0)
\DashArrowLine(10,70)(60,70){5}
\Text(0,70)[c]{$H^-_j$}
\DashLine(60,70)(110,70){5}
\Text(120,70)[c]{$H^0_k$}
\DashArrowLine(60,120)(60,70){5}
\Text(65,115)[l]{$\tilde\nu_L$}
\DashArrowLine(60,70)(60,20){5}
\Text(65,25)[l]{$L^-_i$}
\Vertex(60,70){2}
\end{picture}
&
\raisebox{68\unitlength}{
\begin{minipage}{5cm}
\lefteqn{
{i\over \sqrt{2}} Z_L^{Ii} Z_{\nu}^{IL} \left( {-e^2 \over 2s_W^2}
(Z_H^{1j} Z_R^{1k} + Z_H^{2j} Z_R^{2k}) + (Y_l^I)^2 Z_H^{1j} Z_R^{1k}
\right) 
}
\end{minipage}
}
\end{tabular}

\begin{tabular}{ll}
\begin{picture}(150,110)(0,0)
\DashArrowLine(10,70)(60,70){5}
\Text(0,70)[c]{$H^-_j$}
\DashLine(60,70)(110,70){5}
\Text(120,70)[c]{$A^0_k$}
\DashArrowLine(60,120)(60,70){5}
\Text(65,115)[l]{$\tilde\nu_L$}
\DashArrowLine(60,70)(60,20){5}
\Text(65,25)[l]{$L^-_i$}
\Vertex(60,70){2}
\end{picture}
&
\raisebox{68\unitlength}{
\begin{minipage}{5cm}
\lefteqn{
- {1\over \sqrt{2}} Z_L^{Ii} Z_{\nu}^{IL} \left({e^2 \over 2s_W^2}
A_H^{jk} - (Y_l^I)^2 Z_H^{1j} Z_H^{1k}\right) 
}
\end{minipage}
}
\end{tabular}

\begin{tabular}{ll}
\begin{picture}(150,110)(0,0)
\DashLine(10,70)(60,70){5}
\Text(0,70)[c]{$H^0_k$}
\DashLine(60,70)(110,70){5}
\Text(120,70)[c]{$H^0_l$}
\DashArrowLine(60,120)(60,70){5}
\Text(65,115)[l]{$L^-_i$}
\DashArrowLine(60,70)(60,20){5}
\Text(65,25)[l]{$L^-_j$}
\Vertex(60,70){2}
\end{picture}
&
\raisebox{58\unitlength}{
\begin{minipage}{5cm}
\lefteqn{
i \left( {e^2 \over 2c_W^2} A_R^{kl} \left(\delta^{ij} + {1-4s_W^2
\over 2s_W^2} Z_L^{Ii\star} Z_L^{Ij} \right) \right.
}
\lefteqn{
\left.  - (Y_l^I)^2 Z_R^{1k} Z_R^{1l} (Z_L^{Ii\star} Z_L^{Ij} 
+ Z_L^{(I+3)i\star} Z_L^{(I+3)j})\right) 
}
\end{minipage}
}
\end{tabular}

\begin{tabular}{ll}
\begin{picture}(150,110)(0,0)
\DashLine(10,70)(60,70){5}
\Text(0,70)[c]{$A^0_k$}
\DashLine(60,70)(110,70){5}
\Text(120,70)[c]{$A^0_l$}
\DashArrowLine(60,120)(60,70){5}
\Text(65,115)[l]{$L^-_i$}
\DashArrowLine(60,70)(60,20){5}
\Text(65,25)[l]{$L^-_j$}
\Vertex(60,70){2}
\end{picture}
&
\raisebox{58\unitlength}{
\begin{minipage}{5cm}
\lefteqn{
i \left( {e^2 \over 2c_W^2} A_H^{kl} \left(\delta^{ij} + {1-4s_W^2
\over 2s_W^2} Z_L^{Ii\star} Z_L^{Ij}\right)\right.
}
\lefteqn{
\left.  - (Y_l^I)^2 Z_H^{1k} Z_H^{1l} (Z_L^{Ii\star} Z_L^{Ij} 
+ Z_L^{(I+3)i\star} Z_L^{(I+3)j}) \right) 
}
\end{minipage}
}
\end{tabular}

\begin{tabular}{ll}
\begin{picture}(150,110)(0,0)
\DashArrowLine(10,70)(60,70){5}
\Text(0,70)[c]{$H^-_k$}
\DashArrowLine(60,70)(110,70){5}
\Text(120,70)[c]{$H^-_l$}
\DashArrowLine(60,120)(60,70){5}
\Text(65,115)[l]{$L^-_i$}
\DashArrowLine(60,70)(60,20){5}
\Text(65,25)[l]{$L^-_j$}
\Vertex(60,70){2}
\end{picture}
&
\raisebox{58\unitlength}{
\begin{minipage}{5cm}
\lefteqn{
i \left( {e^2 \over 2c_W^2} A_H^{kl} \left(\delta^{ij} - {1+2s_W^2
\over 2s_W^2} Z_L^{Ii\star} Z_L^{Ij} \right) \right.
}
\lefteqn{
\left.  - (Y_l^I)^2 Z_H^{1k} Z_H^{1l} Z_L^{(I+3)i\star} 
Z_L^{(I+3)j} \right)
}
\end{minipage}
}
\end{tabular}

\lpv Interactions of the squarks and the Higgs particles.

\begin{tabular}{ll}
\begin{picture}(150,70)(0,-10)
\DashLine(10,0)(60,0){5}
\Text(0,0)[c]{$H^0_k$}
\DashArrowLine(60,0)(110,0){5}
\Text(120,0)[c]{$U^-_j$}
\DashArrowLine(60,50)(60,0){5}
\Text(65,45)[l]{$U^-_i$}
\Vertex(60,0){2}
\end{picture}
&
\raisebox{-25\unitlength}{
\begin{minipage}{5cm}
\lefteqn{
i \left( {-e^2 \over 3c_W^2} B_R^k \left(\delta^{ij} + {3-8s_W^2 \over
4s_W^2} Z_U^{Ii\star} Z_U^{Ij}\right)\right.
} 
\lefteqn{
 - v_2 (Y_u^I)^2 Z_R^{2k} (Z_U^{Ii\star} Z_U^{Ij} + Z_U^{(I+3)i\star}
 Z_U^{(I+3)j})
}
\lefteqn{
+ {1\over \sqrt{2}} Z_R^{2k} (A_u^{IJ\star} Z_U^{Ii\star} Z_L^{(J+3)j}
+ A_u^{IJ} Z_U^{Ij} Z_U^{(J+3)i\star})
}
\lefteqn{
+ {1\over \sqrt{2}} Z_R^{1k} (A_u^{'IJ\star} Z_U^{Ii\star}
Z_U^{(J+3)j} + A_u^{'IJ} Z_U^{Ij} Z_U^{(J+3)i\star}) 
}
\lefteqn{
\left.  + {1\over \sqrt{2}} Y_u^I Z_R^{1k} (\mu^{\star} Z_U^{Ij} 
Z_U^{(I+3)i\star} + \mu Z_U^{Ii\star} Z_U^{(I+3)j} )\right)
}
\end{minipage}
}
\end{tabular}

\begin{tabular}{ll}
\begin{picture}(150,70)(0,-10)
\DashLine(10,0)(60,0){5}
\Text(0,0)[c]{$A^0_k$}
\DashArrowLine(60,0)(110,0){5}
\Text(120,0)[c]{$U^-_j$}
\DashArrowLine(60,50)(60,0){5}
\Text(65,45)[l]{$U^-_i$}
\Vertex(60,0){2}
\end{picture}
&
\raisebox{0\unitlength}{
\begin{minipage}{5cm}
\lefteqn{
- {1\over \sqrt{2}} \left(Y_u^I (\mu Z_U^{Ii\star} Z_U^{(I+3)j} -
\mu^{\star} Z_U^{Ij} Z_U^{(I+3)i\star}) Z_H^{1k}\right.
} 
\lefteqn{
+ (A_u^{IJ} Z_U^{Ij} Z_U^{(J+3)i\star} - A_u^{IJ\star} Z_U^{Ii\star}
Z_U^{(J+3)j}) Z_H^{2k}
}
\lefteqn{
\left.  + (A_u^{'IJ\star} Z_U^{Ii\star} Z_U^{(J+3)j} - A_u^{'IJ}
Z_U^{Ij} Z_U^{(J+3)i\star}) Z_H^{1k} \right)
}
\end{minipage}
}
\end{tabular}

\begin{tabular}{ll}
\begin{picture}(150,70)(0,-10)
\DashLine(10,0)(60,0){5}
\Text(0,0)[c]{$H^0_k$}
\DashArrowLine(60,0)(110,0){5}
\Text(120,0)[c]{$D^-_j$}
\DashArrowLine(60,50)(60,0){5}
\Text(65,45)[l]{$D^-_i$}
\Vertex(60,0){2}
\end{picture}
&
\raisebox{-25\unitlength}{
\begin{minipage}{5cm}
\lefteqn{
i \left( {e^2 \over 6c_W^2} B_R^k \left(\delta^{ij} + {3-4s_W^2 \over
2s_W^2} Z_D^{Ii\star} Z_D^{Ij}\right) \right.
} 
\lefteqn{
- v_1 (Y_d^I)^2 Z_R^{1k} (Z_D^{Ii\star} Z_D^{Ij} + Z_D^{(I+3)i\star}
Z_D^{(I+3)j})
}
\lefteqn{
- {1\over \sqrt{2}} Z_R^{1k} (A_d^{IJ\star} Z_D^{Ij} Z_D^{(J+3)i\star}
+ A_d^{IJ} Z_D^{Ii\star} Z_D^{(J+3)j})
}
\lefteqn{
+ {1\over \sqrt{2}} Z_R^{2k} (A_d^{'IJ\star} Z_D^{Ij} Z_D^{(J+3)i\star}
+ A_d^{'IJ} Z_D^{Ii\star} Z_D^{(J+3)j})
}
\lefteqn{
\left.  - {1\over \sqrt{2}} Y_d^I Z_R^{2k} (\mu^{\star} Z_D^{Ii\star}
Z_D^{(I+3)j} + \mu Z_D^{Ij} Z_D^{(I+3)i\star})\right)
}
\end{minipage}
}
\end{tabular}

\begin{tabular}{ll}
\begin{picture}(150,70)(0,-10)
\DashLine(10,0)(60,0){5}
\Text(0,0)[c]{$A^0_k$}
\DashArrowLine(60,0)(110,0){5}
\Text(120,0)[c]{$D^-_j$}
\DashArrowLine(60,50)(60,0){5}
\Text(65,45)[l]{$D^-_i$}
\Vertex(60,0){2}
\end{picture}
&
\raisebox{0\unitlength}{
\begin{minipage}{5cm}
\lefteqn{
- {1\over \sqrt{2}} \left((A_d^{IJ\star} Z_D^{Ij} Z_D^{(J+3)i\star} 
- A_d^{IJ} Z_D^{Ii\star} Z_D^{(J+3)j}) Z_H^{1k}\right. 
}
\lefteqn{
+ (A_d^{'IJ\star} Z_D^{Ij} Z_D^{(J+3)i\star} - A_d^{'IJ} Z_D^{Ii\star}
Z_D^{(J+3)j}) Z_H^{2k}
}
\lefteqn{
\left. + Y_d^I (\mu^{\star} Z_D^{Ii\star} Z_D^{(I+3)j} - \mu Z_D^{Ij}
Z_D^{(I+3)i\star}) Z_H^{2k} \right) 
}
\end{minipage}
}
\end{tabular}

\begin{tabular}{ll}
\begin{picture}(150,70)(0,-10)
\DashArrowLine(10,0)(60,0){5}
\Text(0,0)[c]{$H^+_k$}
\DashArrowLine(60,0)(110,0){5}
\Text(120,0)[c]{$D^+_j$}
\DashArrowLine(60,50)(60,0){5}
\Text(65,45)[l]{$U^-_i$}
\Vertex(60,0){2}
\end{picture}
&
\raisebox{-25\unitlength}{
\begin{minipage}{5cm}
\lefteqn{
i \left[{1\over \sqrt{2}} \left( {-e^2 \over 2s_W^2} v_l Z_H^{lk} +
v_1 (Y_d^I)^2 Z_H^{1k} + v_2 (Y_u^J)^2 Z_H^{2k} \right) K^{JI}
Z_D^{Ij\star} Z_U^{Ji\star} \right.  
}
\lefteqn{
- {\sqrt{2}  M_W s_W \over e} \delta^{1k} Y_u^J Y_d^I K^{JI}
Z_D^{(I+3)j\star} Z_U^{(J+3)i\star}
}
\lefteqn{
+ [Z_H^{1k} \mu^{\star} Y_u^J K^{JI} + (Z_H^{1k} A_u^{'KJ} - Z_H^{2k}
A_u^{KJ}) K^{KI}] Z_U^{(J+3)i\star} Z_D^{Ij\star} 
}
\lefteqn{
\left. + [(Z_H^{1k} A_d^{KI\star}  + Z_H^{2k} A_d^{'KI\star}) K^{JK} 
- Z_H^{2k} \mu Y_d^I K^{JI}] Z_U^{Ji\star} Z_D^{(I+3)j\star}
\right]
}
\end{minipage}
}
\end{tabular}

\vskip 5mm

\begin{tabular}{ll}
\begin{picture}(150,110)(0,0)
\DashLine(10,70)(60,70){5}
\Text(0,70)[c]{$H^0_k$}
\DashLine(60,70)(110,70){5}
\Text(120,70)[c]{$H^0_l$}
\DashArrowLine(60,120)(60,70){5}
\Text(65,115)[l]{$U^-_i$}
\DashArrowLine(60,70)(60,20){5}
\Text(65,25)[l]{$U^-_j$}
\Vertex(60,70){2}
\end{picture}
&
\raisebox{58\unitlength}{
\begin{minipage}{5cm}
\lefteqn{
i \left({-e^2 \over 3c_W^2} A_R^{kl} \left(\delta^{ij} + {3-8s_W^2 \over
4s_W^2} Z_U^{Ii\star} Z_U^{Ij}\right)\right.
}
\lefteqn{
\left.  - (Y_u^I)^2 Z_R^{2k} Z_R^{2l} (Z_U^{Ii\star} Z_U^{Ij} 
+ Z_U^{(I+3)i\star} Z_U^{(I+3)j} )\right)
}
\end{minipage}
}
\end{tabular}

\begin{tabular}{ll}
\begin{picture}(150,110)(0,0)
\DashLine(10,70)(60,70){5}
\Text(0,70)[c]{$A^0_k$}
\DashLine(60,70)(110,70){5}
\Text(120,70)[c]{$A^0_l$}
\DashArrowLine(60,120)(60,70){5}
\Text(65,115)[l]{$U^-_i$}
\DashArrowLine(60,70)(60,20){5}
\Text(65,25)[l]{$U^-_j$}
\Vertex(60,70){2}
\end{picture}
&
\raisebox{58\unitlength}{
\begin{minipage}{5cm}
\lefteqn{
i \left( {-e^2 \over 3c_W^2} A_H^{kl} \left(\delta^{ij} + {3-8s_W^2 \over
4s_W^2} Z_U^{Ii\star} Z_U^{Ij}\right) \right.
}
\lefteqn{
\left.  - (Y_u^I)^2 Z_H^{2k} Z_H^{2l} (Z_U^{Ii\star} Z_U^{Ij} 
+ Z_U^{(I+3)i\star} Z_U^{(I+3)j})\right) 
}
\end{minipage}
}
\end{tabular}

\begin{tabular}{ll}
\begin{picture}(150,110)(0,0)
\DashArrowLine(10,70)(60,70){5}
\Text(0,70)[c]{$H^-_k$}
\DashArrowLine(60,70)(110,70){5}
\Text(120,70)[c]{$H^-_l$}
\DashArrowLine(60,120)(60,70){5}
\Text(65,115)[l]{$U^-_i$}
\DashArrowLine(60,70)(60,20){5}
\Text(65,25)[l]{$U^-_j$}
\Vertex(60,70){2}
\end{picture}
&
\raisebox{48\unitlength}{
\begin{minipage}{5cm}
\lefteqn{
i \left({-e^2 \over 3c_W^2} A_H^{kl} \left(\delta^{ij} - {3+2s_W^2 \over
4s_W^2} Z_U^{Ii\star} Z_U^{Ij}\right)\right. 
}
\lefteqn{
- (Y_u^I)^2 Z_H^{2k} Z_H^{2l} Z_U^{(I+3)i\star} Z_U^{(I+3)j} 
}
\lefteqn{
\left. - (Y_d^I)^2 Z_H^{1k} Z_H^{1l} K^{JI\star} K^{KI} Z_U^{Ki\star}
Z_U^{Jj} \right)
}
\end{minipage}
}
\end{tabular}

\begin{tabular}{ll}
\begin{picture}(150,110)(0,0)
\DashLine(10,70)(60,70){5}
\Text(0,70)[c]{$H^0_k$}
\DashLine(60,70)(110,70){5}
\Text(120,70)[c]{$H^0_l$}
\DashArrowLine(60,120)(60,70){5}
\Text(65,115)[l]{$D^-_i$}
\DashArrowLine(60,70)(60,20){5}
\Text(65,25)[l]{$D^-_j$}
\Vertex(60,70){2}
\end{picture}
&
\raisebox{58\unitlength}{
\begin{minipage}{5cm}
\lefteqn{
i \left( {e^2 \over 6c_W^2} A_R^{kl} \left(\delta^{ij} + {3 - 4s_W^2 
\over 2s_W^2} Z_D^{Ii\star} Z_D^{Ij}\right)\right. 
}
\lefteqn{
\left.  - (Y_d^I)^2 Z_R^{1k} Z_R^{1l} (Z_D^{Ii\star} Z_D^{Ij} 
+ Z_D^{(I+3)i\star} Z_D^{(I+3)j} )\right)
}
\end{minipage}
}
\end{tabular}

\begin{tabular}{ll}
\begin{picture}(150,110)(0,0)
\DashLine(10,70)(60,70){5}
\Text(0,70)[c]{$A^0_k$}
\DashLine(60,70)(110,70){5}
\Text(120,70)[c]{$A^0_l$}
\DashArrowLine(60,120)(60,70){5}
\Text(65,115)[l]{$D^-_i$}
\DashArrowLine(60,70)(60,20){5}
\Text(65,25)[l]{$D^-_j$}
\Vertex(60,70){2}
\end{picture}
&
\raisebox{58\unitlength}{
\begin{minipage}{5cm}
\lefteqn{
i \left({e^2 \over 6c_W^2} A_H^{kl} \left(\delta^{ij} + {3 - 4s_W^2
\over 2s_W^2} Z_D^{Ii\star} Z_D^{Ij} \right)\right.
}
\lefteqn{
\left.  - (Y_d^I)^2 Z_H^{1k} Z_H^{1l} (Z_D^{Ii\star} Z_D^{Ij} 
+ Z_D^{(I+3)i\star} Z_D^{(I+3)j} )\right)
}
\end{minipage}
}
\end{tabular}

\begin{tabular}{ll}
\begin{picture}(150,110)(0,0)
\DashArrowLine(10,70)(60,70){5}
\Text(0,70)[c]{$H^-_k$}
\DashArrowLine(60,70)(110,70){5}
\Text(120,70)[c]{$H^-_l$}
\DashArrowLine(60,120)(60,70){5}
\Text(65,115)[l]{$D^-_i$}
\DashArrowLine(60,70)(60,20){5}
\Text(65,25)[l]{$D^-_j$}
\Vertex(60,70){2}
\end{picture}
&
\raisebox{48\unitlength}{
\begin{minipage}{5cm}
\lefteqn{
i \left({e^2 \over 6c_W^2} A_H^{kl} \left(\delta^{ij} - {3 - 2s_W^2
\over 2s_W^2} Z_D^{Ii\star} Z_D^{Ij} \right)\right. 
}
\lefteqn{
- (Y_d^I)^2 Z_H^{1k} Z_H^{1l} Z_D^{(I+3)i\star} Z_D^{(I+3)j} 
}
\lefteqn{
\left.  -  (Y_u^K)^2 Z_H^{2k} Z_H^{2l} K^{KI\star} K^{KJ}
Z_D^{Ji\star} Z_D^{Ij} \right)
}
\end{minipage}
}
\end{tabular}

\begin{tabular}{ll}
\begin{picture}(150,110)(0,0)
\DashLine(10,70)(60,70){5}
\Text(0,70)[c]{$H^0_l$}
\DashArrowLine(110,70)(60,70){5}
\Text(120,70)[c]{$H^-_k$}
\DashArrowLine(60,120)(60,70){5}
\Text(65,115)[l]{$U^+_i$}
\DashArrowLine(60,70)(60,20){5}
\Text(65,25)[l]{$D^-_j$}
\Vertex(60,70){2}
\end{picture}
&
\raisebox{48\unitlength}{
\begin{minipage}{5cm}
\lefteqn{
{i\over \sqrt{2}} K^{JI\star} \left( {-e^2 \over 2s_W^2} (Z_H^{1k}
Z_R^{1l} + Z_H^{2k} Z_R^{2l}) Z_U^{Ji} Z_D^{Ij} \right.
}
\lefteqn{
- A_P^{lk} Y_u^J Y_d^I Z_U^{(J+3)i} Z_D^{(I+3)j} 
}
\lefteqn{
\left.  + [(Y_u^J)^2 Z_H^{2k} Z_R^{2l} + (Y_d^I)^2 Z_H^{1k} Z_R^{1l}] 
Z_U^{Ji} Z_D^{Ij} \right)
}
\end{minipage}
}
\end{tabular}

\begin{tabular}{ll}
\begin{picture}(150,110)(0,0)
\DashLine(10,70)(60,70){5}
\Text(0,70)[c]{$A^0_l$}
\DashArrowLine(110,70)(60,70){5}
\Text(120,70)[c]{$H^-_k$}
\DashArrowLine(60,120)(60,70){5}
\Text(65,115)[l]{$U^+_i$}
\DashArrowLine(60,70)(60,20){5}
\Text(65,25)[l]{$D^-_j$}
\Vertex(60,70){2}
\end{picture}
&
\raisebox{58\unitlength}{
\begin{minipage}{5cm}
\lefteqn{
- {1\over \sqrt{2}} K^{JI\star}\left( {e^2 \over 2s_W^2} A_H^{kl}
Z_U^{Ji} Z_D^{Ij} + [(Y_u^J)^2 Z_H^{2k} Z_H^{2l} \right.
}
\lefteqn{
\left.  - (Y_d^I)^2 Z_H^{1k} Z_H^{1l}] Z_U^{Ji} Z_D^{Ij} 
- \epsilon_{kl} Y_u^J Y_d^I Z_U^{(J+3)i} Z_D^{(I+3)j} \right)
}
\end{minipage}
}
\end{tabular}

\lpv Self-interactions of the Higgs particles.

\begin{tabular}{ll}
\begin{picture}(150,70)(0,-10)
\DashLine(10,0)(60,0){5}
\Text(0,0)[c]{$H^0_i$}
\DashLine(60,0)(110,0){5}
\Text(120,0)[c]{$H^0_k$}
\DashLine(60,50)(60,0){5}
\Text(65,45)[l]{$H^0_j$}
\Vertex(60,0){2}
\end{picture}
&
\raisebox{35\unitlength}{
\begin{minipage}{5cm}
\lefteqn{
- {ie^2 \over 4s_W^2c_W^2} (A_R^{ij} B_R^k + A_R^{jk} B_R^i + A_R^{ki} B_R^j)
}
\end{minipage}
}
\end{tabular}

\begin{tabular}{ll}
\begin{picture}(150,70)(0,-10)
\DashLine(10,0)(60,0){5}
\Text(0,0)[c]{$A^0_i$}
\DashLine(60,0)(110,0){5}
\Text(120,0)[c]{$A^0_j$}
\DashLine(60,50)(60,0){5}
\Text(65,45)[l]{$H^0_k$}
\Vertex(60,0){2}
\end{picture}
&
\raisebox{35\unitlength}{
\begin{minipage}{5cm}
\lefteqn{
- {ie^2 \over 4s_W^2c_W^2} A_H^{ij} B_R^k 
}
\end{minipage}
}
\end{tabular}

\begin{tabular}{ll}
\begin{picture}(150,70)(0,-10)
\DashArrowLine(10,0)(60,0){5}
\Text(0,0)[c]{$H^-_j$}
\DashArrowLine(60,0)(110,0){5}
\Text(120,0)[c]{$H^-_i$}
\DashLine(60,50)(60,0){5}
\Text(65,45)[l]{$H^0_k$}
\Vertex(60,0){2}
\end{picture}
&
\raisebox{35\unitlength}{
\begin{minipage}{5cm}
\lefteqn{
-i \left( {e^2 \over 4s_W^2c_W^2} A_H^{ij} B_R^k + {e M_W \over 2s_W}
(A_P^{kj} \delta^{1i} + A_P^{ki} \delta^{1j} )\right) 
}
\end{minipage}
}
\end{tabular}

\begin{tabular}{ll}
\begin{picture}(150,70)(0,-10)
\DashArrowLine(10,0)(60,0){5}
\Text(0,0)[c]{$H^-_j$}
\DashArrowLine(60,0)(110,0){5}
\Text(120,0)[c]{$H^-_i$}
\DashLine(60,50)(60,0){5}
\Text(65,45)[l]{$A^0_k$}
\Vertex(60,0){2}
\end{picture}
&
\raisebox{35\unitlength}{
\begin{minipage}{5cm}
\lefteqn{
- {e M_W \over 2s_W} \epsilon_{ij} \delta^{1k} 
}
\end{minipage}
}
\end{tabular}

\vskip 5mm

\begin{tabular}{ll}
\begin{picture}(150,110)(0,0)
\DashLine(10,70)(60,70){5}
\Text(0,70)[c]{$H^0_i$}
\DashLine(60,70)(110,70){5}
\Text(120,70)[c]{$H^0_k$}
\DashLine(60,120)(60,70){5}
\Text(65,115)[l]{$H^0_j$}
\DashLine(60,70)(60,20){5}
\Text(65,25)[l]{$H^0_l$}
\Vertex(60,70){2}
\end{picture}
&
\raisebox{68\unitlength}{
\begin{minipage}{5cm}
\lefteqn{
- {ie^2 \over 4s_W^2c_W^2} (A_R^{ij} A_R^{kl} +A_R^{ik} A_R^{jl} 
+ A_R^{il} A_R^{jk})
}
\end{minipage}
}
\end{tabular}

\begin{tabular}{ll}
\begin{picture}(150,110)(0,0)
\DashLine(10,70)(60,70){5}
\Text(0,70)[c]{$H^0_i$}
\DashLine(60,70)(110,70){5}
\Text(120,70)[c]{$H^0_j$}
\DashLine(60,120)(60,70){5}
\Text(65,115)[l]{$A^0_k$}
\DashLine(60,70)(60,20){5}
\Text(65,25)[l]{$A^0_l$}
\Vertex(60,70){2}
\end{picture}
&
\raisebox{68\unitlength}{
\begin{minipage}{5cm}
\lefteqn{
- {ie^2 \over 4s_W^2c_W^2} A_R^{ij} A_H^{kl}
}
\end{minipage}
}
\end{tabular}

\begin{tabular}{ll}
\begin{picture}(150,110)(0,0)
\DashLine(10,70)(60,70){5}
\Text(0,70)[c]{$A^0_i$}
\DashLine(60,70)(110,70){5}
\Text(120,70)[c]{$A^0_k$}
\DashLine(60,120)(60,70){5}
\Text(65,115)[l]{$A^0_j$}
\DashLine(60,70)(60,20){5}
\Text(65,25)[l]{$A^0_l$}
\Vertex(60,70){2}
\end{picture}
&
\raisebox{68\unitlength}{
\begin{minipage}{5cm}
\lefteqn{
- {ie^2 \over 4s_W^2c_W^2} (A_H^{ij} A_H^{kl} +A_H^{ik} A_H^{jl} 
+ A_H^{il} A_H^{jk})
}
\end{minipage}
}
\end{tabular}

\begin{tabular}{ll}
\begin{picture}(150,110)(0,0)
\DashArrowLine(10,70)(60,70){5}
\Text(0,70)[c]{$H^-_l$}
\DashArrowLine(60,70)(110,70){5}
\Text(120,70)[c]{$H^-_k$}
\DashLine(60,120)(60,70){5}
\Text(65,115)[l]{$H^0_i$}
\DashLine(60,70)(60,20){5}
\Text(65,25)[l]{$H^0_j$}
\Vertex(60,70){2}
\end{picture}
&
\raisebox{68\unitlength}{
\begin{minipage}{5cm}
\lefteqn{
-{ie^2 \over 4 s_W^2} \left( {1 \over c_W^2} A_R^{ij} A_H^{kl} +
A_P^{ik} A_P^{jl} + A_P^{il} A_P^{jk} \right)
}
\end{minipage}
}
\end{tabular}

\begin{tabular}{ll}
\begin{picture}(150,110)(0,0)
\DashArrowLine(10,70)(60,70){5}
\Text(0,70)[c]{$H^-_l$}
\DashArrowLine(60,70)(110,70){5}
\Text(120,70)[c]{$H^-_k$}
\DashLine(60,120)(60,70){5}
\Text(65,115)[l]{$A^0_i$}
\DashLine(60,70)(60,20){5}
\Text(65,25)[l]{$A^0_j$}
\Vertex(60,70){2}
\end{picture}
&
\raisebox{68\unitlength}{
\begin{minipage}{5cm}
\lefteqn{
-{ie^2 \over 4s_W^2} \left( {1 \over c_W^2} A_H^{ij} A_H^{kl} +
\epsilon_{ik} \epsilon_{jl}+ \epsilon_{il} \epsilon_{jk}\right)
}
\end{minipage}
}
\end{tabular}

\begin{tabular}{ll}
\begin{picture}(150,110)(0,0)
\DashArrowLine(10,70)(60,70){5}
\Text(0,70)[c]{$H^-_l$}
\DashArrowLine(60,70)(110,70){5}
\Text(120,70)[c]{$H^-_k$}
\DashLine(60,120)(60,70){5}
\Text(65,115)[l]{$H^0_i$}
\DashLine(60,70)(60,20){5}
\Text(65,25)[l]{$A^0_j$}
\Vertex(60,70){2}
\end{picture}
&
\raisebox{68\unitlength}{
\begin{minipage}{5cm}
\lefteqn{
- {e^2 \over 4s_W^2} A_P^{ij} \epsilon_{kl}
}
\end{minipage}
}
\end{tabular}

\begin{tabular}{ll}
\begin{picture}(150,110)(0,0)
\DashArrowLine(10,70)(60,70){5}
\Text(0,70)[c]{$H^-_l$}
\DashArrowLine(60,70)(110,70){5}
\Text(120,70)[c]{$H^-_k$}
\DashArrowLine(60,120)(60,70){5}
\Text(65,115)[l]{$H^-_j$}
\DashArrowLine(60,70)(60,20){5}
\Text(65,25)[l]{$H^-_i$}
\Vertex(60,70){2}
\end{picture}
&
\raisebox{68\unitlength}{
\begin{minipage}{5cm}
\lefteqn{
- {ie^2 \over 4s_W^2c_W^2} (A_H^{ij} A_H^{kl} + A_H^{il} A_H^{jk})
}
\end{minipage}
}
\end{tabular}

\lpv Interactions of four sleptons or two sleptons and two squarks:

\vskip 7mm

\begin{tabular}{ll}
\begin{picture}(150,110)(0,0)
\DashArrowLine(10,70)(60,70){5}
\Text(0,70)[c]{$\tilde\nu^J$}
\DashArrowLine(60,70)(110,70){5}
\Text(120,70)[c]{$\tilde\nu^I$}
\DashArrowLine(60,120)(60,70){5}
\Text(65,115)[l]{$\tilde\nu^L$}
\DashArrowLine(60,70)(60,20){5}
\Text(65,25)[l]{$\tilde\nu^K$}
\Vertex(60,70){2}
\end{picture}
&
\raisebox{68\unitlength}{
\begin{minipage}{5cm}
\lefteqn{
- {ie^{2} \over 4s_W^2c_W^2} ( \delta^{IJ} \delta^{KL} +
\delta^{IL} \delta^{KJ} )
}
\end{minipage}
}
\end{tabular}

\begin{tabular}{ll}
\begin{picture}(150,110)(0,0)
\DashArrowLine(10,70)(60,70){5}
\Text(0,70)[c]{$L^-_i$}
\DashArrowLine(60,70)(110,70){5}
\Text(120,70)[c]{$L^-_j$}
\DashArrowLine(60,120)(60,70){5}
\Text(65,115)[l]{$\tilde\nu^I$}
\DashArrowLine(60,70)(60,20){5}
\Text(65,25)[l]{$\tilde\nu^J$}
\Vertex(60,70){2}
\end{picture}
&
\raisebox{50\unitlength}{
\begin{minipage}{5cm}
\lefteqn{
-i \left( {- e^2 \over 2c_W^2} \left(\delta^{ij} + {1-4s_W^2 \over
2s_W^2} Z_L^{Ki\star} Z_L^{Kj} \right)\delta^{IJ} \right.
}
\lefteqn{
\left.  + \left( {e^2 \over 2s_W^2} Z_L^{Ki\star} Z_L^{Lj} 
+ Y_l^K Y_l^L Z_L^{(K+3)i\star} Z_L^{(L+3)j}\right) Z_{\nu}^{LI}
Z_{\nu}^{KJ\star} \right)
}
\end{minipage}
}
\end{tabular}

\begin{tabular}{ll}
\begin{picture}(150,110)(0,0)
\DashArrowLine(10,70)(60,70){5}
\Text(0,70)[c]{$U^-_i$}
\DashArrowLine(60,70)(110,70){5}
\Text(120,70)[c]{$U^-_j$}
\DashArrowLine(60,120)(60,70){5}
\Text(65,115)[l]{$\tilde\nu^I$}
\DashArrowLine(60,70)(60,20){5}
\Text(65,25)[l]{$\tilde\nu^J$}
\Vertex(60,70){2}
\end{picture}
&
\raisebox{68\unitlength}{
\begin{minipage}{5cm}
\lefteqn{
- {ie^2 \over 3c_W^2} \left(\delta^{ij} + {3-8s_W^2 \over 4s_W^2}
Z_U^{Ki\star} Z_U^{Kj} \right)\delta^{IJ} 
}
\end{minipage}
}
\end{tabular}

\begin{tabular}{ll}
\begin{picture}(150,110)(0,0)
\DashArrowLine(10,70)(60,70){5}
\Text(0,70)[c]{$D^-_i$}
\DashArrowLine(60,70)(110,70){5}
\Text(120,70)[c]{$D^-_j$}
\DashArrowLine(60,120)(60,70){5}
\Text(65,115)[l]{$\tilde\nu^I$}
\DashArrowLine(60,70)(60,20){5}
\Text(65,25)[l]{$\tilde\nu^J$}
\Vertex(60,70){2}
\end{picture}
&
\raisebox{68\unitlength}{
\begin{minipage}{5cm}
\lefteqn{
{ie^2 \over 6c_W^2} \left(\delta^{ij} + {3-4s_W^2 \over 2s_W^2}
Z_D^{Ki\star} Z_D^{Kj} \right)\delta^{IJ}
}
\end{minipage}
}
\end{tabular}

\begin{tabular}{ll}
\begin{picture}(150,110)(0,0)
\DashArrowLine(10,70)(60,70){5}
\Text(0,70)[c]{$D^-_j$}
\DashArrowLine(60,70)(110,70){5}
\Text(120,70)[c]{$L^-_k$}
\DashArrowLine(60,120)(60,70){5}
\Text(65,115)[l]{$\tilde\nu^I$}
\DashArrowLine(60,70)(60,20){5}
\Text(65,25)[l]{$U^+_i$}
\Vertex(60,70){2}
\end{picture}
&
\raisebox{68\unitlength}{
\begin{minipage}{5cm}
\lefteqn{
- iZ_{\nu}^{JI} Z_U^{Li\star} K^{LK} \left( {e^2 \over 2s_W^2}
Z_D^{Kj\star} Z_L^{Jk} + Y_l^J Y_d^K Z_D^{(K+3)j\star} Z_L^{(J+3)k}
\right)
}
\end{minipage}
}
\end{tabular}

\begin{tabular}{ll}
\begin{picture}(150,110)(0,0)
\DashArrowLine(10,70)(60,70){5}
\Text(0,70)[c]{$L^-_j$}
\DashArrowLine(60,70)(110,70){5}
\Text(120,70)[c]{$L^-_k$}
\DashArrowLine(60,120)(60,70){5}
\Text(65,115)[l]{$L^-_i$}
\DashArrowLine(60,70)(60,20){5}
\Text(65,25)[l]{$L^-_l$}
\Vertex(60,70){2}
\end{picture}
&
\raisebox{30\unitlength}{
\begin{minipage}{5cm}
\lefteqn{
-i \left( {e^2 \over c_W^2} (\delta^{il} \delta^{jk} + \delta^{ik}
\delta^{jl}) \right.
}
\lefteqn{
- {3e^2 \over 2c_W^2}(\delta^{il} Z_L^{Ij\star} Z_L^{Ik} + \delta^{ik}
Z_L^{Ij\star} Z_L^{Il} + \delta^{jl} Z_L^{Ii\star} Z_L^{Ik} +
\delta^{jk} Z_L^{Ii\star} Z_L^{Il} ) 
}
\lefteqn{
+ {e^2(1+8s_W^2) \over 4s_W^2c_W^2} (Z_L^{Ii\star} Z_L^{Jj\star}
Z_L^{Jk} Z_L^{Il} + Z_L^{Ii\star} Z_L^{Jj\star} Z_L^{Jl} Z_L^{Ik} ) 
}
\lefteqn{
\left. + Y_l^I Y_l^J (Z_L^{(I+3)i\star} Z_L^{Jj\star}
+ Z_L^{(I+3)j\star} Z_L^{Ji\star}) (Z_L^{Il} Z_L^{(J+3)k} + Z_L^{Ik}
Z_L^{(J+3)l})\right) 
}
\end{minipage}
}
\end{tabular}

\vskip 4mm

\begin{tabular}{ll}
\begin{picture}(150,110)(0,0)
\DashArrowLine(10,70)(60,70){5}
\Text(0,70)[c]{$L^-_i$}
\DashArrowLine(60,70)(110,70){5}
\Text(120,70)[c]{$L^-_j$}
\DashArrowLine(60,120)(60,70){5}
\Text(65,115)[l]{$U^-_k$}
\DashArrowLine(60,70)(60,20){5}
\Text(65,25)[l]{$U^-_l$}
\Vertex(60,70){2}
\end{picture}
&
\raisebox{58\unitlength}{
\begin{minipage}{5cm}
\lefteqn{
{ie^2 \over 6c_W^2} \left( {3+12s_W^2 \over 2s_W^2} Z_L^{Ii\star}
Z_L^{Ij} Z_U^{Jk\star} Z_U^{Jl} \right.
}
\lefteqn{
\left.  - 6 \delta^{kl} Z_L^{Ii\star} Z_L^{Ij} - 5\delta^{ij} 
Z_U^{Ik\star} Z_U^{Il} + 4\delta^{ij} \delta^{kl} \right) 
}
\end{minipage}
}
\end{tabular}

\begin{tabular}{ll}
\begin{picture}(150,110)(0,0)
\DashArrowLine(10,70)(60,70){5}
\Text(0,70)[c]{$L^-_i$}
\DashArrowLine(60,70)(110,70){5}
\Text(120,70)[c]{$L^-_j$}
\DashArrowLine(60,120)(60,70){5}
\Text(65,115)[l]{$D^-_k$}
\DashArrowLine(60,70)(60,20){5}
\Text(65,25)[l]{$D^-_l$}
\Vertex(60,70){2}
\end{picture}
&
\raisebox{50\unitlength}{
\begin{minipage}{5cm}
\lefteqn{
i \left[ {e^2 \over 6c_W^2} \left( -{3 \over 2s_W^2} Z_L^{Ii\star}
Z_L^{Ij} Z_D^{Jk\star} Z_D^{Jl} \right. \right.  
}
\lefteqn{
\left.  +3\delta^{kl} Z_L^{Ii\star} Z_L^{Ij} + \delta^{ij} 
Z_D^{Ik\star} Z_D^{Il} - 2\delta^{ij} \delta^{kl} \right) 
}
\lefteqn{
\left.  - Y_l^I Y_d^J (Z_L^{(I+3)i\star} Z_L^{Ij} Z_D^{Jk\star} 
Z_D^{(J+3)l} + Z_L^{Ii\star} Z_L^{(I+3)j} Z_D^{(J+3)k\star} Z_D^{Jl} )
\right]
}
\end{minipage}
}
\end{tabular}

\lpv Strong interactions of the quarks, squarks, gluons and gluinos.

\input qcd


\begin{thebibliography}{99}

\bibitem{PRD41} J. Rosiek, {\em Phys. Rev.}  {\bf D41} (1990) 3464.

\bibitem{CPR}  P. Chankowski, S. Pokorski, J. Rosiek {\sl Nucl. Phys.}
{\bf B423} (1994) 437; {\sl Nucl. Phys.} {\bf B423} (1994) 497;
V. Driesen, W. Hollik, J. Rosiek, {\em Zeit. Phys.} {\bf C71} (1996)
259; S. Heinemeyer, W. Hollik, J. Rosiek, G. Weiglein, {\em
Eur.Phys.J.} {\bf C19} (2001) 535-546.

\bibitem{DELTAR} P. Chankowski et al. {\sl Nucl. Phys.} {\bf B417} (1994) 101.

\bibitem{POROSA} S. Pokorski, J. Rosiek, C. Savoy {\em Nucl. Phys.} 
{\bf B570} (2000) 81-116.

\bibitem{MIPORO} M. Misiak, S. Pokorski, J. Rosiek {\sl
``Supersymmetry and the FCNC effects''}, {\em hep-ph}/9703442,
published in {\em Adv.Ser.Direct.High Energy Phys.} {\bf 15} (1998)
795-828.

\bibitem{BCRS} A. Buras, P. Chankowski, J. Rosiek, {\L}. S{\l}awianowska,
{\em Nucl.Phys.} {\bf B619} (2001) 434-466.

\bibitem{ref1} H. E. Haber and  G. L. Kane, {\em Phys. Rep.}  
{\bf 117} (1985) 75.

\bibitem{ref2} J. F. Gunion and  H. E. Haber, {\em Nucl. Phys.} 
{\bf B272} (1986) 1.

\bibitem{ref4}  P. Fayet and J. Iliopoulos, {\em Phys. Lett.}  
{\bf 51B} (1974) 461.

\bibitem{ref5}  L. Girardello and M. T. Grisaru, {\em Nucl. Phys.}
{\bf B194} (1982) 65.

\bibitem{rparity} H. Dreiner, {\sl ``An Introduction to explicit
R-parity violation''}, {\em hep-ph}/9707435, published in *Kane,
G.L. (ed.): Perspectives on supersymmetry* 462-479.

\bibitem{QCD_BREAK} J. A. Casas, {\sl ``Charge and color
breaking''}, {\em hep-ph}/9707475, published in *Kane, G.L. (ed.):
Perspectives on supersymmetry* 378-401.

\bibitem{denner} A. Denner , H. Eck, O. Hahn, J. Kublbeck, {\em
Phys.Lett.} {\bf B291} (1992) 278-280; A. Denner, H. Eck, O. Hahn,
J. Kublbeck, {\em Nucl.Phys.} {\bf B387} (1992) 467-484.

\end{thebibliography}
\end{document}